\documentclass[fleqn,usenatbib]{mnras}

\usepackage{newtxtext,newtxmath}

\usepackage[T1]{fontenc}

\DeclareRobustCommand{\VAN}[3]{#2}
\let\VANthebibliography\thebibliography
\def\thebibliography{\DeclareRobustCommand{\VAN}[3]{##3}\VANthebibliography}

\usepackage{graphicx}	
\usepackage{amsmath}	
\usepackage{xcolor}
\usepackage{enumitem}
\usepackage[T1]{fontenc}
\usepackage{textcomp}

\title[Prospects for DM observations in dSphs with CTAO]{Prospects for dark matter observations in dwarf spheroidal galaxies with the Cherenkov Telescope Array Observatory}

\author[K.~Abe et al.]{\parbox{\textwidth}{\raggedright\small%
  K.~Abe,$^{\ref{AFFIL::JapanUTokai}}$
  S.~Abe,$^{\ref{AFFIL::JapanUTokyoICRR}}$
  J.~Abhir,$^{\ref{AFFIL::SwitzerlandETHZurich}}$
  A.~Abhishek,$^{\ref{AFFIL::ItalyUSienaandINFN}}$
  F.~Acero,$^{\ref{AFFIL::FranceCEAIRFUDAp},\ref{AFFIL::SpainFSLACIRLCNRSIAC}}$
  A.~Acharyya,$^{\ref{AFFIL::USAUAlabamaTuscaloosa}}$
  R.~Adam,$^{\ref{AFFIL::FranceOCotedAzur},\ref{AFFIL::FranceLLREcolePolytechnique}}$
  A.~Aguasca-Cabot,$^{\ref{AFFIL::SpainICCUB}}$
  I.~Agudo,$^{\ref{AFFIL::SpainIAACSIC}}$
  A.~Aguirre-Santaella,$^{\ref{AFFIL::UnitedKingdomICCUDurham}}$
  J.~Alfaro,$^{\ref{AFFIL::ChileUPontificiaCatolicadeChile}}$
  R.~Alfaro,$^{\ref{AFFIL::MexicoUNAMMexico}}$
  C.~Alispach,$^{\ref{AFFIL::SwitzerlandUGenevaDPNC}}$
  R.~Alves~Batista,$^{\ref{AFFIL::FranceLPNHEUSorbonne}}$
  J.-P.~Amans,$^{\ref{AFFIL::FranceObservatoiredeParis}}$
  E.~Amato,$^{\ref{AFFIL::ItalyOArcetri}}$
  G.~Ambrosi,$^{\ref{AFFIL::ItalyUPerugiaandINFN}}$
  D.~Ambrosino,$^{\ref{AFFIL::ItalyINFNNapoli},\ref{AFFIL::ItalyUNapoli}}$
  F.~Ambrosino,$^{\ref{AFFIL::ItalyORoma}}$
  L.~Angel,$^{\ref{AFFIL::BrazilURioGrandedoNorteIIP},\ref{AFFIL::BrazilURioGrandedoNortePhys}}$
  L.~A.~Antonelli,$^{\ref{AFFIL::ItalyORoma}}$
  C.~Aramo,$^{\ref{AFFIL::ItalyINFNNapoli}}$
  C.~Arcaro,$^{\ref{AFFIL::ItalyINFNPadova}}$
  K.~Asano,$^{\ref{AFFIL::JapanUTokyoICRR}}$
  Y.~Ascasibar,$^{\ref{AFFIL::SpainIFTUAMCSIC}}$
  L.~Augusto~Stuani,$^{\ref{AFFIL::BrazilIFUSaoPaulo}}$
  M.~Backes,$^{\ref{AFFIL::NamibiaUNamibia},\ref{AFFIL::SouthAfricaNWU}}$
  C.~Balazs,$^{\ref{AFFIL::AustraliaUMonash}}$
  M.~Balbo,$^{\ref{AFFIL::SwitzerlandUGenevaDPNC}}$
  A.~Baquero~Larriva,$^{\ref{AFFIL::SpainUCMAltasEnergias},\ref{AFFIL::EcuadorUAzuay}}$
  V.~Barbosa~Martins,$^{\ref{AFFIL::GermanyUBochumPhysAst}}$
  J.~A.~Barrio,$^{\ref{AFFIL::SpainUCMAltasEnergias}}$
  L.~Barrios-Jim\'enez,$^{\ref{AFFIL::SpainIAC}}$
  C.~Bartolini,$^{\ref{AFFIL::ItalyUTrento},\ref{AFFIL::ItalyUandINFNBari}}$
  P.~I.~Batista,$^{\ref{AFFIL::GermanyUErlangenECAP}}$
  I.~Batkovi\'c,$^{\ref{AFFIL::ItalyUPadovaandINFN}}$
  R.~Batzofin,$^{\ref{AFFIL::GermanyUPotsdam}}$
  J.~Becerra~Gonz\'alez,$^{\ref{AFFIL::SpainIAC}}$
  G.~Beck,$^{\ref{AFFIL::SouthAfricaUWitwatersrand}}$
  J.~Becker~Tjus,$^{\ref{AFFIL::GermanyUBochum}}$
  W.~Benbow,$^{\ref{AFFIL::USACfAHarvardSmithsonian}}$
  D.~Berge,$^{\ref{AFFIL::GermanyUBerlin},\ref{AFFIL::GermanyDESY}}$
  E.~Bernardini,$^{\ref{AFFIL::ItalyUPadovaandINFN}}$
  J.~Bernete,$^{\ref{AFFIL::SpainCIEMAT}}$
  A.~Berti,$^{\ref{AFFIL::GermanyMPP}}$
  B.~Bertucci,$^{\ref{AFFIL::ItalyUPerugiaandINFN}}$
  V.~Beshley,$^{\ref{AFFIL::UkraineIAPMMLviv}}$
  P.~Bhattacharjee,$^{\ref{AFFIL::SloveniaUNovaGoricaCAC},\ref{AFFIL::FranceLAPPUSavoieMontBlanc}}$
  S.~Bhattacharyya,$^{\ref{AFFIL::SloveniaUNovaGoricaCAC}}$
  C.~Bigongiari,$^{\ref{AFFIL::ItalyORoma},\ref{AFFIL::ASISpaceScienceDataCenter}}$
  A.~Biland,$^{\ref{AFFIL::SwitzerlandETHZurich}}$
  E.~Bissaldi,$^{\ref{AFFIL::ItalyPolitecnicoBari},\ref{AFFIL::ItalyINFNBari}}$
  O.~Blanch,$^{\ref{AFFIL::SpainIFAEBIST}}$
  J.~Blazek,$^{\ref{AFFIL::CzechRepublicFZU}}$
  G.~Bonnoli,$^{\ref{AFFIL::ItalyOBrera},\ref{AFFIL::ItalyINFNPisa}}$
  A.~Bonollo,$^{\ref{AFFIL::ItalyIUSSPaviaINAF},\ref{AFFIL::ItalyUTrento}}$
  Z.~Bosnjak,$^{\ref{AFFIL::CroatiaUZagreb}}$
  E.~Bottacini,$^{\ref{AFFIL::ItalyUPadovaandINFN}}$
  M.~B\"ottcher,$^{\ref{AFFIL::SouthAfricaNWU}}$
  T.~Bringmann,$^{\ref{AFFIL::NorwayUOslo}}$
  E.~Bronzini,$^{\ref{AFFIL::ItalyOASBologna}}$
  R.~Brose,$^{\ref{AFFIL::IrelandDCU},\ref{AFFIL::IrelandDIAS}}$
  G.~Brunelli,$^{\ref{AFFIL::ItalyOASBologna}}$
  J.~Buces~S\'aez,$^{\ref{AFFIL::SpainUCMAltasEnergias}}$
  M.~Bunse,$^{\ref{AFFIL::LamarrInstituteGermany}}$
  L.~Burmistrov,$^{\ref{AFFIL::SwitzerlandUGenevaDPNC}}$
  M.~Burton,$^{\ref{AFFIL::UnitedKingdomArmaghObservatoryandPlanetarium},\ref{AFFIL::AustraliaUNewSouthWales}}$
  P.~G.~Calisse,$^{\ref{AFFIL::GermanyCTAOHeidelberg}}$
  A.~Campoy-Ordaz,$^{\ref{AFFIL::SpainUABandCERESIEEC}}$
  B.~K.~Cantlay,$^{\ref{AFFIL::ThailandUKasetsart},\ref{AFFIL::ThailandNARIT}}$
  G.~Capasso,$^{\ref{AFFIL::ItalyOCapodimonte}}$
  A.~Caproni,$^{\ref{AFFIL::BrazilUCidadeSPaulo}}$
  R.~Capuzzo-Dolcetta,$^{\ref{AFFIL::ItalyORoma},\ref{AFFIL::ItalyURomaSapienza}}$
  P.~Caraveo,$^{\ref{AFFIL::ItalyIASFMilano}}$
  S.~Caroff,$^{\ref{AFFIL::FranceLAPPUSavoieMontBlanc}}$
  R.~Carosi,$^{\ref{AFFIL::ItalyINFNPisa}}$
  E.~Carquin,$^{\ref{AFFIL::ChileUTecnicaFedericoSantaMaria}}$
  M.-S.~Carrasco,$^{\ref{AFFIL::FranceCPPMUAixMarseille}}$
  E.~Cascone,$^{\ref{AFFIL::ItalyOCapodimonte}}$
  G.~Castignani,$^{\ref{AFFIL::ItalyOASBologna}}$
  A.~J.~Castro-Tirado,$^{\ref{AFFIL::SpainIAACSIC}}$
  D.~Cerasole,$^{\ref{AFFIL::ItalyUandINFNBari}}$
  M.~Cerruti,$^{\ref{AFFIL::FranceAPCUParisCite}}$
  A.~Cervi\~no~Cort{\'\i}nez,$^{\ref{AFFIL::SpainUCMAltasEnergias}}$
  P.~M.~Chadwick,$^{\ref{AFFIL::UnitedKingdomUDurham}}$
  Y.~Chai,$^{\ref{AFFIL::JapanUTokyoICRR}}$
  S.~Chaty,$^{\ref{AFFIL::FranceAPCUParisCite}}$
  A.~W.~Chen,$^{\ref{AFFIL::SouthAfricaUWitwatersrand}}$
  Y.~Chen,$^{\ref{AFFIL::USAUCLA}}$
  M.~Chernyakova,$^{\ref{AFFIL::IrelandDCU}}$
  A.~Chiavassa,$^{\ref{AFFIL::ItalyINFNTorino},\ref{AFFIL::ItalyUTorino}}$
  G.~Chon,$^{\ref{AFFIL::GermanyMPP}}$
  J.~Chudoba,$^{\ref{AFFIL::CzechRepublicFZU}}$
  G.~M.~Cicciari,$^{\ref{AFFIL::ItalyUPalermo},\ref{AFFIL::ItalyINFNCatania}}$
  A.~Cifuentes,$^{\ref{AFFIL::SpainCIEMAT}}$
  C.~H.~Coimbra~Araujo,$^{\ref{AFFIL::BrazilUFPR}}$
  M.~Colapietro,$^{\ref{AFFIL::ItalyOCapodimonte}}$
  V.~Conforti,$^{\ref{AFFIL::ItalyOASBologna}}$
  J.~L.~Contreras,$^{\ref{AFFIL::SpainUCMAltasEnergias}}$
  B.~Cornejo,$^{\ref{AFFIL::FranceCEAIRFUDPhP}}$
  J.~Cortina,$^{\ref{AFFIL::SpainCIEMAT}}$
  A.~Costa,$^{\ref{AFFIL::ItalyOCatania}}$
  H.~Costantini,$^{\ref{AFFIL::FranceCPPMUAixMarseille}}$
  G.~Cotter,$^{\ref{AFFIL::UnitedKingdomUOxford}}$
  P.~Cristofari,$^{\ref{AFFIL::FranceObservatoiredeParis}}$
  O.~Cuevas,$^{\ref{AFFIL::ChileUdeValparaiso}}$
  Z.~Curtis-Ginsberg,$^{\ref{AFFIL::USAUWisconsin}}$
  G.~D'Amico,$^{\ref{AFFIL::SpainIFAEBIST}}$
  F.~D'Ammando,$^{\ref{AFFIL::ItalyRadioastronomiaINAF}}$
  L.~David,$^{\ref{AFFIL::GermanyDESY}}$
  F.~Dazzi,$^{\ref{AFFIL::ItalyINAF}}$
  A.~De~Angelis,$^{\ref{AFFIL::ItalyUPadovaandINFN}}$
  M.~de~Bony~de~Lavergne,$^{\ref{AFFIL::FranceCPPMUAixMarseille}}$
  F.~De~Frondat~Laadim,$^{\ref{AFFIL::FranceObservatoiredeParis}}$
  E.~M.~de~Gouveia~Dal~Pino,$^{\ref{AFFIL::BrazilIAGUSaoPaulo}}$
  B.~De~Lotto,$^{\ref{AFFIL::ItalyUUdineandINFNTrieste}}$
  M.~de~Naurois,$^{\ref{AFFIL::FranceLLREcolePolytechnique}}$
  G.~De~Palma,$^{\ref{AFFIL::ItalyUandINFNBari}}$
  V.~de~Souza,$^{\ref{AFFIL::BrazilIFSCUSaoPaulo}}$
  R.~Del~Burgo,$^{\ref{AFFIL::ItalyINFNNapoli}}$
  L.~del~Peral,$^{\ref{AFFIL::SpainUAlcala}}$
  M.~V.~del~Valle,$^{\ref{AFFIL::BrazilIAGUSaoPaulo}}$
  C.~Delgado,$^{\ref{AFFIL::SpainCIEMAT}}$
  D.~della~Volpe,$^{\ref{AFFIL::SwitzerlandUGenevaDPNC}}$
  D.~Depaoli,$^{\ref{AFFIL::GermanyMPIK}}$
  T.~Di~Girolamo,$^{\ref{AFFIL::ItalyUNapoli},\ref{AFFIL::ItalyINFNNapoli}}$
  A.~Di~Piano,$^{\ref{AFFIL::ItalyOASBologna}}$
  F.~Di~Pierro,$^{\ref{AFFIL::ItalyINFNTorino}}$
  R.~Di~Tria,$^{\ref{AFFIL::ItalyINFNBari}}$
  L.~Di~Venere,$^{\ref{AFFIL::ItalyINFNBari}}$
  S.~Diebold,$^{\ref{AFFIL::GermanyIAAT}}$
  A.~Dinesh,$^{\ref{AFFIL::SpainUCMAltasEnergias}}$
  J.~Djuvsland,$^{\ref{AFFIL::NorwayUBergen}}$
  A.~Donini,$^{\ref{AFFIL::ItalyORoma}}$
  J.~D\"orner,$^{\ref{AFFIL::GermanyUBochum}}$
  M.~Doro,$^{\ref{AFFIL::ItalyUPadovaandINFN}}$\ref{CONTACTAUTHOR::1}
  C.~Duangchan,$^{\ref{AFFIL::GermanyUPotsdam}}$
  L.~Ducci,$^{\ref{AFFIL::GermanyIAAT}}$
  V.~V.~Dwarkadas,$^{\ref{AFFIL::USAUChicagoDAA}}$
  J.~Ebr,$^{\ref{AFFIL::CzechRepublicFZU}}$
  C.~Eckner,$^{\ref{AFFIL::SloveniaUNovaGoricaCAC}}$
  K.~Egberts,$^{\ref{AFFIL::GermanyUPotsdam}}$
  D.~Els\"asser,$^{\ref{AFFIL::GermanyUDortmundTU}}$
  G.~Emery,$^{\ref{AFFIL::SpainIAACSIC}}$
  C.~Esca\~nuela~Nieves,$^{\ref{AFFIL::GermanyMPIK}}$
  P.~Escarate,$^{\ref{AFFIL::ChileEscIngElec}}$
  M.~Escobar~Godoy,$^{\ref{AFFIL::USASCIPP}}$
  P.~Esposito,$^{\ref{AFFIL::ItalyIUSSPaviaINAF},\ref{AFFIL::ItalyIASFMilano}}$
  S.~Ettori,$^{\ref{AFFIL::ItalyOASBologna}}$
  D.~Falceta-Goncalves,$^{\ref{AFFIL::BrazilEACHUSaoPaulo}}$
  E.~Fedorova,$^{\ref{AFFIL::ItalyORoma},\ref{AFFIL::UkraineAstObsofUKyiv}}$
  S.~Fegan,$^{\ref{AFFIL::FranceLLREcolePolytechnique}}$
  Q.~Feng,$^{\ref{AFFIL::USAUUtah}}$
  G.~Ferrand,$^{\ref{AFFIL::CanadaUManitoba},\ref{AFFIL::JapanRIKEN}}$
  F.~Ferrarotto,$^{\ref{AFFIL::ItalyINFNRomaLaSapienza}}$
  E.~Fiandrini,$^{\ref{AFFIL::ItalyUPerugiaandINFN}}$
  A.~Fiasson,$^{\ref{AFFIL::FranceLAPPUSavoieMontBlanc}}$
  M.~Filipovic,$^{\ref{AFFIL::AustraliaUWesternSydney}}$
  V.~Fioretti,$^{\ref{AFFIL::ItalyOASBologna}}$
  L.~Foffano,$^{\ref{AFFIL::ItalyIAPS}}$
  G.~Fontaine,$^{\ref{AFFIL::FranceLLREcolePolytechnique}}$
  F.~Fr{\'\i}as~Garc{\'\i}a-Lago,$^{\ref{AFFIL::SpainIAC}}$
  Y.~Fukazawa,$^{\ref{AFFIL::JapanUHiroshima}}$
  Y.~Fukui,$^{\ref{AFFIL::JapanUNagoya}}$
  A.~Furniss,$^{\ref{AFFIL::USASCIPP}}$
  G.~Galanti,$^{\ref{AFFIL::ItalyIASFMilano}}$
  G.~Galaz,$^{\ref{AFFIL::ChileUPontificiaCatolicadeChile}}$
  S.~Gallozzi,$^{\ref{AFFIL::ItalyORoma}}$
  V.~Gammaldi,$^{\ref{AFFIL::SpainUniversidadSanPabloCEU},\ref{AFFIL::SpainIFTUAMCSIC}}$
  S.~Garc{\'\i}a~Soto,$^{\ref{AFFIL::SpainCIEMAT}}$
  M.~Garczarczyk,$^{\ref{AFFIL::GermanyDESY}}$
  C.~Gasbarra,$^{\ref{AFFIL::ItalyORoma},\ref{AFFIL::ItalyINFNRomaTorVergata}}$
  D.~Gasparrini,$^{\ref{AFFIL::ItalyINFNRomaTorVergata}}$
  M.~Gaug,$^{\ref{AFFIL::SpainUABandCERESIEEC}}$
  S.~Germani,$^{\ref{AFFIL::ItalyUPerugiaandINFN}}$
  A.~Ghalumyan,$^{\ref{AFFIL::ArmeniaNSLAlikhanyan}}$
  F.~Gianotti,$^{\ref{AFFIL::ItalyOASBologna}}$
  J.~G.~Giesbrecht~Formiga~Paiva,$^{\ref{AFFIL::BrazilCBPF}}$
  N.~Giglietto,$^{\ref{AFFIL::ItalyPolitecnicoBari},\ref{AFFIL::ItalyINFNBari}}$
  F.~Giordano,$^{\ref{AFFIL::ItalyUandINFNBari}}$
  R.~Giuffrida,$^{\ref{AFFIL::FranceCEAIRFUDAp},\ref{AFFIL::ItalyOPalermo}}$
  J.-F.~Glicenstein,$^{\ref{AFFIL::FranceCEAIRFUDPhP}}$
  J.~Glombitza,$^{\ref{AFFIL::GermanyUErlangenECAP}}$
  P.~Goldoni,$^{\ref{AFFIL::FranceAPCUParisCite}}$
  J.~M.~Gonz\'alez,$^{\ref{AFFIL::ChileUAndresBello}}$
  J.~Goulart~Coelho,$^{\ref{AFFIL::BrazilUFES}}$
  T.~Gradetzke,$^{\ref{AFFIL::GermanyUDortmundTU}}$
  J.~Granot,$^{\ref{AFFIL::IsraelOpenUniversityofIsrael},\ref{AFFIL::USAGWUWashingtonDC}}$
  L.~Gr\'eaux,$^{\ref{AFFIL::GermanyUBochumPhysAst}}$
  D.~Green,$^{\ref{AFFIL::ItalyCTAOBologna}}$
  J.~G.~Green,$^{\ref{AFFIL::GermanyMPP}}$
  G.~Grolleron,$^{\ref{AFFIL::FranceLAPPUSavoieMontBlanc}}$
  L.~M.~V.~Guedes,$^{\ref{AFFIL::BrazilURioGrandedoNortePhys}}$
  O.~Gueta,$^{\ref{AFFIL::GermanyCTAOSDMC}}$
  D.~Hadasch,$^{\ref{AFFIL::SpainICECSIC}}$
  P.~Hamal,$^{\ref{AFFIL::CzechRepublicFZU}}$
  W.~Hanlon,$^{\ref{AFFIL::USACfAHarvardSmithsonian}}$
  S.~Hara,$^{\ref{AFFIL::JapanUYamanashiGakuin}}$
  V.~M.~Harvey,$^{\ref{AFFIL::AustraliaUAdelaide}}$
  T.~Hassan,$^{\ref{AFFIL::SpainCIEMAT}}$
  K.~Hayashi,$^{\ref{AFFIL::JapanNITSendaiHirose},\ref{AFFIL::JapanUTokyoICRR}}$
  B.~He{\ss},$^{\ref{AFFIL::GermanyIAAT}}$
  L.~Heckmann,$^{\ref{AFFIL::FranceAPCUParisCite},\ref{AFFIL::GermanyMPP}}$
  M.~Heller,$^{\ref{AFFIL::SwitzerlandUGenevaDPNC}}$
  N.~Hiroshima,$^{\ref{AFFIL::JapanUTokyoICRR},\ref{AFFIL::JapanUYokohamaNational}}$
  B.~Hnatyk,$^{\ref{AFFIL::UkraineAstObsofUKyiv}}$
  R.~Hnatyk,$^{\ref{AFFIL::UkraineAstObsofUKyiv}}$
  D.~Hoffmann,$^{\ref{AFFIL::FranceCPPMUAixMarseille}}$
  W.~Hofmann,$^{\ref{AFFIL::GermanyMPIK}}$
  D.~Horan,$^{\ref{AFFIL::FranceLLREcolePolytechnique}}$
  P.~Horvath,$^{\ref{AFFIL::CzechRepublicUOlomouc}}$
  D.~Hrupec,$^{\ref{AFFIL::CroatiaUOsijek}}$
  S.~Hussain,$^{\ref{AFFIL::SloveniaUNovaGoricaCAC}}$
  M.~Iarlori,$^{\ref{AFFIL::ItalyCETEMPSandUandINFNAquila}}$
  T.~Inada,$^{\ref{AFFIL::JapanUTokyoICRR},\ref{AFFIL::JapanUKyushu}}$
  F.~Incardona,$^{\ref{AFFIL::ItalyOCatania}}$
  S.~Inoue,$^{\ref{AFFIL::JapanUChiba},\ref{AFFIL::JapanUTokyoICRR}}$
  Y.~Inoue,$^{\ref{AFFIL::JapanUOsaka},\ref{AFFIL::JapanRIKEN}}$
  F.~Iocco,$^{\ref{AFFIL::ItalyUNapoli},\ref{AFFIL::ItalyINFNNapoli}}$
  A.~Iuliano,$^{\ref{AFFIL::ItalyINFNNapoli}}$
  ~Jahanvi,$^{\ref{AFFIL::ItalyUUdineandINFNTrieste}}$
  M.~Jamrozy,$^{\ref{AFFIL::PolandUJagiellonian}}$
  P.~Janecek,$^{\ref{AFFIL::CzechRepublicFZU}}$
  F.~Jankowsky,$^{\ref{AFFIL::GermanyLSW}}$
  C.~Jarnot,$^{\ref{AFFIL::FranceIRAPUToulouse}}$
  I.~Jaroschewski,$^{\ref{AFFIL::FranceCEAIRFUDPhP}}$
  P.~Jean,$^{\ref{AFFIL::FranceIRAPUToulouse}}$
  I.~Jim\'enez~Mart{\'\i}nez,$^{\ref{AFFIL::GermanyMPP}}$
  W.~Jin,$^{\ref{AFFIL::USAUCLA}}$
  J.~Jurysek,$^{\ref{AFFIL::CzechRepublicFZU}}$
  O.~Kalekin,$^{\ref{AFFIL::GermanyUErlangenECAP}}$
  V.~Karas,$^{\ref{AFFIL::CzechRepublicASU}}$
  J.~Kataoka,$^{\ref{AFFIL::JapanUWaseda}}$
  S.~Kaufmann,$^{\ref{AFFIL::UnitedKingdomUDurham}}$
  D.~Kazanas,$^{\ref{AFFIL::GreeceUThessaloniki}}$
  T.~Keita,$^{\ref{AFFIL::FranceCEAIRFUDAp}}$
  D.~Kerszberg,$^{\ref{AFFIL::FranceLPNHEUSorbonne}}$
  D.~B.~Kieda,$^{\ref{AFFIL::USAUUtah}}$
  R.~Kissmann,$^{\ref{AFFIL::AustriaUInnsbruck}}$
  W.~Klu\'zniak,$^{\ref{AFFIL::PolandNicolausCopernicusAstronomicalCenter}}$
  K.~Kohri,$^{\ref{AFFIL::JapanNAOJ},\ref{AFFIL::JapanKEK}}$
  D.~Kolar,$^{\ref{AFFIL::SloveniaUNovaGoricaCAC}}$
  N.~Komin,$^{\ref{AFFIL::SouthAfricaUWitwatersrand}}$
  P.~Kornecki,$^{\ref{AFFIL::SpainIAACSIC}}$
  G.~Kowal,$^{\ref{AFFIL::BrazilEACHUSaoPaulo}}$
  H.~Kubo,$^{\ref{AFFIL::JapanUTokyoICRR}}$
  J.~Kushida,$^{\ref{AFFIL::JapanUTokai}}$
  A.~La~Barbera,$^{\ref{AFFIL::ItalyIASFPalermo}}$
  N.~La~Palombara,$^{\ref{AFFIL::ItalyIASFMilano}}$
  B.~Lacave,$^{\ref{AFFIL::SwitzerlandUGenevaDPNC}}$
  M.~L\'ainez,$^{\ref{AFFIL::SpainUCMAltasEnergias}}$
  A.~Lamastra,$^{\ref{AFFIL::ItalyORoma}}$
  J.~Lapington,$^{\ref{AFFIL::UnitedKingdomULeicester}}$
  S.~Lazarevi\'c,$^{\ref{AFFIL::AustraliaUWesternSydney}}$
  J.-P.~Lenain,$^{\ref{AFFIL::FranceLPNHEUSorbonne}}$
  F.~Leone,$^{\ref{AFFIL::ItalyUCatania}}$
  E.~Leonora,$^{\ref{AFFIL::ItalyINFNCatania}}$
  Y.~Li,$^{\ref{AFFIL::NetherlandsUAmsterdam}}$
  E.~Lindfors,$^{\ref{AFFIL::FinlandUTurku}}$
  M.~Linhoff,$^{\ref{AFFIL::GermanyCTAOSDMC},\ref{AFFIL::GermanyUDortmundTU}}$
  S.~Lombardi,$^{\ref{AFFIL::ItalyORoma},\ref{AFFIL::ASISpaceScienceDataCenter}}$
  F.~Longo,$^{\ref{AFFIL::ItalyUandINFNTrieste}}$
  R.~L\'opez-Coto,$^{\ref{AFFIL::SpainIAACSIC}}$
  M.~L\'opez-Moya,$^{\ref{AFFIL::SpainUCMAltasEnergias}}$
  A.~L\'opez-Oramas,$^{\ref{AFFIL::SpainIAC}}$
  S.~Loporchio,$^{\ref{AFFIL::ItalyUandINFNBari}}$
  J.~Lozano~Bahilo,$^{\ref{AFFIL::SpainUAlcala}}$
  H.~Luciani,$^{\ref{AFFIL::ItalyUandINFNTrieste}}$
  P.~L.~Luque-Escamilla,$^{\ref{AFFIL::SpainUJaen}}$
  E.~Lyard,$^{\ref{AFFIL::SwitzerlandUGenevaISDC}}$
  O.~Macias,$^{\ref{AFFIL::NetherlandsUAmsterdam}}$
  J.~Mackey,$^{\ref{AFFIL::IrelandDIAS}}$
  P.~Majumdar,$^{\ref{AFFIL::IndiaSahaInstitute}}$
  D.~Malyshev,$^{\ref{AFFIL::GermanyIAAT}}$
  D.~Mandat,$^{\ref{AFFIL::CzechRepublicFZU}}$
  S.~Mangano,$^{\ref{AFFIL::SpainCIEMAT}}$
  G.~Manic\`o,$^{\ref{AFFIL::ItalyINFNCatania},\ref{AFFIL::ItalyUCatania}}$
  A.~Marchetti,$^{\ref{AFFIL::ItalyOBrera}}$
  M.~Mariotti,$^{\ref{AFFIL::ItalyUPadovaandINFN}}$
  I.~M\'arquez,$^{\ref{AFFIL::SpainIAACSIC}}$
  G.~Marsella,$^{\ref{AFFIL::ItalyUPalermo},\ref{AFFIL::ItalyINFNCatania}}$
  D.~Mart{\'\i}n~Dom{\'\i}nguez,$^{\ref{AFFIL::SpainUCMAltasEnergias}}$
  G.~A.~Mart{\'\i}nez,$^{\ref{AFFIL::SpainCIEMAT}}$
  M.~Mart{\'\i}nez,$^{\ref{AFFIL::SpainIFAEBIST}}$
  O.~Martinez,$^{\ref{AFFIL::SpainReyJuanCarlosUniversity},\ref{AFFIL::SpainUCMElectronica}}$
  D.~Mazin,$^{\ref{AFFIL::JapanUTokyoICRR},\ref{AFFIL::GermanyMPP}}$
  A.~J.~T.~S.~Mello,$^{\ref{AFFIL::BrazilUFPR},\ref{AFFIL::BrazilUTFPR}}$
  J.~M\'endez~Gallego,$^{\ref{AFFIL::SpainIAACSIC}}$
  S.~Menon,$^{\ref{AFFIL::ItalyORoma},\ref{AFFIL::ItalyURomaTorVegata}}$
  S.~Mereghetti,$^{\ref{AFFIL::ItalyIASFMilano}}$
  D.~M.-A.~Meyer,$^{\ref{AFFIL::SpainICECSIC}}$
  M.~Meyer,$^{\ref{AFFIL::GermanyUHamburg}}$
  D.~Miceli,$^{\ref{AFFIL::ItalyINFNPadova}}$
  M.~Miceli,$^{\ref{AFFIL::ItalyUPalermo},\ref{AFFIL::ItalyOPalermo}}$
  M.~Michailidis,$^{\ref{AFFIL::USAStanford},\ref{AFFIL::USASLAC}}$
  T.~Miener,$^{\ref{AFFIL::SwitzerlandUGenevaDPNC}}$
  J.~M.~Miranda,$^{\ref{AFFIL::SpainUCMElectronica},\ref{AFFIL::SpainIPARCOSInstitute}}$
  A.~Mitchell,$^{\ref{AFFIL::GermanyUErlangenECAP}}$
  M.~Molero,$^{\ref{AFFIL::SpainIAC}}$
  C.~Molfese,$^{\ref{AFFIL::ItalyINAF}}$
  M.~Molina~Delicado,$^{\ref{AFFIL::SpainUCMAltasEnergias}}$
  E.~Molina,$^{\ref{AFFIL::SpainIAC}}$
  T.~Montaruli,$^{\ref{AFFIL::SwitzerlandUGenevaDPNC}}$
  A.~Moralejo,$^{\ref{AFFIL::SpainIFAEBIST}}$
  A.~Moreno~Ramos,$^{\ref{AFFIL::SpainUCMElectronica}}$
  A.~Morselli,$^{\ref{AFFIL::ItalyINFNRomaTorVergata}}$\ref{CONTACTAUTHOR::2}
  E.~Moulin,$^{\ref{AFFIL::FranceCEAIRFUDPhP}}$
  V.~Moya~Zamanillo,$^{\ref{AFFIL::SpainUCMAltasEnergias}}$
  K.~Munari,$^{\ref{AFFIL::ItalyOCatania}}$
  T.~Murach,$^{\ref{AFFIL::GermanyDESY}}$
  A.~Muraczewski,$^{\ref{AFFIL::PolandNicolausCopernicusAstronomicalCenter}}$
  H.~Muraishi,$^{\ref{AFFIL::JapanUKitasato}}$
  T.~Nakamori,$^{\ref{AFFIL::JapanUYamagata}}$
  R.~Nemmen,$^{\ref{AFFIL::BrazilIAGUSaoPaulo},\ref{AFFIL::USAStanford}}$
  J.~P.~Neto,$^{\ref{AFFIL::BrazilURioGrandedoNortePhys},\ref{AFFIL::BrazilURioGrandedoNorteIIP}}$
  J.~Niemiec,$^{\ref{AFFIL::PolandIFJ}}$
  D.~Nieto,$^{\ref{AFFIL::SpainUCMAltasEnergias}}$
  M.~Nievas~Rosillo,$^{\ref{AFFIL::SpainIAC}}$
  M.~Niko{\l}ajuk,$^{\ref{AFFIL::PolandUBiaystok}}$
  K.~Nishijima,$^{\ref{AFFIL::JapanUKanagawaFacofEngineering}}$
  K.~Noda,$^{\ref{AFFIL::JapanUChiba},\ref{AFFIL::JapanUTokyoICRR}}$
  D.~Nosek,$^{\ref{AFFIL::CzechRepublicUPrague}}$
  V.~Novotny,$^{\ref{AFFIL::CzechRepublicUPrague}}$
  S.~Nozaki,$^{\ref{AFFIL::JapanUTokyoICRR}}$
  A.~Okumura,$^{\ref{AFFIL::JapanUNagoyaISEE},\ref{AFFIL::JapanUNagoyaKMI}}$
  J.-F.~Olive,$^{\ref{AFFIL::FranceIRAPUToulouse}}$
  R.~A.~Ong,$^{\ref{AFFIL::USAUCLA}}$
  R.~Orito,$^{\ref{AFFIL::JapanUTokushima}}$
  M.~Orlandini,$^{\ref{AFFIL::ItalyOASBologna}}$
  E.~Orlando,$^{\ref{AFFIL::ItalyUandINFNTrieste}}$
  S.~Orlando,$^{\ref{AFFIL::ItalyOPalermo}}$
  J.~Otero-Santos,$^{\ref{AFFIL::SpainIAACSIC}}$
  I.~Oya,$^{\ref{AFFIL::SpainCIEMAT}}$
  M.~Ozlati~Moghadam,$^{\ref{AFFIL::GermanyUPotsdam}}$
  A.~Pagliaro,$^{\ref{AFFIL::ItalyIASFPalermo}}$
  M.~Palatiello,$^{\ref{AFFIL::ItalyORoma}}$
  A.~Pandey,$^{\ref{AFFIL::USAUUtah}}$
  G.~Panebianco,$^{\ref{AFFIL::ItalyOASBologna}}$
  D.~Paneque,$^{\ref{AFFIL::GermanyMPP}}$
  F.~R.~Pantaleo,$^{\ref{AFFIL::ItalyINFNBari},\ref{AFFIL::ItalyPolitecnicoBari}}$
  R.~Paoletti,$^{\ref{AFFIL::ItalyUSienaandINFN}}$
  J.~M.~Paredes,$^{\ref{AFFIL::SpainICCUB}}$
  N.~Parmiggiani,$^{\ref{AFFIL::ItalyOASBologna}}$
  B.~Patricelli,$^{\ref{AFFIL::ItalyORoma},\ref{AFFIL::ItalyUPisa}}$
  M.~Pech,$^{\ref{AFFIL::CzechRepublicFZU}}$
  M.~Pecimotika,$^{\ref{AFFIL::SpainIFAEBIST}}$
  M.~Peresano,$^{\ref{AFFIL::GermanyMPP}}$
  A.~P\'erez~Aguilera,$^{\ref{AFFIL::SpainUCMAltasEnergias}}$
  J.~P\'erez-Romero,$^{\ref{AFFIL::SloveniaUNovaGoricaCAC}}$
  G.~Peron,$^{\ref{AFFIL::ItalyOArcetri}}$
  F.~Perrotta,$^{\ref{AFFIL::ItalyOCapodimonte}}$
  M.~Persic,$^{\ref{AFFIL::ItalyOPadova},\ref{AFFIL::ItalyOandINFNTrieste}}$
  O.~Petruk,$^{\ref{AFFIL::UkraineIAPMMLviv},\ref{AFFIL::ItalyOPalermo}}$
  F.~Pfeifle,$^{\ref{AFFIL::GermanyUWurzburg}}$
  E.~Pietropaolo,$^{\ref{AFFIL::ItalyUandINFNAquila}}$
  L.~Pinchbeck,$^{\ref{AFFIL::AustraliaUMonash}}$
  F.~Pintore,$^{\ref{AFFIL::ItalyIASFPalermo}}$
  G.~Pirola,$^{\ref{AFFIL::GermanyMPP}}$
  C.~Pittori,$^{\ref{AFFIL::ItalyORoma}}$
  F.~Podobnik,$^{\ref{AFFIL::ItalyUSienaandINFN}}$
  M.~Pohl,$^{\ref{AFFIL::GermanyUPotsdam}}$
  V.~Poireau,$^{\ref{AFFIL::FranceLAPPUSavoieMontBlanc}}$
  V.~Pollet,$^{\ref{AFFIL::FranceLAPPUSavoieMontBlanc}}$
  G.~Ponti,$^{\ref{AFFIL::ItalyOBrera}}$
  C.~Pozo~Gonz\'alez,$^{\ref{AFFIL::SpainIAACSIC}}$
  E.~Prandini,$^{\ref{AFFIL::ItalyUPadovaandINFN}}$
  G.~Principe,$^{\ref{AFFIL::ItalyUandINFNTrieste}}$
  M.~Prouza,$^{\ref{AFFIL::CzechRepublicFZU}}$
  E.~Pueschel,$^{\ref{AFFIL::GermanyUBochumPhysAst}}$
  G.~P\"uhlhofer,$^{\ref{AFFIL::GermanyIAAT}}$
  M.~L.~Pumo,$^{\ref{AFFIL::ItalyUCatania},\ref{AFFIL::ItalyINFNCatania}}$
  A.~Quirrenbach,$^{\ref{AFFIL::GermanyLSW}}$
  S.~Rain\`o,$^{\ref{AFFIL::ItalyUandINFNBari}}$
  R.~Rando,$^{\ref{AFFIL::ItalyUPadovaandINFN}}$
  S.~Recchia,$^{\ref{AFFIL::ItalyOArcetri}}$
  M.~Regeard,$^{\ref{AFFIL::FranceAPCUParisCite}}$
  A.~Reimer,$^{\ref{AFFIL::AustriaUInnsbruck}}$
  O.~Reimer,$^{\ref{AFFIL::AustriaUInnsbruck}}$
  I.~Reis,$^{\ref{AFFIL::BrazilIFSCUSaoPaulo},\ref{AFFIL::FranceCEAIRFUDPhP}}$
  A.~Reisenegger,$^{\ref{AFFIL::ChileUPontificiaCatolicadeChile},\ref{AFFIL::ChileUMCE}}$
  W.~Rhode,$^{\ref{AFFIL::GermanyUDortmundTU}}$
  M.~Rib\'o,$^{\ref{AFFIL::SpainICCUB}}$
  C.~Ricci,$^{\ref{AFFIL::SwitzerlandUGenevaISDC},\ref{AFFIL::ChileUniversidadDiegoPortales}}$
  T.~Richtler,$^{\ref{AFFIL::ChileUdeConcepcion}}$
  J.~Rico,$^{\ref{AFFIL::SpainIFAEBIST}}$
  F.~Rieger,$^{\ref{AFFIL::GermanyMPIK}}$
  L.~Riitano,$^{\ref{AFFIL::USAUWisconsin}}$
  C.~R{\'\i}os,$^{}$
  V.~Rizi,$^{\ref{AFFIL::ItalyUandINFNAquila}}$
  E.~Roache,$^{\ref{AFFIL::USACfAHarvardSmithsonian}}$
  G.~Rodr{\'i}guez-Fern{\'a}ndez,$^{\ref{AFFIL::ItalyINFNRomaTorVergata}}$\ref{CONTACTAUTHOR::3}
  M.~D.~Rodr{\'\i}guez~Fr{\'\i}as,$^{\ref{AFFIL::SpainUAlcala}}$
  J.~J.~Rodr{\'\i}guez-V\'azquez,$^{\ref{AFFIL::SpainCIEMAT}}$
  P.~Romano,$^{\ref{AFFIL::ItalyOBrera}}$
  G.~Romeo,$^{\ref{AFFIL::ItalyOCatania}}$
  J.~Rosado,$^{\ref{AFFIL::SpainUCMAltasEnergias}}$
  A.~Rosales~de~Leon,$^{\ref{AFFIL::FranceLPNHEUSorbonne}}$
  G.~Rowell,$^{\ref{AFFIL::AustraliaUAdelaide}}$
  A.~Roy,$^{\ref{AFFIL::JapanHASC}}$
  B.~Rudak,$^{\ref{AFFIL::PolandNicolausCopernicusAstronomicalCenter}}$
  A.~Ruina,$^{\ref{AFFIL::ItalyINFNPadova}}$
  E.~Ruiz-Velasco,$^{\ref{AFFIL::FranceLAPPUSavoieMontBlanc}}$
  F.~S.~Queiroz,$^{\ref{AFFIL::BrazilURioGrandedoNorteIIP},\ref{AFFIL::BrazilURioGrandedoNortePhys}}$
  I.~Sadeh,$^{\ref{AFFIL::GermanyDESY}}$
  L.~Saha,$^{\ref{AFFIL::USACfAHarvardSmithsonian}}$
  H.~Salzmann,$^{\ref{AFFIL::USASCIPP}}$
  M.~S\'anchez-Conde,$^{\ref{AFFIL::SpainIFTUAMCSIC}}$
  P.~Sangiorgi,$^{\ref{AFFIL::ItalyIASFPalermo}}$
  H.~Sano,$^{\ref{AFFIL::JapanUGifu},\ref{AFFIL::JapanUTokyoICRR}}$
  R.~Santos-Lima,$^{\ref{AFFIL::BrazilIAGUSaoPaulo}}$
  V.~Sapienza,$^{\ref{AFFIL::ItalyOPalermo},\ref{AFFIL::ItalyUPalermo}}$
  S.~Sarkar,$^{\ref{AFFIL::UnitedKingdomUOxford}}$
  F.~G.~Saturni,$^{\ref{AFFIL::ItalyORoma}}$\ref{CONTACTAUTHOR::4}
  S.~Savarese,$^{\ref{AFFIL::ItalyOCapodimonte}}$
  V.~Savchenko,$^{\ref{AFFIL::SwitzerlandEPFLausanne}}$
  A.~Scherer,$^{\ref{AFFIL::ChileUniversidaddeSantiagodeChile}}$
  F.~Schiavone,$^{\ref{AFFIL::ItalyUandINFNBari}}$
  P.~Schipani,$^{\ref{AFFIL::ItalyOCapodimonte}}$
  F.~Schussler,$^{\ref{AFFIL::FranceCEAIRFUDPhP}}$
  D.~Sengupta,$^{\ref{AFFIL::AustraliaUNewSouthWales}}$
  O.~Sergijenko,$^{\ref{AFFIL::UkraineAstObsofUKyiv},\ref{AFFIL::UkraineObsNASUkraine},\ref{AFFIL::PolandAGHCracowSTC}}$
  V.~Sguera,$^{\ref{AFFIL::ItalyOASBologna}}$
  H.~Siejkowski,$^{\ref{AFFIL::PolandCYFRONETAGH}}$
  G.~Silvestri,$^{\ref{AFFIL::ItalyUPadovaandINFN}}$
  A.~Simongini,$^{\ref{AFFIL::ItalyINAF},\ref{AFFIL::ItalyURomaTorVegata}}$
  C.~Siqueira,$^{\ref{AFFIL::BrazilNationalObservatoryRiodeJaneiro}}$
  V.~Sliusar,$^{\ref{AFFIL::SwitzerlandUGenevaISDC}}$
  I.~Sofia,$^{\ref{AFFIL::ItalyINFNTorino},\ref{AFFIL::GermanyMPIK}}$
  H.~Sol,$^{\ref{AFFIL::FranceObservatoiredeParis}}$
  D.~Spiga,$^{\ref{AFFIL::ItalyOBrera}}$
  S.~Spinello,$^{\ref{AFFIL::ItalyOCatania}}$
  A.~Stamerra,$^{\ref{AFFIL::ItalyORoma},\ref{AFFIL::ItalyCTAOBologna}}$
  S.~Stani\v{c},$^{\ref{AFFIL::SloveniaUNovaGoricaCAC}}$
  T.~Starecki,$^{\ref{AFFIL::PolandWUTElectronics}}$
  R.~Starling,$^{\ref{AFFIL::UnitedKingdomULeicester}}$
  T.~Stolarczyk,$^{\ref{AFFIL::FranceCEAIRFUDAp}}$
  Y.~Suda,$^{\ref{AFFIL::JapanUHiroshima}}$
  H.~Tajima,$^{\ref{AFFIL::JapanUNagoyaISEE},\ref{AFFIL::JapanUNagoyaKMI}}$
  M.~Takahashi,$^{\ref{AFFIL::JapanUNagoyaISEE}}$
  R.~Takeishi,$^{\ref{AFFIL::JapanUTokyoICRR}}$
  S.~J.~Tanaka,$^{\ref{AFFIL::JapanUAoyamaGakuin}}$
  L.~A.~Tejedor,$^{\ref{AFFIL::SpainUCMAltasEnergias}}$
  M.~Teshima,$^{\ref{AFFIL::GermanyMPP}}$
  V.~Testa,$^{\ref{AFFIL::ItalyORoma}}$
  W.~W.~Tian,$^{\ref{AFFIL::JapanUTokyoICRR},\ref{AFFIL::ChinaUSunYatsen}}$
  L.~Tibaldo,$^{\ref{AFFIL::FranceIRAPUToulouse}}$
  O.~Tibolla,$^{\ref{AFFIL::UnitedKingdomUDurham}}$
  S.~J.~Tingay,$^{\ref{AFFIL::AustraliaUCurtin}}$
  F.~Tombesi,$^{\ref{AFFIL::ItalyURomaTorVegata},\ref{AFFIL::ItalyINFNRomaTorVergata}}$
  D.~Tonev,$^{\ref{AFFIL::BulgariaINRNEBAS}}$
  F.~Torradeflot,$^{\ref{AFFIL::SpainPIC},\ref{AFFIL::SpainCIEMAT}}$
  D.~F.~Torres,$^{\ref{AFFIL::SpainICECSIC}}$
  N.~Tothill,$^{\ref{AFFIL::AustraliaUWesternSydney}}$
  A.~Tramacere,$^{\ref{AFFIL::SwitzerlandUGenevaISDC}}$
  P.~Travnicek,$^{\ref{AFFIL::CzechRepublicFZU}}$
  A.~Trois,$^{\ref{AFFIL::ItalyINAFCagliari}}$
  S.~Truzzi,$^{\ref{AFFIL::ItalyUSienaandINFN}}$
  A.~Tutone,$^{\ref{AFFIL::ItalyIASFPalermo}}$
  M.~Vacula,$^{\ref{AFFIL::CzechRepublicUOlomouc},\ref{AFFIL::CzechRepublicFZU}}$
  C.~van~Eldik,$^{\ref{AFFIL::GermanyUErlangenECAP}}$
  J.~Vandenbroucke,$^{\ref{AFFIL::USAUWisconsin}}$
  V.~Vassiliev,$^{\ref{AFFIL::USAUCLA}}$
  M.~V\'azquez~Acosta,$^{\ref{AFFIL::SpainIAC}}$
  M.~Vecchi,$^{\ref{AFFIL::NetherlandsUGroningen}}$
  S.~Ventura,$^{\ref{AFFIL::ItalyINFNPisa}}$
  S.~Vercellone,$^{\ref{AFFIL::ItalyOBrera}}$
  G.~Verna,$^{\ref{AFFIL::ItalyUSienaandINFN}}$
  I.~Viale,$^{\ref{AFFIL::ItalyINFNTorino}}$
  A.~Viana,$^{\ref{AFFIL::BrazilIFSCUSaoPaulo}}$
  N.~Viaux,$^{\ref{AFFIL::ChileDepFisUTecnicaFedericoSantaMaria}}$
  A.~Vigliano,$^{\ref{AFFIL::ItalyUUdineandINFNTrieste}}$
  J.~Vignatti,$^{\ref{AFFIL::ChileDepFisUTecnicaFedericoSantaMaria}}$
  C.~F.~Vigorito,$^{\ref{AFFIL::ItalyINFNTorino},\ref{AFFIL::ItalyUTorino}}$
  E.~Visentin,$^{\ref{AFFIL::ItalyINFNTorino},\ref{AFFIL::ItalyUTorino}}$
  V.~Vitale,$^{\ref{AFFIL::ItalyINFNRomaTorVergata}}$
  V.~Voitsekhovskyi,$^{\ref{AFFIL::SwitzerlandUGenevaDPNC}}$
  S.~Vorobiov,$^{\ref{AFFIL::SloveniaUNovaGoricaCAC}}$
  G.~Voutsinas,$^{\ref{AFFIL::SwitzerlandUGenevaDPNC}}$
  R.~Walter,$^{\ref{AFFIL::SwitzerlandUGenevaISDC}}$
  M.~Wechakama,$^{\ref{AFFIL::ThailandUKasetsart},\ref{AFFIL::ThailandNARIT}}$
  M.~White,$^{\ref{AFFIL::AustraliaUAdelaide}}$
  A.~Wierzcholska,$^{\ref{AFFIL::PolandIFJ}}$
  F.~Wohlleben,$^{\ref{AFFIL::GermanyMPIK}}$
  F.~Xotta,$^{\ref{AFFIL::SloveniaUNovaGoricaCAC}}$
  R.~Yamazaki,$^{\ref{AFFIL::JapanUAoyamaGakuin}}$
  Y.~Yao,$^{\ref{AFFIL::JapanUTokai}}$
  T.~Yoshikoshi,$^{\ref{AFFIL::JapanUTokyoICRR}}$
  D.~Zavrtanik,$^{\ref{AFFIL::SloveniaUNovaGoricaCAC}}$
  M.~Zavrtanik,$^{\ref{AFFIL::SloveniaUNovaGoricaCAC}}$
  A.~Zech,$^{\ref{AFFIL::FranceObservatoiredeParis}}$
  W.~Zhang,$^{\ref{AFFIL::SpainICECSIC}}$
  V.~I.~Zhdanov,$^{\ref{AFFIL::UkraineAstObsofUKyiv}}$
  M.~\v{Z}ivec,$^{\ref{AFFIL::SloveniaUNovaGoricaCAC}}$
  and J.~Zuriaga-Puig$^{\ref{AFFIL::SpainIFTUAMCSIC}}$
\newline\newline
\emph{Affiliations can be found at the end of the article}}}

\date{Accepted XXX. Received YYY; in original form ZZZ}

\pubyear{2025}

\begin{document}
\label{firstpage}
\pagerange{\pageref{firstpage}--\pageref{lastpage}}
\maketitle

\footnotetext[1]{\url{michele.doro@unipd.it} (M.~Doro)\label{CONTACTAUTHOR::1}}
\footnotetext[4]{\url{francesco.saturni@inaf.it} (F.~G.~Saturni)\label{CONTACTAUTHOR::4}} 
\footnotetext[3]{\url{gonzalo.rodriguez.fdez@gmail.com} (G.~Rodr{\'i}guez-Fern{\'a}ndez)\label{CONTACTAUTHOR::3}}
\footnotetext[2]{\url{aldo.morselli@roma2.infn.it} (A.~Morselli)\label{CONTACTAUTHOR::2}}

\begin{abstract}
The dwarf spheroidal galaxies (dSphs) orbiting the Milky Way are widely regarded as systems supported by velocity dispersion against self-gravity, and as prime targets for the search for indirect dark matter (DM) signatures in the GeV–to-TeV $\gamma$-ray range owing to their lack of astrophysical $\gamma$-ray background. We present forecasts of the sensitivity of the forthcoming Cherenkov Telescope Array Observatory (CTAO) to annihilating or decaying DM signals in these targets. An original selection of candidates is performed from the current catalogue of known objects, including both classical and ultra-faint dSphs. For each, the expected DM content is derived using the most comprehensive photometric and spectroscopic data available, within a consistent framework of analysis. This approach enables the derivation of novel astrophysical factor profiles for indirect DM searches, which are compared with results from the literature. From an initial sample of 64 dSphs, eight promising targets are identified -- Draco~I, Coma Berenices, Ursa Major~II, Ursa Minor and Willman~1 in the North, Reticulum~II, Sculptor and Sagittarius~II in the South -- for which different DM density models yield consistent expectations, leading to robust predictions. CTAO is expected to provide the strongest limits above $\sim$10 TeV, reaching velocity-averaged annihilation cross sections of $\sim$5$\times$10$^{-25}$ cm$^3$ s$^{-1}$ and decay lifetimes up to $\sim$10$^{26}$ s for combined limits. The dominant uncertainties arise from the imprecise determination of the DM content, particularly for ultra-faint dSphs. Observation strategies are proposed that optimise either deep exposures of the best candidates or diversified target selections.
\end{abstract}
 
\begin{keywords}
radiation mechanisms: non-thermal -- methods: numerical -- galaxies: dwarf -- galaxies: kinematics and dynamics -- galaxies: Local Group -- cosmology: dark matter -- gamma-rays: galaxies
\end{keywords}



\section{Introduction}
\label{sec:intro}
The existence of dark matter \citep[DM; e.g.][for an historical perspective]{Bertone:2016nfn} in our Universe is well established, but its nature is at present still unknown. Astrophysical observations suggest that the matter in the Universe consists not only of particles of the Standard Model (SM) of particle physics, but also new non-baryonic particle candidates that must be advocated~\citep[e.g.][]{Feng:2010gw}. Observations of galaxy cluster dynamics and mergers \citep[e.g.,][]{clo06,har15} and galactic rotation curves \citep[e.g.,][]{van86,lel16}, primordial light element abundances left over by the Big Bang nucleosynthesis (\citealt{PDG}; see also \citealt{ioc09}), or the spectral shape of the cosmic microwave background \citep{pla14} point to a substantial fraction -- around 85\% \citep{pla16} of the Universe's matter density -- being in a form of matter that does not interact significantly with SM particles. Numerical simulations of large-scale structures also support this conclusion, requiring non-relativistic (cold) DM in order to be consistent with observations \citep[see e.g.][and refs. therein]{cir24}.

For a long time, super-symmetry \citep[SUSY, e.g.][]{mal98} has been generally accepted as a promising extension of the SM, since it naturally solves the hierarchy problem at the same time providing a \textit{natural} DM candidate with its lightest super-symmetric particle (LSP). The LSP has the characteristics of a weakly interacting massive particle (WIMP). One prototypical WIMP particle is the lightest neutralino. The  characteristic feature of WIMPs is that they were in thermal equilibrium with ordinary matter in the early universe, but acquired an abundance from thermal freeze-out that matches the observed DM abundance. The minimal SUSY standard model \citep[MSSM; e.g.][]{bae22} is a useful framework with which to test current ideas about detection, both in DM direct- \unskip\parfillskip 0pt \par
\vfill\break
\noindent or indirect-detection experiments. This model contains many features which are expected to be universal for any WIMP model. WIMP DM candidates also appear in many non-SUSY extensions to the SM -- e.g., little Higgs models \citep[e.g.][]{Birkedal:2006fz} or universal extra dimensions \citep[e.g.][]{Hooper:2007qk}. However, simplified templates that rely only on a few free parameters, like the constrained minimal SUSY model \citep[cMSSM;][]{jun96} or the minimal super-gravity \citep[mSUGRA;][]{alv83}, are already strongly constrained with LHC data \citep{aal12,kha15,aad16}. These results hint at pushing the $\mathrm{DM}$ mass scale up to the TeV and multi-TeV range, where particles belonging to the WIMP family such as the wino \citep{Hisano:2003ec,Beneke:2016ync} or the Higgsino \citep{ark05} are plausible DM candidates.

In these frameworks, WIMPs can pair-annihilate \citep{zel80} or decay \citep{iba13}. Both scenarios would lead to the generation of SM products that in turn generate final-state photons in the $\gamma$-ray domain \citep{ber98}. Coupled with the hypothesized DM thermal production mechanisms and the limits on the age of the Universe, respectively, this would yield a DM velocity-averaged annihilation cross section $\langle \sigma_{\rm ann} v \rangle \sim 2\div3 \times 10^{-26}$ cm$^3$ s$^{-1}$ \citep[the so-called ``thermal relic'' cross section;][]{Lee77,Gondolo:1990dk,Steigman:2012nb,Smi19}, respectively a lifetime at least larger than the Hubble time for decaying DM. A debated claim for detection of $\gamma$-rays from DM interactions towards the Galactic Center (GC) direction has been made in the $1 - 60$ GeV energy range, thanks to the availability of deep {\it Fermi}-LAT data \citep{day16,ack17}; however, despite a thorough modeling of the expected DM emission from the innermost regions of the Milky Way (MW) halo \citep{cal14,cho22}, the interpretation of this GeV excess as a WIMP signal is still controversial \citep{car14,pet14,yua14,cho15,bar16,lee16,abd17,mac18}.

The potential production of $\gamma$-ray signals thus makes DM indirectly detectable through appropriate astronomical facilities, including satellite-borne pair-production experiments (e.g. \textit{Fermi}-LAT in the MeV-GeV regime), ground-based imaging atmospheric Cherenkov telescopes in the TeV regime \citep[IACTs; e.g.][]{den15} and shower front detector in the TeV-PeV regime (e.g. HAWC, LHAASO or the future SWGO). Focus of this work are IACTs, which  detect very-high energy (VHE; $>$100 GeV) $\gamma$-rays through the Cherenkov emission in particle showers generated in the Earth's atmosphere by through-going $\gamma$-rays \citep{wee89}. DM searches in VHE $\gamma$-rays have been carried out in the past decades with all of the major existing IACTs -- MAGIC, H.E.S.S. and VERITAS -- in the energy range $0.1 \div 100$ TeV \citep[see][for a recent comprehensive review of all these observations]{Doro:2021dzh,Hutten:2022hud,mon24}. The construction of the next-generation IACT array, the Cherenkov Telescope Array Observatory (CTAO\footnote{See \url{www.ctao.org} for details.}), will allow more sensitive searches  for $\gamma$-ray emission from annihilating or decaying DM for WIMP in this range. 

A wide DM--oriented program is considered for CTAO~\citep{cta19}, diversified on several targets:
\begin{itemize}
   \item the GC, and in general the entire MW halo with a very high DM content ($\gtrsim$10$^{12}$ M$_\odot$; \citealt{ioc17,kar20}, but see \citealt{lac16,ior23}), is the closest known DM target; these characteristics point towards a relatively high DM signal (see \autoref{sec:dm}) and were investigated in \citep{ach21};
    \item the most nearby galaxy clusters are suited for the searches of $\gamma$-ray signals from DM decay; CTAO prospects were investigated in \citep{CTAConsortium:2023yak};
    \item the Large Magellanic Cloud (LMC), the largest disrupted satellite galaxy of the MW and one of the closest DM dominated targets, whose CTAO prospects were investigated in \citep{ach23}
    \item narrow-line emission associated to specific DM annihilation or decay modes that are incontrovertible signatures of DM, for which CTAO prospects were investigated in \citep{CTA:DM_line}.
\end{itemize}

In this work, we focus on the dwarf spheroidal galaxies \citep[dSphs; see, e.g.][]{Strigari:2008ib,Mayer:2009au,McConnachie:2012vd}. In contrast to regions already known for hosting $\gamma$-ray emitting sources from ordinary astrophysics processes, such as the GC, galaxy clusters and the LMC, dSphs are in fact almost completely free of any significant background \citep[but see][]{cro22}, thus making a $\gamma$-ray detection from their direction a highly compelling signature for evidence of DM annihilation or decay in their DM haloes.

The $\Lambda$-CDM model \citep{rie04}, where CDM stands for ``cold dark matter'', predicts that the formation of visible structures has been guided by gravitational accretion of baryons onto previously formed hierarchical DM over-densities -- from MW-sized haloes to satellite sub-haloes -- whose subsequent evolution occurred in different ways. In particular, at galactic and sub-galactic scales, the process depends on the initial DM halo parameters, such as the halo mass and the mass density profile, the evolution history, and conditions set by the local galactic environment in which this evolution takes place \citep[see e.g.][and refs. therein]{bert98}. As a result, less massive DM clumps could have evolved into invisible objects, constituting at present DM over-densities purely observable in $\gamma$-rays or secondary CRs emerging from CDM interactions. While in principle secondary emission from inverse Compton or synchrotron scattering processes (dependent on ambient photons and magnetic fields, respectively) could produce diffuse multi-wavelength (MWL) signatures, classic MWL survey techniques have not been proven sufficiently sensitive for their detection yet~\citep[see e.g.][and refs. therein]{Kar:2019hnj}. On the other hand, the sub-haloes that were sufficiently massive for accreting enough baryons to initiate star formation have evolved into the variety of satellite galaxies that we actually observe in the MW halo; in this scenario, in-falling baryonic matter settling into the innermost parts of the most massive DM sub-haloes could have originated the dSphs \citep{str08}.

\begin{table}
\caption{Parameters of the Northern (CTAO-N) and Southern (CTAO-S) baseline installations of CTAO (the so-called \texttt{Alpha} configuration) of relevance for this work. ORM stands for {\it Observatorio del Roque de los Muchachos} (La Palma, Spain), whereas Paranal stands for the {\it Cerro Paranal} observing site (Chile). The information presented here, retrieved from publicly available data \citep[the so-called \texttt{Prod5} performance release;][]{cta_irf}, may be subject to updates as the telescope layout and data pipelines are being fine-tuned.}
\label{tab:cta_arrays}
    \centering
    \resizebox{\columnwidth}{!}{
    \begin{tabular}{l|c|c}
    \hline
    \hline
         & CTAO-N & CTAO-S \\
         \hline
Location & ORM  & Cerro Paranal \\
Latitude  & $28^\circ 45'$~N & $24^\circ 41'$~S\\
Altitude   & $\sim$2200 m & $\sim$2150 m \\
\hline
Energy Threshold & $0.02$ TeV & $0.07$ TeV\\
Angular resolution at 1~TeV & 0.058$^\circ$ & 0.055$^\circ$\\
Energy resolution at 1~TeV & 8\% & 7\%\\
Sensitivity at 1~TeV (50 h) & $1.8\times10^{-13}$  & $1.3\times10^{-13}$\\
& erg cm$^{-2}$ s$^{-1}$ & erg cm$^{-2}$ s$^{-1}$ \\ 
\hline
Total no. of telescopes    & 13 & 51 \\
\;\;Number of LSTs & 4 & 0 \\
\;\;Number of MSTs & 9 & 14 \\
\;\;Number of SSTs & 0 & 37 \\
\hline
\end{tabular}
}
\end{table}

First discovered in 1938 \citep{sha38}, the dSphs are galaxies supported against gravity by the stellar velocity dispersion, possessing a spherical or elliptical appearance and containing $\mathcal{O}(10^2)$ stars (the so-called ``ultra-faint'' dSphs) up to $\mathcal{O}(10^7)$ stars \citep[the ``classical'' dSphs;][]{McConnachie:2012vd}. The most striking property of the dSphs is related to their matter content, which appears strongly dominated by the DM content, as opposed to the baryonic content. If in fact such systems are old enough to be in gravitational equilibrium -- typically reached in $\sim$100 Myr~\citep{Walker:2009zp} -- and following the virial theorem\footnote{Note, however, that the large velocity dispersion measured in these objects might be biased, leading to significant overestimates of the virial mass \citep[see e.g.][]{wal09b,jon21,pia22}.} and assuming spherical symmetry and isotropic velocity distribution, one gets a total gravitational mass $M_{\rm tot}$ that is proportional to both the system's virial size $R_{\rm vir}$ and the measured radial velocity dispersion\footnote{Assuming a perfectly spherical symmetry, one gets $\sigma_{\rm tot}^2 = 3\sigma_r^2$.} $\sigma_r$ of their components:
\begin{equation}\label{eqn:mvir}
M_{\rm tot} = 3 \frac{R_{\rm vir} \sigma_r^2}{G}\, ,
\end{equation}
where $G$ is the gravitational constant. The mass derived in this way is $\sim$10 to $\gtrsim$1000 times larger than that obtainable from the integrated dSph luminosity, which implies mass-to-light ratios 10 M$_\odot$/L$_\odot$ $\lesssim M/L \lesssim 10^4$ M$_\odot$/L$_\odot$; in contrast, a baryon-dominated system would exhibit $M/L < 10$ M$_\odot$/L$_\odot$ at most~\citep[see][and refs. therein]{McConnachie:2012vd}. This places the dSphs among the most DM-dominated objects in the local Universe; however, they are rather light in terms of absolute amount of DM hosted in their haloes ($M_{\rm tot} \lesssim 10^8$ M$_\odot$) if compared to e.g. the GC, and their distances are typically larger than that of the GC ($\sim$10 kpc), ranging from few tens to hundreds of kpc.  Ultra-faint dSphs share properties with globular clusters, but with more compact size and lower luminosity. Due to their strong gravitational potential, they retain material ejected by stars, resulting in a stellar population with wider metallicity spread than globular clusters~\citep{Willman:2012uj}.

To date, there are more than $50$ MW satellites classified as dSphs, with many of them already targeted as DM sources by current IACTs \citep{Doro:2021dzh,Hutten:2022hud}. Some of them were observed over several hundreds of hours \citep[see e.g.][]{acc22}, with no detection achieved. Thanks to the availability of data from recently concluded \citep{san16} and current survey projects \citep{drl16}, the number of known confirmed dSphs and candidates is expected to increase with time to $>$200 objects in the next years. In this work, we present updated determinations of the CTAO performance in detecting the expected photon yield emitted by DM annihilation or decay in dSph haloes. Such prospects are based on novel calculations of the DM amount in the most promising dSphs, taking advantage of the availability of stellar velocity measurements and robust mathematical procedures such as the Jeans analysis of the kinematics of their member stars \citep{eva04}, which we apply following \citet{bon15a}.

CTAO will be a world-class instrument to perform VHE $\gamma$-ray astronomy and will be composed of two arrays of telescopes, one in the Northern hemisphere (CTAO-N) and one in the Southern hemisphere (CTAO-S). \autoref{tab:cta_arrays} reports some basic facts about these two installations which are hereafter shortly discussed\footnote{The interested reader may find a complete report on CTAO in \citet{CTAConsortium:2013ofs}.}. 
CTAO-N will be located in the {\it Roque de los Muchachos} Observatory (ORM; $28^\circ 45'$ N $17^\circ 53'$ W, 2396 m a.s.l) in La Palma (Canary Islands, Spain). The site already hosts the MAGIC telescopes as well as several other installations\footnote{See \url{https://www.iac.es/en/observatorios-de-canarias/roque-de-los-muchachos-observatory} for details.}. Due to its location, CTAO-N is best suited for off-Galactic plane observations, in particular of extra-Galactic targets. For this reason, it is sensitive to the lowest energies -- of the order of $\sim$20 GeV -- achievable with the Cherenkov technique. This is obtained with two classes of telescopes of different size: the Large-Sized Telescopes (LSTs), which have 23-m diameter parabolic dishes \citep{cor19}, and the Medium-Sized Telescopes (MSTs), equipped with Davies-Cotton dishes of 12-m diameter \citep{bar20}. 
Both telescopes have tessellated primary mirrors and photo-multiplier (PMT) based focal plane instrumentation. There are currently 4 LSTs and 9 MSTs planned for construction in CTAO-N~\citep[so-called alpha configuration,][]{Zanin:2022mvt}.

CTAO-S will be located close to the ESO Paranal site in Chile ($24^\circ 41'$ S $70^\circ 18'$ W, 2635 m a.s.l.), near the ESO VLT and E-ELT facilities. The location grants an optimal view of the GC and Galactic plane. To detect $\gamma$-rays from PeV cosmic-ray and $\gamma$-ray sources in the Galaxy \citep[the so-called PeVatrons; e.g][]{Cao2021,deo24}, CTAO-S requires an improved sensitivity in the highest achievable energies -- up to $\mathcal{O}(100)$ TeV: this is obtained through a sparse array, covering $\sim$km$^2$ area, of Small-Sized Telescopes (SSTs), 4-m diameter Schwarzschild-Couder telescopes with segmented primary mirrors, monolithic secondary mirrors and silicon-photomultiplier (SiPM) based focal plane instrumentation \citep{tag22}. 

The exact characteristic of the arrays are still being fine-tuned. This, along with evolving event reconstruction tools, determines its expected performance \footnote{We recommend to check \url{https://www.cta-observatory.org/science/cta-performance/} for updates.}. The most up-to-date expected performance of CTAO is publicly released in the form of FITS and ROOT instrument response functions \citep[IRFs;][]{cta_irf} that are intended to be used together with the CTAO public analysis pipeline software \texttt{gammapy} \citep{gammapy} for estimating the CTAO capabilities in detecting $\gamma$-ray signals from several classes of VHE emitters. Besides the sensitivity, the most important instrumental characteristics which affects the searches for DM signals are: (i) the energy resolution, that determines the ability to discriminate a potential DM spectrum from an astrophysical one, and (ii) the angular resolution, which allows better discrimination of regions with higher DM content. Above 1 TeV, both CTAO-N and CTAO-S have energy resolutions of the order of $6$\%$- 7$\% and angular resolutions of the order of $0.055 - 0.057$ deg; both resolutions tend to worsen toward lower energies.

\begin{figure*}
    \centering
    \includegraphics[width=0.99\textwidth]{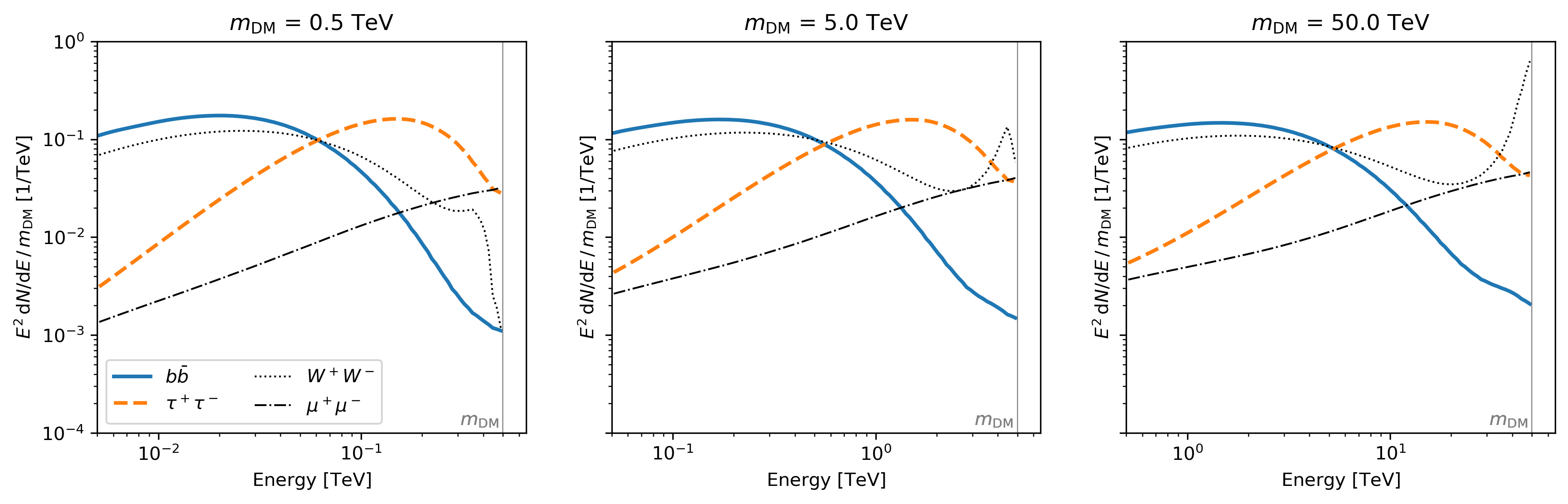}
    \caption{DM $\gamma$-ray spectra as a function of the DM mass for pure WIMP annihilation into specific channels, obtained with \texttt{gammapy}~\citep{gammapy} and based on the PPPC parametrization by \citet{Cirelli:2010xx}. We compare three values of $m_{\mathrm{DM}}=0.5,5,50$~TeV ({\it gray solid lines}) to show that the photon yield and spectral differences as a function of the mass are minor, with the exception of the $W^+W^-$ where electroweak loop corrections and final-state radiation significantly affect the high-energy tail of the spectrum. The $b\bar{b}$ and $\tau^+\tau^-$ channels are selected as representative examples of the theoretical DM photon yields, the former being the softest with a signal peak at $m_{\mathrm{DM}}/20$ and the latter being the hardest with a peak at $m_\mathrm{DM}/3$. In the case of DM decay, the major difference is that the cut-off happens at $m_\mathrm{DM}/2$ -- while preserving the spectral shape.}
    \label{fig:spectra}
\end{figure*}

This paper is organized as follows: in \autoref{sec:dm}, we present the physical framework for the computation of the expected $\gamma$-ray flux from DM annihilation and decay, including a novel computation of the DM content for a selection of best candidate dSphs; in \autoref{sec:methods}, we describe the adopted methodology for determining the CTAO sensitivity to indirect DM searches towards in these targets; in \autoref{sec:results}, we present our results for several combinations of important features -- spectral models, DM density profile shapes, combining different sources; in \autoref{sec:discussion}, we compare our findings with the current results from other facilities for $\gamma$-ray observations, and discuss them in the context of future surveys aimed at discovering new dSphs; finally, in \autoref{sec:conc} we give a  summary of our work and its main conclusions. Furthermore, \autoref{app:astro} reports further details on the computation of the DM content in dSphs and \autoref{app:systematics} outlines a quantitative study of the systematics.

\section{The $\gamma$-ray flux from annihilating and decaying dark matter in dSph haloes}
\label{sec:dm}
The expected $\gamma$-ray flux ($\Phi_\gamma$) from DM annihilation and decay processes depends on the microscopic nature of the DM particle, often called the particle-physics factor (discussed in  \autoref{sec:dm-particle}), and on a model for the astrophysical DM distribution -- the so-called astrophysical factor\footnote{\textbf{This is true in the case of velocity-independent DM interactions, and should be modified if e.g. the Sommerfeld effect~\citep{ark09,Feng:2010zp} or higher partial waves ($p$- and $d$-wave) are relevant. In such cases, \autoref{eqn:dm_flux} should be replaced by a formulation with a velocity-weighed astrophysical factor.}} \citep[discussed in \autoref{sec:dm-astro}]{eva04}. While the former factor is calculable within a general particle-physics DM framework, the computation of the astrophysical factor requires supporting astrophysical data. We briefly describe our method for estimating the amount of DM in dSph haloes in \autoref{sec:dm-astro}, focusing on the main concepts; further details are deferred to \autoref{app:astro}. 

Depending on the case of annihilating or decaying DM particles, two different expressions define the expected $\Phi_\gamma$:
\begin{equation}\label{eqn:dm_flux}
\resizebox{.9\columnwidth}{!}{$\frac{d\Phi_\gamma}{dE_\gamma}=
\begin{cases}
 \dfrac{\langle \sigma v \rangle}{8 k\pi \; m_{\mathrm{DM}}^2} \sum_i {\rm BR}_i \dfrac{dN^{i}_\gamma}{dE_\gamma} \;\cdot\; J_{\rm ann}(\Delta\Omega) \qquad \rm{Annihilating~DM} \\
\\
\dfrac{1}{4 \pi\, m_{\mathrm{DM}}} \sum_i \Gamma_i \dfrac{dN^{i}_\gamma}{dE_\gamma}\;\cdot\; J_{\rm dec}(\Delta\Omega) \qquad\qquad \rm{Decaying~DM}
\end{cases}$
}
\end{equation}
In the above expressions, $m_{\mathrm{DM}}$ the DM particle mass, $\langle \sigma v \rangle$ is the velocity-averaged DM annihilation cross section in case of annihilating DM, ${\rm BR}_i$ is the branching ratio (BR) for the $i$-th SM channel and $dN^{i}_\gamma/dE_\gamma$ is the number of photons (yield) produced during one annihilation at a given energy $E_\gamma$ into that channel; $k=1(2)$ for a Majorana (Dirac) DM particle. In the decay case, $\Gamma_i = 1/\tau_i$ are the decay amplitudes -- i.e. the inverse of the particle lifetimes $\tau_i$ -- for the specific $i$-th  channel. The astrophysical factor $J(\Delta\Omega)$ accounts for the signal intensity observable on Earth depending on the density of DM toward a target, integrated over the line-of-sight (l.o.s.) and the aperture $\Delta\Omega$. We label as $J_{\rm ann}$ the parameter for annihilating DM and $J_{\rm dec}$ that for decaying DM. The two, in turn, take the explicit form:

\begin{equation}\label{eqn:jfac}
\resizebox{.9\columnwidth}{!}{$J(\Delta \Omega)=
\begin{cases}
J_{\rm ann}(\Delta \Omega) = \int_{\Delta \Omega}  \int_{\rm l.o.s.} \rho_{\rm DM}^2(\ell, \Omega) \;d\ell\,d\Omega \qquad \rm{Annihilating~DM}\\
\\
J_{\rm dec}(\Delta \Omega) = \int_{\Delta \Omega}  \int_{\rm l.o.s.} \rho_{\rm DM}(\ell, \Omega) \; d\ell\,d\Omega \qquad \rm{Decaying~DM}
\end{cases}$}
\end{equation}

For distant objects, the two integrals in \autoref{eqn:jfac} reduce to volume integrals over the source halo divided by the squared distance \citep[see e.g. Eq.~4.3 from][]{bri09}. There are therefore four factors that affect the signal for DM annihilation or decay: $m_{\mathrm{DM}}$, $\langle \sigma v \rangle$ (or $1/\tau_i$), $dN^i_\gamma/dE_\gamma$ and $J_{\rm ann}$ (or $J_{\rm dec}$). Out of these, the WIMP mass $m_{\mathrm{DM}}$ and the (differential) number of particles per annihilation or decay process $dN_\gamma/dE_\gamma$ are defined by the DM particle model. Since $J_{\rm ann}$ or $J_{\rm dec}$ linearly increases the signal yield, it is of utmost relevance to clearly assess the magnitude of the astrophysical factor in order to infer DM properties from positive detection, or to provide strong constraints in case of null detection. We discuss the computation of these factors in \autoref{sec:dm-particle} (particle-physics factor) and \autoref{sec:dm-astro} (astrophysical factor), respectively.

\subsection{Dark matter particle models}
\label{sec:dm-particle}
Primary products of the WIMP annihilation and decay processes are SM particles such as quarks, gauge bosons or leptons. After production, quark and bosons hadronize and then decay, generating neutral pions which in turn decay into photons with continuous energies up to $m_\mathrm{DM}$ for DM annihilation, or up to half this value for DM decay. Regardless the specific SM channel, the $\gamma$-ray spectra expected in these cases are very similar. For leptonic decay modes, $\pi^0$ are only found for $\tau^+\tau^-$ final states, and most photons are produced as final-state radiation directly off the leptons. We note that the BR$_i$ defined in \autoref{eqn:dm_flux} may in principle depend on the incoming DM kinetic energy, but this contribution is subdominant and is neglected here\footnote{In the case of cold DM with typical velocities \( v \sim 10^{-3}c \) and s-wave annihilation, the kinetic energy of the incoming DM particles is of order \( \sim 10^{-6} m_{\mathrm{DM}} \), resulting in changes to the branching ratios of $\lesssim$1\% except near sharp kinematic thresholds or resonances.}
.

\begin{figure*}
    \centering
    \includegraphics[width=\textwidth]{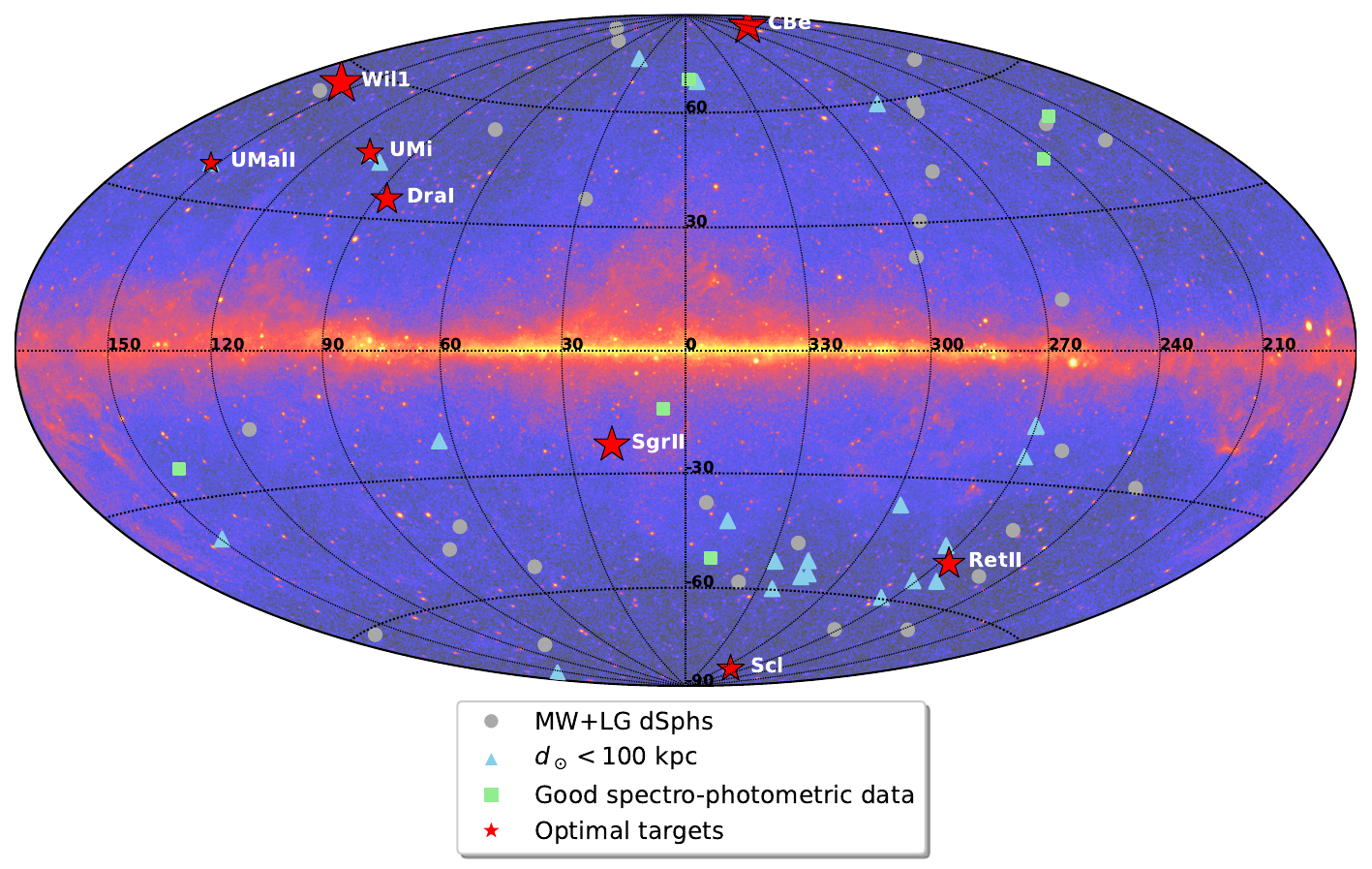}
    \caption{Sky distribution of known MW satellites and Local Group dSphs, superimposed to the {\itshape Fermi}-LAT $\gamma$-ray background (credits: NASA/DOE/{\itshape Fermi}-LAT Collaboration). All of the targets from \autoref{tab:dsph_excluded} ({\itshape gray dots}) are reported, along with the sources from \autoref{tab:dsph_firstcut} (i.e. those passing the first selection cut on distance; {\itshape blue triangles}) and those objects passing the second selection (availability of good spectral and photometric data; {\itshape green squares}). The optimal targets ({\itshape red stars}) are highlighted with symbols of increasing size, proportional to the value of their $\log{J_{\rm ann}(<0.1^\circ)}$ (see \autoref{tab:dsph_jfactors}).}
    \label{fig:dsph_skymap}
\end{figure*}

A widely used parameterization of DM $\gamma$-ray spectra is that provided by \citet[also called PPPC4 DM ID, hereafter PPPC]{Cirelli:2010xx}, that computes $\gamma$-ray spectra for different DM masses including electroweak corrections. PPPC is adopted as the reference parametrization for this work; however, one should notice that, for WIMPs different from an electroweak singlet or where virtual internal Bremsstrahlung (VIB) effects are relevant \citep{Bringmann:2017sko}, the PPPC parameterization is less accurate. The same is also true for DM particles with $m_\mathrm{DM} \gg 10$ TeV. Furthermore, PPPC relies on the \texttt{PYTHIA} event generator, which exhibits known discrepancies with respect to other generators such as \texttt{Herwig} \citep{Cembranos:2013cfa}. Recently, refined computations of $\gamma-$ray spectra were explored~\citep[e.g.,][]{bauer2021dark,Arina:2023eic}; however, these new parameterization should have a minor impact on our results, and limited to DM mass values above 10 TeV. 

The resulting spectral energy distribution is normally discriminated between a continuous part and a line-like one, the latter arising in processes like $\mathrm{DM} \mathrm{DM} \to \gamma \gamma$ or $\mathrm{DM} \mathrm{DM} \to \gamma Z_0$. Such a line-like emission would produce an  almost monochromatic signal, broadened by the mass of the neutral particle but mostly by the finite energy resolution of the instrument. Line-like signals constitute a clear DM identification, because there is hardly an astrophysical process able to mimic such spectral shapes of the emission, at these energies. However,  primary $\gamma$-rays can only be produced via second-order loop processes, which are typically reduced by $\alpha^2$ ($\alpha$ being the fine-structure constant) with respect to the cross-section expected from thermal freeze-out \citep{bri12}. The search for narrow-line DM signatures in CTAO is investigated in another work \citep{CTA:DM_line}; here, we focus on continuous $\gamma$-ray spectra.

The validity of the WIMP freeze-out scenario relies upon the assumption of explaining the observed DM abundance at present day through an (inverse) annihilation process \citep{sre88,Gondolo:1990dk}; nevertheless, here we test both the WIMP annihilation and decay possibilities. Furthermore, since each DM model has its $\gamma$-ray spectrum defined by \autoref{eqn:dm_flux}, the overall spectral shape of the annihilation (decay) signal is determined by the branching ratios BR$_i$ (decaying amplitudes $\Gamma_i$) to the specific SM channels. In principle, to cast a limit on such models one should know all BR$_i$ (or $\Gamma_i$) together with the $\gamma$-ray yields for each channel. Knowing all  BR$_i$ requires a complete modeling of the DM particle interactions, which is in general useful only if a specific DM particle rather than a class are searched for; it is therefore convenient, and has become a practice in the field, to use a `benchmark' approach in which one selectively sets BR$_i=1$ for the $i$-th channel and zero for all other channels (or $\Gamma_j = 0$ for all channels $j \neq i$ in the case of DM decay).

\begin{table*}
\caption{Basic properties of the MW dSph satellites and those of the LG not associated to major galaxies that satisfy our first selection cut (heliocentric distance $<$100 kpc), separated between Northern ({\itshape upper section}) and Southern hemisphere ({\itshape lower section}). In both sections, objects indicated with ``cls'' are ``classical'' dSphs whereas ``uft'' stands for ``ultra-faint'', and dashes in the culmination columns indicate objects that do not rise above the horizon for the CTAO array in the relevant hemisphere.}\vspace{-5mm}
\label{tab:dsph_firstcut}
\begin{center}
\resizebox{.99\textwidth}{!}{
\begin{tabular}{llllllllll}
\hline
\hline
Name & Abbr. & Type & R.A. & dec. & Distance  & ZA$_{\rm culm}$ & ZA$_{\rm culm}$ & Month & Ref.\\
 & & & (hh mm ss) & (dd mm ss) & (kpc) & N (deg) & S (deg) & & \\
\hline
\multicolumn{10}{l}{\textbf{CTAO-N candidate dSphs}}\\
\hline
Bo{\"o}tes I & Bo{\"o}I & uft & 14 00 06.0 & +14 30 00 & $65 \pm 3$ & 14.3 & 39.1 & Apr & 1,2\\
Bo{\"o}tes II & Bo{\"o}II & uft & 13 58 00.0 & +12 51 00 & $39 \pm 2$ & 15.9 & 37.5 & Apr & 1,3\\
Bo{\"o}tes III & Bo{\"o}III & uft & 13 57 12.0 & +26 48 00 & $46 \pm 2$ & 2.0 & 51.4 & Apr & 1,3\\
Coma Berenices & CBe & uft & 12 26 59.0 & +23 54 15 & $42 \pm 2$ & 4.9 & 48.5 & Mar & 1,4\\
Draco I & DraI & cls & 17 20 12.4 & +57 54 55 & $75 \pm 4$ & 29.2 & 82.5 & Jun & 1,5\\
Draco II & DraII & uft & 15 52 47.6 & +64 33 55 & $20 \pm 3$ & 35.8 & 89.2 & May & 6\\
Laevens 3 & Lae3 & uft & 21 06 54.3 & +14 58 48 & $67 \pm 3$ & 13.8 & 39.6 & Aug & 7\\
Segue 1 & Seg1 & uft & 10 07 04.0 & +16 04 55 & $23 \pm 2$ & 12.7 & 40.7 & Feb & 1,8\\
Segue 2 & Seg2 & uft & 02 19 16.0 & +20 10 31 & $36 \pm 2$ & 8.6 & 44.8 & Oct & 1,9\\
Triangulum II & TriII & uft & 02 13 17.4 & +36 10 42 & $30 \pm 2$ & 7.4 & 60.8 & Oct & 10\\
Ursa Major II & UMaII & uft & 08 51 30.0 & +63 07 48 & $35 \pm 2$ & 34.4 & 87.8 & Feb & 1,11\\
Ursa Minor & UMi & cls & 15 09 08.5 & +67 13 21 & $68 \pm 2$ & 38.5 & --- & May & 1,12\\
Willman 1 & Wil1 & uft & 10 49 21.0 & +51 03 00 & $38 \pm 7$ & 22.3 & 75.7 & Mar & 1,8\\
\hline
\multicolumn{10}{l}{\textbf{CTAO-S candidate dSphs}}\\
\hline
Carina II & CarII & uft & 07 36 26.3 & $-$58 00 00 & $36 \pm 1$ & 86.7 & 33.3 & Jan & 13\\
Carina III & CarIII & uft & 07 38 31.2 & $-$57 54 00 & $28 \pm 2$ & 86.7 & 33.3 & Jan & 13\\
Cetus II & CetII & uft & 01 17 52.8 & $-$17 25 12 & $30 \pm 3$ & 46.2 & 7.2 & Oct & 14\\
Eridanus III & EriIII & uft & 02 22 45.5 & $-$52 16 48 & $95 \pm 27$ & 81.0 & 27.7 & Oct & 15\\
Grus II & GruII & uft & 22 04 04.8 & $-$46 26 24 & $53 \pm 5$ & 75.2 & 21.8 & Aug & 14\\
Horologium I & HorI & uft & 02 55 28.9 & $-$54 06 36 & $87 \pm 13$ & 82.9 & 29.5 & Oct & 15\\
Horologium II & HorII & uft & 03 16 26.4 & $-$50 03 00 & $78 \pm 8$ & 77.5 & 26.7 & Nov & 16\\
Hydrus I & HyiI & uft & 02 29 33.7 & $-$79 18 36 & $28 \pm 1$ & --- & 53.3 & Oct & 17\\
Indus I & IndI & uft & 21 08 48.1 & $-$51 09 36 & $69 \pm 16$ & 79.9 & 26.5 & Aug & 15\\
Phoenix II & PheII & uft & 23 39 57.6 & $-$54 24 36 & $95 \pm 18$ & 83.2 & 29.8 & Sep & 15\\
Pictor II & PicII & uft & 06 44 43.1 & $-$59 54 00 & $45 \pm 5$ & 88.3 & 35.8 & Jan & 18\\
Reticulum II & RetII & uft & 03 35 40.9 & $-$54 03 00 & $32 \pm 2$ & 82.8 & 29.4 & Nov & 15\\
Reticulum III & RetIII & uft & 03 45 26.3 & $-$60 27 00 & $92 \pm 13$ & 89.2 & 35.8 & Nov & 19\\
Sagittarius I & SgrI & cls & 18 55 19.5 & $-$30 32 43 & $31 \pm 1$ & 59.3 & 5.9 & Jul & 1,20\\
Sagittarius II & SgrII & uft & 19 52 40.5 & $-$22 04 05 & $67 \pm 5$ & 50.8 & 2.6 & Jul & 7\\
Sculptor & Scl & cls & 01 00 09.4 & $-$33 42 33 & $84 \pm 2$ & 62.5 & 9.1 & Oct & 1,21\\
Sextans & Sex & cls & 10 13 03.0 & $-$01 36 53 & $84 \pm 3$ & 30.4 & 23.0 & Feb & 1,22\\
Tucana II & TucII & uft & 22 52 16.7 & $-$58 33 36 & $58 \pm 6$ & 87.3 & 33.9 & Sep & 15\\
Tucana III & TucIII & uft & 23 56 35.9 & $-$59 36 00 & $25 \pm 2$ & 88.4 & 35.0 & Sep & 14\\
Tucana IV & TucIV & uft & 00 02 55.3 & $-$60 51 00 & $48 \pm 4$ & 89.6 & 36.2 & Sep & 14\\
Tucana V & TucV & uft & 23 37 23.9 & $-$63 16 12 & $55 \pm 9$ & --- & 38.3 & Sep & 23\\
Virgo I & VirI & uft & 12 00 09.1 & $-$00 40 52 & $87 \pm 11$ & 40.0 & 24.2 & Mar & 24\\
\hline
\end{tabular}
}
\end{center}
\scriptsize{
References: $^1$\citet{McConnachie:2012vd}, $^2$\citet{oka12}, $^3$\citet{ses14}, $^4$\citet{mus09}, $^5$\citet{her16}, $^{6}$\citet{lae15a}, $^{7}$\citet{lae15a}, $^{8}$\citet{dej08}, $^{9}$\citet{boe13}, $^{10}$\citet{lae15b}, $^{11}$\citet{dal12}, $^{12}$\citet{ruh11}, $^{13}$\citet{tor18}, $^{14}$\citet{drl15}, $^{15}$\citet{bec15}, $^{16}$\citet{kim15}, $^{17}$\citet{kop18}, $^{18}$\citet{drl16}, $^{19}$\citet{drl15}, $^{20}$\citet{val15}, $^{21}$\citet{vaz15}, $^{22}$\citet{med18}, $^{23}$\citet{con18}, $^{24}$\citet{hom16}.
}
\end{table*}

\begin{table*}
\caption{Kinematic properties of the dSphs surviving our first and second selection cuts. In the table, we report: the dSph short name; the best CTAO site for observation; the adopted method for the stellar classification -- ``EM'' denotes the application of the expectation-maximization algorithm by \citet{wal09} to determine individual stellar membership probabilities, whereas ``bin'' indicates the adoption of the binary memberships reported in the literature (see \autoref{app:astro}); the number of stars surviving the selection over the total number of input stars; the average radial velocity and dispersion; the reference for the stellar kinematic data. We also report the tidal radii computed with \texttt{CLUMPY} assuming two shapes for the DM density profiles.}
\label{tab:dsph_parameters_2}
\begin{center}
\begin{tabular}{lc|cccccc|cc}
\hline\hline
Name & Site & Membership & $N_{\rm mem}/N_{\rm tot}$ & $\langle v_r \rangle$ & $\sigma_v$ & Ref. & & $R_{\rm tid}^{\rm (Ein)}$ & $R_{\rm tid}^{\rm (Bur)}$\\
 & & & & (km s$^{-1}$) & (km s$^{-1}$) & & & (kpc) & (kpc)\\
\hline
Bo{\"o}I & N & bin & 37/113 & 100.6 & 4.3 & 1 & & $5.1^{+10.7}_{-2.2}$ & $15.1^{+30.4}_{-9.6}$\\
CBe & N & bin & 59/102 & 97.8 & 5.8 & 2 & & $6.3^{+9.5}_{-3.4}$ & $19^{+55}_{-16}$\\
DraI & N & EM & 466/1565 & $-$292.4 & 9.5 & 3 & & $4.83^{+1.16}_{-0.84}$ & $4.30^{+0.86}_{-0.54}$\\
GruII & S & bin & 21/235 & $-$109.8 & 1.8 & 4 & & $0.35^{+1.01}_{-0.32}$ & $\lesssim$9.5\\
RetII & S &  bin & 18/38 & 64.0 & 3.6 & 5 & & $1.66^{+4.46}_{-0.97}$ & $5.8^{+19.3}_{-5.3}$\\
Scl & S & EM & 1120/1541 & 111.5 & 9.1 & 6 & & $2.95^{+0.55}_{-0.30}$ & $3.71^{+0.30}_{-0.18}$\\
Seg1 & N &  EM & 154/522 & 206 & 15 & 7 & & $0.43^{+3.23}_{-0.35}$ & $\lesssim$28\\
Sex & S &  EM & 356/947 & 224 & 11 & 6 & & $7.8^{+4.4}_{-2.9}$ & $9.9^{+5.7}_{-3.4}$\\
SgrI & S &  EM & 288/503 & 140 & 17 & 8 & & $1.56^{+0.34}_{-0.73}$ & $\lesssim$1.7\\
SgrII & S & bin & 21/26 & $-$175.7 & 5.0 & 9 & & $3.7^{+13.9}_{-2.7}$ & $4.2^{+36.4}_{-2.8}$\\
TriII & N & bin & 13/33 & $-$381.7 & 2.5 & 10 & & $0.36^{+3.20}_{-0.35}$ & $\lesssim$56\\
UMaII & N &  bin & 20/54 & $-$116.1 & 8.1 & 2 & & $2.15^{+1.69}_{-0.99}$ & $2.23^{+6.48}_{-0.98}$\\
UMi & N &  EM & 467/973 & $-$247 & 12 & 11 & & $14.7^{+6.6}_{-4.1}$ & $15.3^{+8.6}_{-3.9}$\\
Wil1 & N &  bin & 40/97 & $-$13.6 & 6.3 & 12 & & $1.20^{+4.08}_{-0.51}$ & $1.35^{+26.35}_{-0.48}$\\
\hline
\end{tabular}
\end{center}
\scriptsize{References: 
$^{1}$\citet{kop11}; $^{2}$\citet{sim07}; $^{3}$\citet{wal15}; $^{4}$\citet{sim20}; $^{5}$\citet{wal15b}; $^{6}$\citet{wal09b}; $^{7}$\citet{sim11}; $^{8}$\citet{iba97}; $^{9}$\citet{lon20}; $^{10}$\citet{kir17}; $^{11}$\citet{spe18}; $^{12}$\citet{wil11}.}
\end{table*}

Widely used spectra in the literature are those corresponding to DM annihilation (or decay) into the $b\bar{b}$, $W^+W^-$, $\tau^+\tau^-$, and $\mu^+\mu^-$ channels; we display some of the corresponding photon yields for the annihilating DM case for a DM particle of 0.5, 5, 50~TeV mass in \autoref{fig:spectra}. Changes in the WIMP mass  introduce small modifications in the spectral shape of the resulting DM photon yield, although this may not be general. Two of the spectra plausibly bracket the uncertainty on the true annihilation: the $b\bar{b}$ spectrum is a prototype of a soft DM spectrum that peaks at $E_\gamma \sim m_{\mathrm{DM}}/20$, while the $\tau^+\tau^-$ spectrum is a prototype of a hard DM spectrum which peaks at $E_\gamma \sim m_{\mathrm{DM}}/3$. Some other spectra are shown in \autoref{fig:spectra} for comparison. PPPC provides such $\gamma$-ray spectra for a large set of DM masses. Similar considerations hold for the decaying DM case where the cut-off happens at $m_\mathrm{DM}/2$ but the spectral shape is preserved.

\subsection{Modeling of the astrophysical factor and selection of the optimal dSphs}
\label{sec:dm-astro}

The procedure for estimating the DM astrophysical factor of a dSph has an ample literature. Different methods and assumptions lead to mild incompatibilities (in most cases) or vigorous ones for some debated targets. This computation becomes particularly relevant for the case of CTAO which, as opposed to other wide-field of view (FoV) instruments, will only be able point to a selection of dSphs for deep observations, very likely starting with those with the highest -- and possibly most accurately determined -- predicted values of the astrophysical factor. Considering the relevance of the astrophysical factor for the computation of detection limits of $\gamma$-ray signals from DM annihilation or decay (see \autoref{eqn:dm_flux}), we provide our own estimates based on a consistent treatment of all the dSphs considered here.

Previous attempts to compute the astrophysical factor for large samples of dSphs can be found in the literature \citep{eva04,bon15c,ger15,mar15,hay16}. In these works, the authors adopted a variety of methods to determine $J_{\rm ann}$ and $J_{\rm dec}$ of dSphs from observed quantities (e.g., surface brightness and stellar kinematics). For instance, \citet{str08} perform maximum-likelihood fits to the dSph stellar line-of-sight velocities to derive scale densities, radii and indexes of the DM haloes; \citet{ger15} apply the agnostic modeling of DM density profiles from stellar spectro-photometric data also described in \citet{cha11}; \citet{mar15} computes DM halo parameters through hierarchical mass modeling of dSphs; finally, \citet{bon15c} and \citet{hay16} rely on different applications of the Jeans analysis to compute posterior distributions of halo parameters from observable data.

Others like e.g. the \textit{Fermi}-LAT based-studies~\citep{ack15,drl15b,Fermi-LAT:2016uux,mcd23}, make use of astrophysical factor estimates primarily based on \textit{distance} measurement, using the universality of dSph properties discussed by e.g. \citet{Strigari:2008ib}. This is also motivated by the fact that several dSphs in the FoV of \textit{Fermi}-LAT did not have astrometry-based measurements available. \citet{ack15}, \citet{drl15b} and \citet{Fermi-LAT:2016uux} used a relation between $J_{\rm ann}$ and the dSph distance $d_\odot$ with uncertainties up to $\sim$0.8~dex. More recently, \citet{pac19} introduced a novel scaling relation which not only included the distance, but also the line-of-sight velocity spread $\sigma_v$ as well as the stellar half-light radius $r_{1/2}$ to more accurately compute the $J_{\rm ann}$ enclosed within $0.5^\circ$, obtaining:
\begin{equation}\label{eq:J-distance2}
 J_{\rm ann}(<0.5^\circ)=10^{17.87}\left(\frac{\sigma_v}{5~\mathrm{km}\mathrm{s}^{-1}}\right)^4\left(\frac{d_{\odot}}{100~\mathrm{kpc}}\right)^{-2}\left(\frac{r_{1/2}}{100~\mathrm{pc}}\right)^{-1}
\end{equation}
with claimed uncertainties on the level of $\lesssim$0.1~dex. Recently, \citet{mcd23} used such scaling relations to provide \textit{Fermi}-LAT legacy limits on 50~dSphs, one of the largest sample ever used. In figure 1 of \citet{mcd23}, they classify the dSphs in 3 groups: `Measured' includes dSphs for which astrometric data are available in the literature, `Benchmark' includes also the dSphs for which only an estimate of $J_{\rm ann}$ from \autoref{eq:J-distance2} can be obtained, and `Inclusive' also contains the controversial dSphs. We compare our own catalog with this work throughout this paper. 

Our calculations of the dSph astrophysical factors are based on the procedure described in \citet{bon15c}, that makes use of the publicly available \texttt{CLUMPY} code \citep{cha12,bon16,hut19}. \texttt{CLUMPY} allows the execution of a Markov-Chain Monte Carlo (MCMC) dynamical analysis of the dSphs' DM haloes. In detail, the galaxy is treated as a steady-state, collisionless systems in spherical symmetry (but not necessarily isotropic) and with negligible rotation, in which the contribution of the stellar component to the total mass can be also neglected \citep[see][for a comprehensive discussion on these assumptions]{bon15a}. The MCMC analysis relies on the solution of the second-order spherical Jeans equation~\citep{bin08}:


\begin{align}\label{eqn:jeans}
    & \frac{1}{n^*(r)}\left\{
    \frac{d}{dr}\left[
    n^*(r)\overline{v_r^2}
    \right]
    \right\} + 2\beta_{\rm ani}(r)\frac{\overline{v_r^2}}{r} 
     = -\frac{G}{r^2}\left[
    M^*(r) + M_{\rm DM}(r)
    \right] \nonumber \\
    & \simeq -\frac{GM_{\rm DM}(r)}{r^2}\, ,
\end{align}

where $n^*(r)$ is the stellar number density, $\overline{v_r^2}$ is the average squared radial velocity and $\beta_{\rm ani}(r) = 1 - \overline{v_\theta^2}/\overline{v_r^2}$ is the velocity anisotropy of the dSph (with $\overline{v_\theta^2}$ the average squared tangential velocity). 

Such quantities are fed to \texttt{CLUMPY} as either a parametric fixed input (the stellar number density), a set of discrete input values with which the MCMC is computed (the stellar kinematics data that determine the velocity dispersion) or a set of free parameters that describe the adopted profile (the DM density $\rho_{\rm DM}$ and the velocity anisotropy profile). In such conditions, the Jeans equation can be solved to obtain $\rho_{\rm DM}(r)$ along the object's 3D radial coordinate $r$. The complete procedure that we adopted is described in more details in \autoref{app:astro}; here, we report the main results, that are based on the use of the best input of dSph currently available in the literature and of the latest \texttt{CLUMPY} version (\texttt{v3}).

\subsubsection{Initial catalog of dSphs and selection criteria}
We collect from the literature the basic positional data (equatorial coordinates and distance) for a complete sample of MW and Local Group (LG) classical and ultra-faint dSph satellites known to date, merging catalogs of already known targets \citep{McConnachie:2012vd} and results from recent surveys \citep{bec15,drl15,kop15,lae15a,lae15b,mar15b,drl20}. From their equatorial coordinates, we compute the culmination zenith angles (ZAs) of each target in both hemispheres at the CTAO-N and CTAO-S sites. In this way, we identify 64 targets, whose sky distribution is shown in \autoref{fig:dsph_skymap}. All dSphs analyzed in \citet{mcd23} appear in our catalog. Since it is not possible to accurately model the DM content of all the 64 dSphs (see \autoref{app:astro} for a discussion), we narrow down the best candidates following consecutive selection steps:
\begin{enumerate}
    \item  We apply a distance selection $d_\odot \leq 100$ kpc to remove those dSphs for which we expect very low values of $J_{\rm ann}$ based on \autoref{eq:J-distance2} (35 remaining targets).
    \item  For the surviving dSphs, we investigate the availability and quality of spectro-photometric data sets in the literature and discard those with missing or unreliable data (14 remaining targets). 
    \item  We select the dSphs with the highest values of $J_{\rm ann}$ for further signal modeling and computation of prospects with CTAO observations (8 remaining targets).
\end{enumerate}

\begin{figure*}
    \centering
    \includegraphics[width=\textwidth]{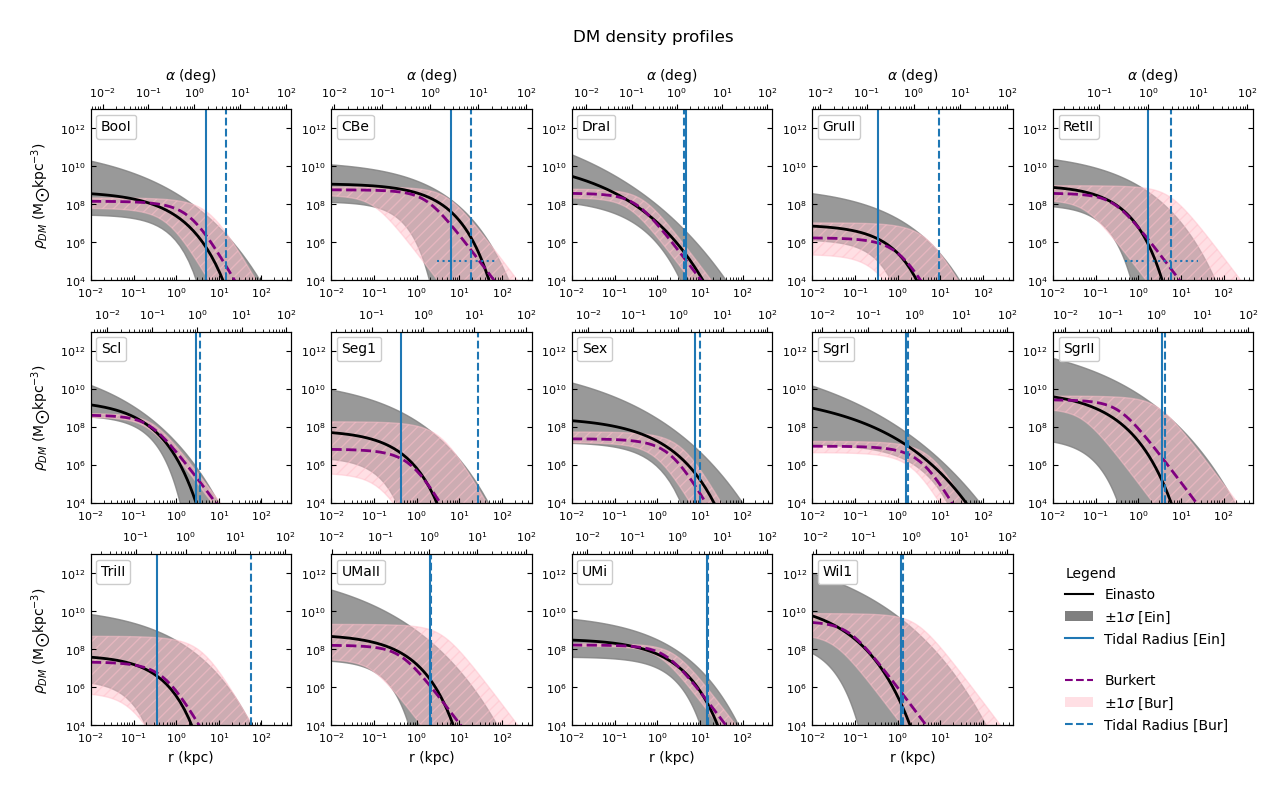}
    \caption{DM density profiles of the 14 optimal dSphs for both cored \citep[{\it purple dashed lines}]{bur95} and cuspy models \citep[{\it black solid lines}]{ein65}, along with the corresponding uncertainties at 68\% CL ({\it pink and gray shaded areas}). In all panels, the tidal radius (see text for details) for the Einasto ({\it blue  solid line}) and Burkert profile ({\it blue dashed line}) is also indicated, along with the typical uncertainty ({\it blue dotted line}) for the representative cases of CBe and RetII.}
    \label{fig:profile_density_14}
\end{figure*}

\begin{figure*}
\centering
\includegraphics[width=0.99\linewidth]{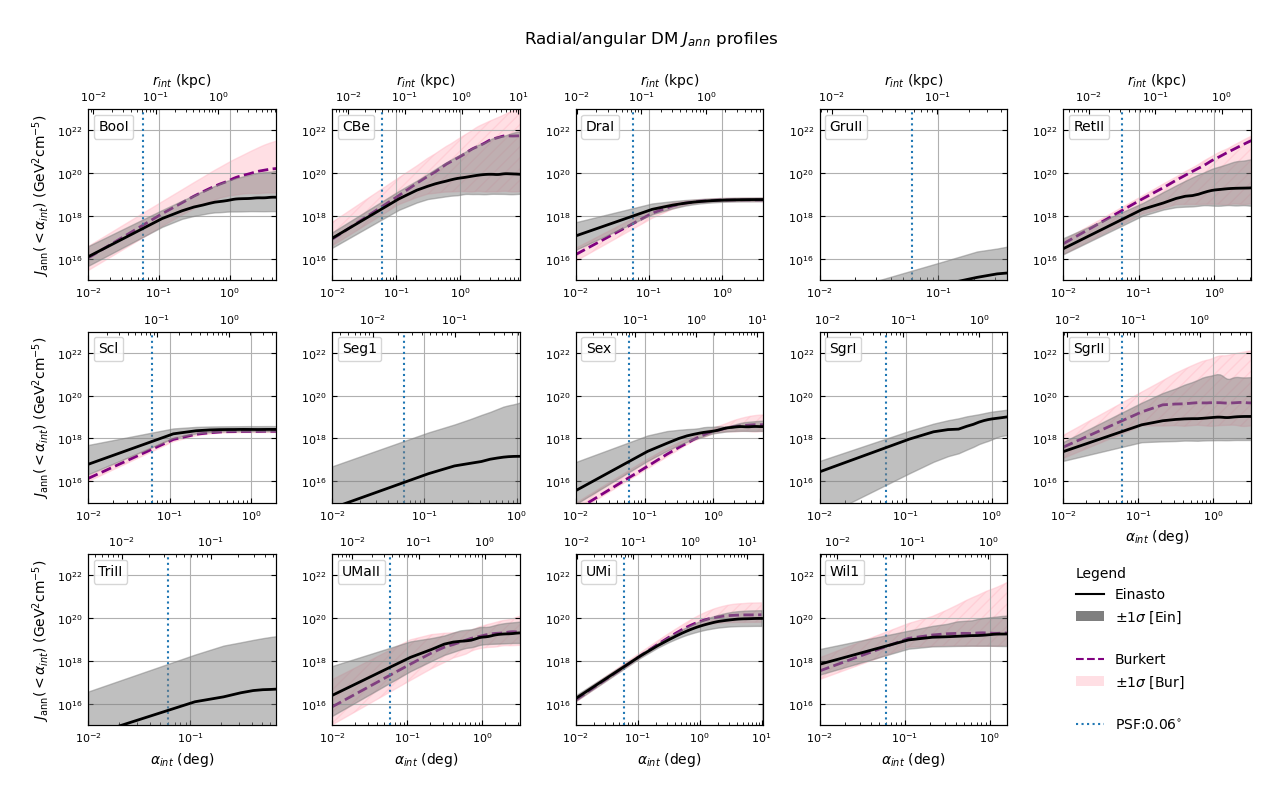}
\caption{Astrophysical factors for DM annihilation $J_{\rm ann}(<\alpha_{\rm int})$ as functions of the integration angle $\alpha_{\rm int}$ (or the equivalent integration distance from the dSph centroid $r_{\rm int}$) for the best Northern and Southern dSphs. In all panels, the median astrophysical factor profiles for both cuspy (Einasto; {\itshape black solid lines}) and cored (Burkert; {\itshape purple dashed lines}) DM density profiles are plotted alongside the relative uncertainties at 1$\sigma$ CL ({\itshape gray shaded areas} and {\itshape pink shaded areas}). The integration angles corresponding to an instrumental PSF of $0.06^\circ$ ({\it blue dotted lines}) are also indicated. Each profile is truncated at the corresponding dSph tidal radius for the Einasto profile (see \autoref{eq:tidal_radius} and \autoref{tab:dsph_parameters}). The Burkert profiles for GruII, Seg1, SgrI and TriII are not reported because no finite integration could be obtained.}
\label{fig:dsph_profile_ann}
\end{figure*}

\subsubsection{Distance-based selection}
The first selection is done according to the target distance $d_\odot\leq100$~kpc, on the basis of the scaling relation~\autoref{eq:J-distance2}. After the application of this first cut, we end up with 35 remaining candidates, that we report in \autoref{tab:dsph_firstcut}, and 29 discarded dSphs that we report in \autoref{tab:dsph_excluded}. We indicate their names with the corresponding abbreviation, the dSph type (`{\itshape cls}' for the classical dSphs and `{\itshape uft}' for the ultra-faint ones\footnote{This classification is not physical, since it is mostly based on the discovery of dSphs before the Sloan Digital Sky Survey \citep[SDSS;][]{yor00} for the classical objects and after for the ultra-faint targets. However, it is useful to use a crude separation between objects with hundreds to thousands of member stars (the classical dSphs), and few to tens (the ultra-faint case).}), their J2000 right ascension (R.A.) and declination (dec), their distance, and their culmination zenith angle (ZA$_{\rm culm}$) at the CTAO-N and CTAO-S sites, together with the culmination month. Out of these, 13 are visible for CTAO-N and 22 for CTAO-S. All the dSphs in our catalog that were not reported in \citet{mcd23} do not pass the distance cut, except for the Sculptor dSph which is excluded by \citet{mcd23} due to source confusion with 4FGL J0059.5$-$3338, and Triangulum~II for the tension between the value of $J_{\rm ann}$ derived from its available astrometric data and that obtained from the distance relations. From Fig.~1 of \citet{mcd23}, we check the conservativeness of this cut by noting that the targets lying right beyond the distance limit (Aquarius II, Carina I, and Ursa Major I) have values of $J_{\rm ann} \lesssim 3 \times 10^{18}$ GeV$^2$ cm$^{-5}$. All the farther ones either have $J_{\rm ann} < 10^{18}$~GeV$^2$ cm$^{-5}$, or do not posses kinematic data sets. These would not have passed our further selection criteria.

\subsubsection{Stellar data quality selection}
A second step is based  on the quality and availability of photometric and spectroscopic data sets, which are required in order to compute well-defined astrophysical factors. The complete methodology is reported in \autoref{app:astro} and here we only shortly recall that the starting point is obtaining the  brightness density profiles $n^*(r)$ of each dSph for which literature data exist. During this step, we discard 16 targets: Bo{\"o}tes III, Carina II, Carina III, Draco II, Laevens 3, Cetus II, Eridanus III, Horologium I, Horologium II, Hydrus I, Indus I, Pictor I, Phoenix II, Reticulum III, Tucana V and Virgo I, since such targets have no spectroscopic measurements over adequately populated samples of member stars \citep[e.g.,][]{koc09,kop15b,drl15b} and we end up with 19 dSphs. The stellar brightness density profiles of these surviving dSphs are then fitted with a 3D Zhao-Hernquist profile \citep[see \autoref{app:astro}]{her90,zha96}, projected onto the corresponding circularized 2D surface brightness profile (see \autoref{fig:dsph_brightness_profile}); at this level, we remove 5 additional dSphs -- Segue 2, Tucana II, Tucana III, Tucana IV and Bo{\"o}tes II -- for lacking 2D brightness data (see \autoref{app:astro} for further details). 

We perform such fits using the IDL package \texttt{MPFIT} \citep{mar09}; in this way, we are left with a list composed by 14 candidates, reported now in \autoref{tab:dsph_parameters_2} along with the number of member stars, the average velocity and spread, and the references for data. We also report our estimation of their tidal radius, discussed below (see also \autoref{app:rtid}). We note that these objects are well distributed between both hemispheres, thus providing CTAO with a balanced pool of choice for both sites. The stellar surface brightness profiles for these targets are reported in \autoref{fig:dsph_brightness_profile}. Next, we study the stellar kinematics of the remaining dSphs: to this aim, we collect the most updated and complete samples of stars for each source that are provided in the literature (see references reported in \autoref{tab:dsph_parameters_2}). The advantage of using such samples lies in their cleanliness from problematic data, e.g. binary stars \citep{spe17}, thanks to the analyses performed by their respective authors \citep[see e.g.][]{kir17}. The distribution of stellar velocities for each sample of stars that fall within the sky extension of a given dSph is shown in \autoref{fig:dsph_brightness_profile}. For the treatment of the velocity anisotropy, we use the Baes \& van Hese profile \citep[see \autoref{eqn:velanis};][]{bae07}. In \autoref{tab:dsph_parameters}, we also report the $V$-band integrated dSph magnitude, eccentricity, brightness scale radius and density, and the stellar membership statistics for each target.

\begin{figure*}
\centering
\includegraphics[width=0.99\linewidth]{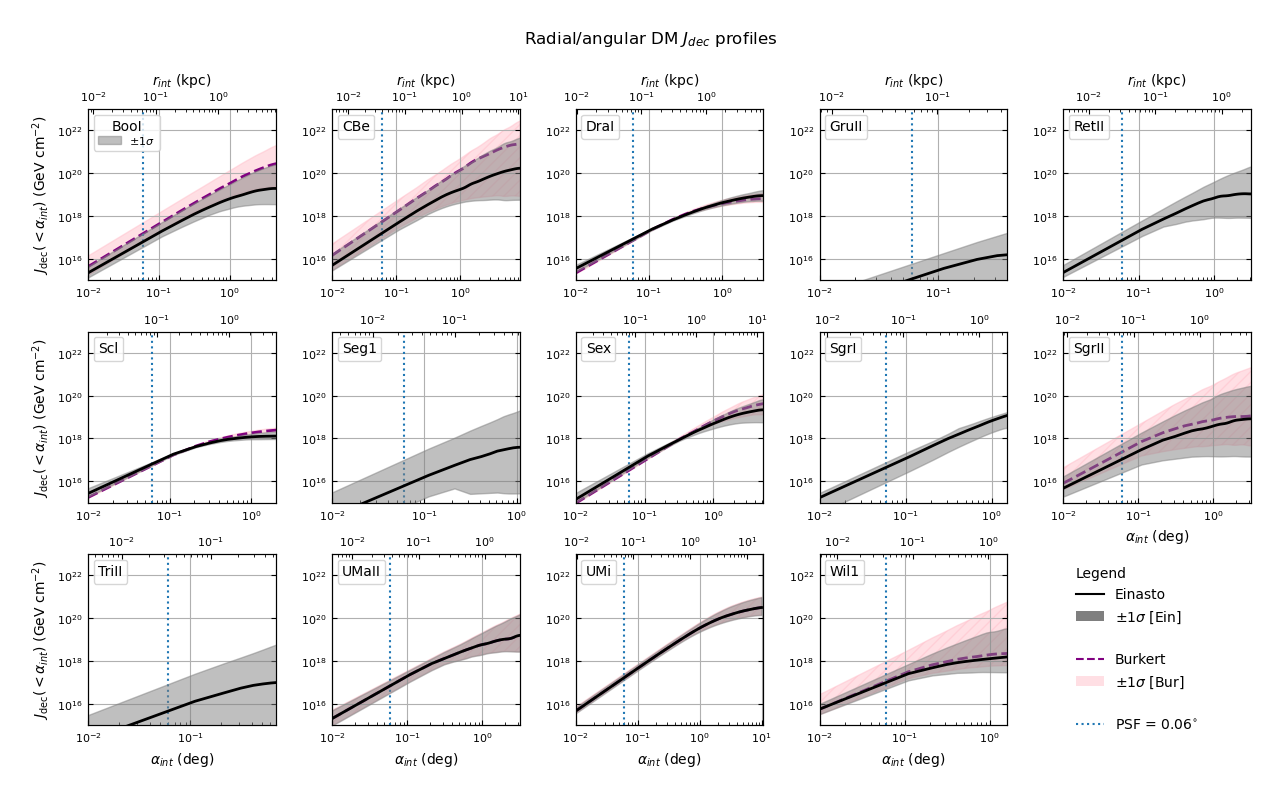}
\caption{Same as \autoref{fig:dsph_profile_ann}, but for the case of DM decay. In this case, also the Burkert profile for RetII is excluded due to a non-finite integration.}
\label{fig:dsph_profile_dec}
\end{figure*}

Lacking any direct information on the DM density distribution in each target, an underlying functional form must be assumed to model its profile. Broadly speaking, there are two classes derived from $N$-body simulations: a cuspy DM profile -- strongly peaked towards the center -- and a cored DM profile -- flat towards the center. We investigate both scenarios, by adopting the Einasto profile \citep{ein65} with 3 free parameters (DM scale density $\rho_s$, DM scale radius $r_s$ and DM inner slope $\alpha$) for the cuspy case, and the Burkert profile \citep{bur95} with 2 free parameters (scale density and radius) for the cored case:
\begin{equation}\label{eq:dm_profiles}
    \rho_{\rm DM}(r) = 
    \begin{cases}
    ~\rho_s \exp{\left\{-\dfrac{2}{\alpha}\left[
    \left(
    \dfrac{r}{r_s}
    \right)^\alpha - 1
    \right]
    \right\}} \qquad \text{Einasto} \\
    \\
    ~\dfrac{\rho_s}{\left(
    1 + \dfrac{r}{r_s}
    \right) \left[
    1 + \left(
    \dfrac{r}{r_s}
    \right)^2
    \right]} \qquad\qquad \text{Burkert}
    \end{cases}
\end{equation}
We opt for the Einasto profile as representative of the class of cuspy DM density profiles, since it is known from the literature that different choices of cuspy parameterizations -- e.g., the Navarro-Frenk-White (NFW) shape \citep{nav97} -- have no impact on the calculation of $J_{\rm ann}$ and $J_{\rm dec}$ for the case of dSphs \citep{bon15c} and, in general, when integrating signals up to angular sizes comparable to the instrumental point-spread function (PSF) in the $\gamma$-ray energy regime \citep{ack15}.

Due to interaction with the gravitational field of the MW, dSphs are expected to lose the outer rims of the DM halo due to tidal interaction \citep[see, e.g.][]{Errani:2022aru}. The exact value of the tidal radius is known with some uncertainties and subject to assumptions. As made by \citet{bon15c}, we compute for each dSph the tidal radius $R_{\rm tid}$, iteratively solving the tidal equation \citep[see \autoref{eq:tidal_radius}]{Springel:2008cc,mol15}, and report it for both the Einasto and the Burkert profiles in \autoref{tab:dsph_parameters_2}. Finally, within the main DM halo of the dSph, smaller DM substructures called DM sub-haloes can retain significant amounts of clumped DM. The contribution of DM sub-haloes to the total astrophysical factor has been subject to strong debates in the past \citep{san11,san14,and19}; such a contribution can be factorized into a  boost factor $\mathcal{B}$ to $J_{\rm ann}$ and $J_{\rm dec}$. $\mathcal{B}$ has the effect of increasing the expected DM $\gamma$-ray flux due to inclusion of the contributions from the DM sub-haloes within the main halo (see \autoref{eq:jboost}). Following results in the literature \citep{san14,mol17}, we conservatively assume $\mathcal{B}=0$. A more in-depth discussion is deferred to \autoref{sec:boost}.  

\subsubsection{Dark matter density profiles for selected dSphs}
With these ingredients, we are able to compute the parametrized DM density profiles $\rho_{\mathrm{DM}}(r)$ for each of the selected dSphs by running 200 independent \texttt{CLUMPY} MC chains of $10^5$ realizations each for every target; for all the free parameters considered in the MCMC Jeans analysis, we adopt the conservative priors determined by \citet{bon15a}. Out of the resulting posterior distributions of the fitted parameters, we derive the astrophysical factor profiles for both the Einasto and Burkert DM profiles using \autoref{eqn:jfac}, along with the median value and the corresponding uncertainties at 68\% confidence level (CL). The Einasto and Burkert profiles for annihilating and decaying DM can be found in the Online Material~\citep{zenodo_dsph}. In \autoref{fig:profile_density_14}, we present such DM density profiles for the 14 selected dSphs, for both the Einasto \citep{ein65} and Burkert \citep{bur95} profiles, also displaying the corresponding tidal radii (see \autoref{eq:tidal_radius}).

Overall, there is a good agreement between the two profiles in the central region before the halo scale radius,
with larger discrepancies in the inner regions for the classical dSphs DraI, Sex and SgrI as well as for the ultra-faint GruII and Seg1  -- where the Einasto profile deviates from the central plateau of the Burkert profile -- whereas SgrII displays larger differences at the outer radii.  We also note that the tidal radii computed for both DM profiles reported in \autoref{tab:dsph_parameters_2} always 
agree well for all of the classical dSphs, and within uncertainties for the ultra-faint ones. In \autoref{fig:dsph_profile_ann}, we report the uncertainty on the tidal radius of CBe and RetII to graphically show this compatibility. The future availability of expanded stellar samples will be crucial to allow more accurate estimates of the gravitational properties of such objects; however, we remark that inaccuracies on the determination of quantities such as tidal radii only have a minor impact on the calculation of $J_{\rm ann}$ and $J_{\rm dec}$, given that the bulk of the expected $\gamma$-ray flux comes from the innermost regions of the dSph haloes and the signal saturation has already been reached at the tidal radius.

\subsubsection{Signal robustness selection}
In order to select the most promising dSphs as CTAO targets, we now compute the astrophysical factor following \autoref{eqn:jfac}, which requires a double integral over the DM density profile (square for annihilating DM). We use the \texttt{CLUMPY} methods also to derive these quantities. This allow us to have a sample of astrophysical factor (median and 68\% containment values) for 30 equally spaced angular distances from $0.01^\circ$ to $10^\circ$. All such tables are reported in the Online Material \citep{zenodo_dsph}. The astrophysical factor profiles of the 14 targets are reported as a function of the instrumental integration angle\footnote{Equivalent to a solid angle element $\Delta\Omega = 2\pi (1 - \cos{\alpha_{\rm int}})$ in \autoref{eqn:dm_flux} and \autoref{eqn:jfac}.} $\alpha_{\rm int}$ -- for both Einasto (cuspy) and Burkert (cored) DM density profiles -- in \autoref{fig:dsph_profile_ann} for the annihilation mode and \autoref{fig:dsph_profile_dec} for the decay mode. In these figures, we also report the CTAO average angular resolution to show that all targets appear between moderately and very extended if integrated out of their rims.

\autoref{fig:dsph_profile_ann} and \autoref{fig:dsph_profile_dec} show that the median profiles and also their uncertainties do not significantly vary between the two cases of cored and cuspy DM distribution, at least for the classical dSphs DraI and Scl and for the ultra-faint dSphs UMaII, UMi and Wil1. The similarity between the resulting DM density profiles computed with \texttt{CLUMPY} for different choices of the DM density functional profile shape was already highlighted by \citet{bon15c} for the class of cuspy profiles -- that also comprises those of Navarro-Frenk-White \citep[NFW;][]{nav97} and the Zhao-Hernquist one \citep{her90,zha96} -- and by \citet{fer15} for the class of cored profiles (see their Fig.~7). For GruII, Seg1, SgrI and TriII in the $J_{\rm ann}$ case, with the addition of RetII in the $J_{\rm dec}$ one, we do not report the Burkert profile because of numerical integration failures in the MC chains that prevented us to obtain finite non-null median values of the astrophysical factor. In addition, for CBe and SgrII, we find a bigger discrepancy between the Einasto and Burkert profiles than for the other targets. We argue that such issues could be caused by the MCMC Jeans analysis being unable to correctly fit the stellar velocity dispersion in these targets with a Burkert density profile, due to either choices of non-optimal priors on the free profile parameters \citep[see][]{bon15a}, or to the data actually preferring a cuspy DM distribution over a cored one (see also \autoref{sec:result_cusp}). A detailed study of the optimal parameter priors to be associated with each choice of fitting DM density profile is out of scope in this paper, and is therefore deferred to a future publication; here, we focus on those targets whose $J_{\rm ann}$ and $J_{\rm dec}$ profiles do not exhibit features of major integration issues.

Out of the 14 dSphs selected in the previous steps, we further narrow down our choice to the top candidates per site. Such a choice is justified by the reasons that (i) CTAO will not likely observe more than a few candidates in the initial years of observation; (ii) the meaningful targets to be observed are those with the highest -- and possibly more precisely determined -- astrophysical factors; and (iii) the DM detection limits scale linearly with the astrophysical factor\footnote{This is true only at a first approximation, i.e. only in case the acceptance is the same for different targets. Such an assumption is not completely true in case of e.g. targets observed at different ZAs or with different spatial extensions.} -- it is therefore easy to estimate obtainable limits for alternative dSphs by considering the appropriate difference in $J_{\rm ann}$ or $J_{\rm dec}$. Motivated by the fact that the dSphs appear as extended targets with respect to the CTAO angular resolution (see \autoref{tab:cta_arrays}), we select those dSphs with the highest $J_{\rm ann}(<0.1^\circ)$ and $J_{\rm ann}(<0.5^\circ)$ using the $J_{\rm ann}(<\alpha_{\rm int})$ profiles; out of the two possibilities, we prioritize those objects with $J_{\rm ann}(<0.1^\circ) \gtrsim 10^{18}$ GeV$^2$ cm$^{-5}$.

\begin{table}
\centering
\caption{Signal fraction enclosed within a determined integration angle for annihilating and decaying DM profiles computed with the Einasto model. The ratio between the astrophysical factor at a given angle and that at the `saturation' level computed at the tidal radius (see \autoref{eq:tidal_radius}) is reported.}
\label{tab:extension}
\begin{small}
\begin{tabular}{c|ccc|ccc}
\hline
\hline
\multicolumn{1}{l}{ } & \multicolumn{3}{c}{A{\scriptsize NNIHILATING} DM} & \multicolumn{3}{c}{D{\scriptsize ECAYING} DM} \\
\cline{2-7}
Name & $<$0.1$^\circ$ & $<$0.5$^\circ$ & $<$1.0$^\circ$ & $<$0.1$^\circ$ & $<$0.5$^\circ$ & $<$1.0$^\circ$ \\
\hline
CBe   & 0.5\%  & 8.2\%  & 19.4\%  & 0.2\%  & 4.3\%  & 10.2\%  \\   
DraI  & 3.5\%  & 34.5\% & 70.3\%  & 2.3\%  & 22.6\% & 46.0\%  \\  
RetII & 1.0\%  & 14.2\% & 33.6\%  &  1.8\%  & 27.3\% & 64.5\%  \\ 
Scl   & 16.0\%  & 33.5\% & 44.7\% & 2.3\% & 68.9\% & 91.8\%  \\   
SgrII & 2.4\%  & 20.6\% & 35.7\%  & 3.0\%  & 26.5\% & 46.0\%  \\  
UMaII & 0.9\%  & 11.2\% & 24.8\%  & 1.4\%  & 17.6\% & 39.3\%  \\  
UMi   & 0.4\%  & 8.7\%  & 28.2\%  & 0.1\%  & 2.9\%  & 9.5\%   \\   
Wil1  & 1.2\%  & 5.1\%  & 7.0\%  & 14.5\% & 60.1\% & 82.8\%  \\  
\hline
\end{tabular}
\end{small}
\end{table}

As a result, we find 8 dSphs satisfying this criterion; the final sample of most promising dSphs to be observed with CTAO is therefore composed by:
\begin{itemize}
    \item DraI, UMi (classical), CBe, UMaII and Wil1 (ultra-faints) for CTAO-N
    \item Scl (classical), RetII and SgrII (ultra-faints) for CTAO-S
\end{itemize}
This selection yields the same objects when selecting for either $J_{\rm ann}(<0.1^\circ)$ or $J_{\rm ann}(<0.5^\circ)$, except for BooI -- that has $J_{\rm ann}(<0.1^\circ) < 10^{18}$ GeV$^2$ cm$^{-5}$ -- and SgrI -- whose $J_{\rm ann}$ profile is however affected by large uncertainties due to its altered stellar dynamics by several gravitational interactions with the innermost regions of the MW potential well. Due to the inclusion of additional targets when integrating up to larger angular sizes, we further investigate the robustness of this selection criterion by performing a scan of additional integration angles, namely 1$^\circ$ and $\alpha_{\rm tid} = \arctan{(R_{\rm tid}/d_\odot)}$. We discuss the ranking for different integration angles in \autoref{sec:ranking_extension}. In \autoref{app:stripping}, we also report the amount of change in the expected DM signal intensity when the full DM density profile is integrated taking into account a model for tidal stripping in the dSph halo outer rims \citep[e.g.,][]{pen10,err21}.

\begin{figure*}
\centering
\includegraphics[scale=0.55,angle=-90]{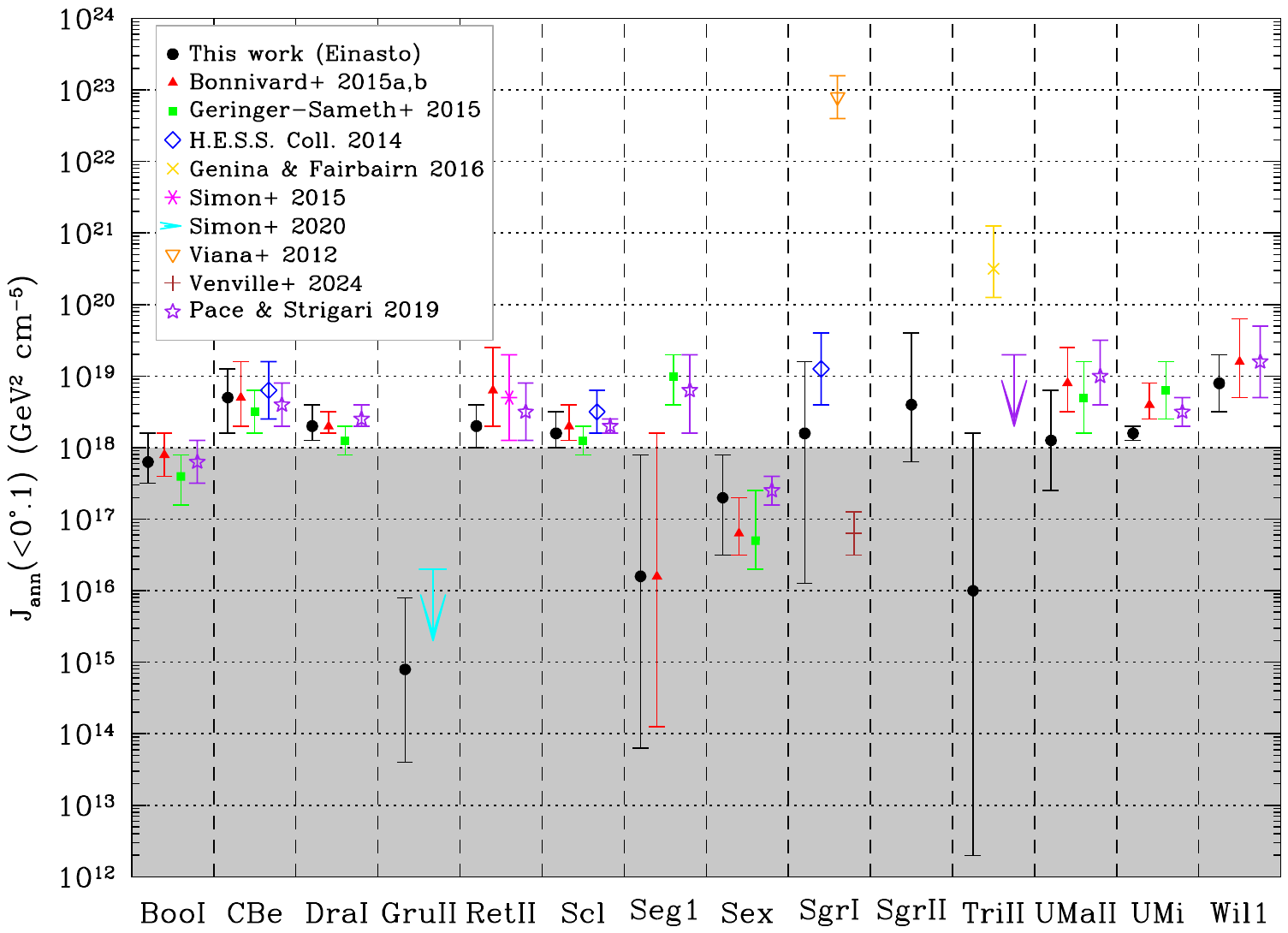}
\caption{Comparison of the astrophysical factors computed in this work for DM annihilation (Einasto density profile) within $0.1^\circ$ of integration $J_{\rm ann}($<$0.1^\circ)$ ({\itshape black dots}) with similar estimates from the literature \citep[see legend]{via12,hes14,bon15c,ger15,sim15,gen16,pac19,sim20,ven24}. The range that has been excluded for identifying optimal targets for CTAO based on their $J_{\rm ann}($<$0.1^\circ)$ values ({\itshape gray area}) is marked. A similar comparison for $J_{\rm ann}($<$0.5^\circ)$ is made in \autoref{fig:dsph_jfactor_05}.}
\label{fig:dsph_comparison_01deg}
\end{figure*}

\subsubsection{Comparison with literature results}
In order to validate our results, we perform a comparison with previous literature estimates. Often, solely the astrophysical factors computed through \autoref{eqn:jfac} at specific integration radii -- normally $J($<$0.1^\circ)$ or $J($<$0.5^\circ)$ -- are reported. Only in some cases \citep[e.g.][]{Bonnivard:2014kza} the full astrophysical profile is reported. For such reasons we show the comparison between our obtained values of $J_{\rm ann}($<$0.1^\circ)$ and those from the literature\footnote{A comparison between our $J_{\rm ann}(<0.5^\circ)$ estimates and the literature is reported in \autoref{fig:dsph_jfactor_05}, and leads to similar conclusions.} in \autoref{fig:dsph_comparison_01deg}. This figure reveals a very good agreement between our determinations and those from the literature; furthermore, we are able to provide calculations of $J_{\rm ann}$ and $J_{\rm dec}$ for the recently discovered SgrII dSph that are based on its stellar kinematics -- whereas only estimates from scaling relations were previously available for this target \citep{mcd23}.

We also confirm that, prior to the analysis by \citet{bon15b}, the astrophysical factor of Seg1 was overestimated by a factor of $>$100 due to the inclusion in its member sample of the spurious stellar population with $\langle v_r \rangle \sim 300$ km s$^{-1}$ (see \autoref{fig:dsph_brightness_profile}). An even more severe overestimation by $>$4 orders of magnitude was made for TriII, due to poorly determined kinematics of its member stars \citep{kir17}. The need for selecting clean kinematic samples in dSph haloes to obtain a reliable measurement of the DM amount is well exemplified by the case of SgrI, which would be classified as a DM-dominated source \citep[$J_{\rm ann} \gtrsim 10^{18}$ GeV$^2$ cm$^{-5}$;][]{via12,hes14} if the gravitational disturbance due to its proximity to the dense Galactic bulge were not known \citep[e.g.,][]{ven24}.

\begin{figure*}
    \centering
    \includegraphics[width=\textwidth]{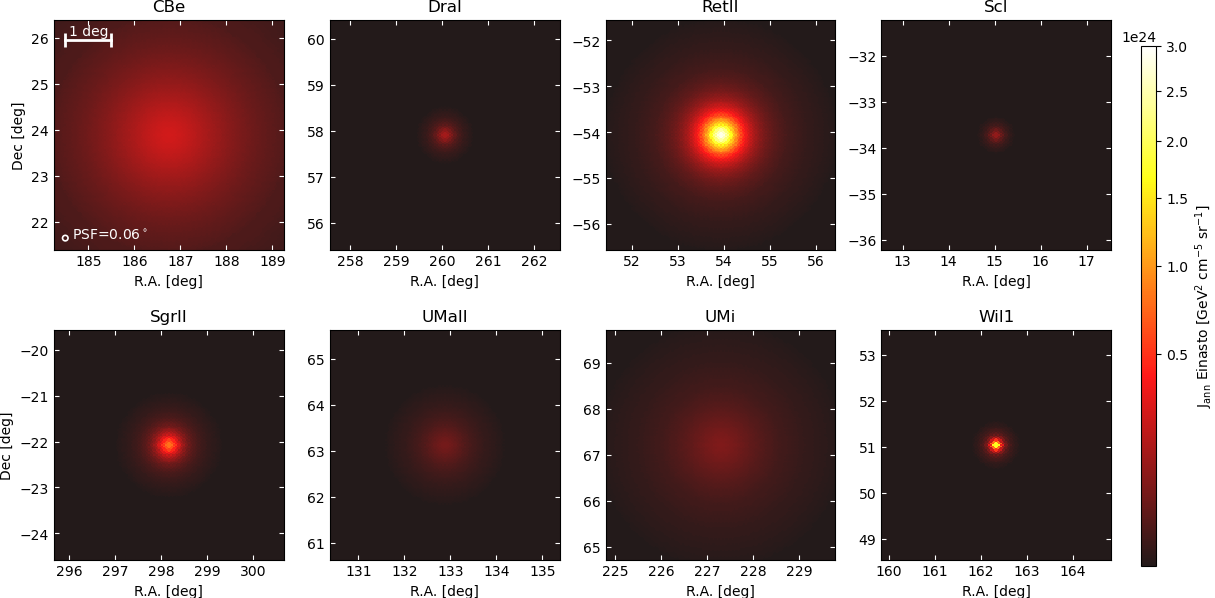}
    \caption{2D distributions of the astrophysical factors of the selected dSphs obtained with \texttt{CLUMPY} for an Einasto profile. The CTAO PSF ({\it white circle}) is indicated in the first panel.}
    \label{fig:jfactor_2Dmaps_einasto}
\end{figure*}

The outer rims of dSphs can in fact be tidally disrupted by the gravitational field of the MW \citep{Errani:2022aru}; in particular, the interaction of the dSph haloes with the deepest regions of the Galactic potential spuriously increases the stellar velocity dispersion, often causing its misinterpretation as due to a large DM content \citep{ger15}. In detail, \autoref{tab:dsph_jfactors} contains at least three debated dSphs, namely SgrI \citep{iba97}, UMaII \citep{mun10} and Wil1 \citep{wil11}, in which tidal disruption is potentially underway. Although we are aware that the inclusion of objects with disturbed kinematics due to this process may lead to severely overestimating their DM content, we decide to keep such targets in our final sample to show how this bias affects the analysis in case of disrupted dSphs, with the only exception of SgrI due to its (weak) $\gamma$-ray background emission\footnote{See \citet{CTA:DM_line} for a detailed discussion of the $\gamma$-ray emission from this target. The CTAO capabilities in observing sources lying close to the Galactic plane are presented in \citet{cta24}.} potentially associated with canonical astrophysical sources \citep{cro22}.

Finally, compared to \citet{mcd23}, our catalog includes TriII and Scl as mentioned before, but excludes HorI, TucII, DraII, TucIV, HyiI, BooII and CarII from among their `Measured' sample. Nevertheless, all these dSphs show low or uncertain values of $J_{\rm ann}$ in their work. We also exclude all of their `Benchmark' dSphs due to their lack of astrometric data, with the exception of GruII (data from \citealt{drl15} and \citealt{sim20}). Finally, we remove BooIII, HorII and VirI with respect to their `Inclusive' sample again due to missing astrometric data, but we keep SgrI and Wil1 that satisfy our selection cuts (data from \citealt{maj03,iba97} and \citealt{mar08,wil11}, respectively).

\section{Analysis methodology}
\label{sec:methods}

In order to predict the significance of the $\gamma$-ray emission for CTAO pointing at the selected dSphs, we follow methodologies that are commonly adopted in the literature \citep[see e.g.][]{fer15}. We make use of the open source code \texttt{gammapy v1.2}~\citep{gammapy,Gammapy:2023gvb}, which is the official CTAO analysis code and is publicly available together with the up-to-date instrument response functions (IRFs)~\citep{cta_irf}. For this work, we use the \texttt{prod5-v0.1} release. IRFs encompass the energy resolution $f_E(E'|E,P)$, the angular resolution $f_P(P'|E,P)$, the effective area $A_{\rm eff}(E,P)$ and the estimated residual background rate. In these definitions, $(E,E')$ stand for true and reconstructed energy, and $(P,P')$ for the true and reconstructed direction. At the moment, all these functions and parameters are estimated via Monte Carlo (MC) simulations of the CTAO detectors, before the actual instruments are built. We also present a cross-check of our results made with the public codes \texttt{ctools}~\citep{ctools} and \texttt{Swordfish} \citep{Edwards:2017kqw,Edwards:2017mnf}, obtaining compatible results (see \autoref{app:systematics}). The results obtained in this paper are reproducible from the Online Material~\citep{zenodo_dsph}. 

Our analysis defines a signal model following \autoref{eqn:dm_flux}, assuming $\langle\sigma v\rangle$ or $\tau$ as parameter of interest for annihilating or decaying DM models and log-normal uncertainty distributions associated to the astrophysical factors computed with \texttt{CLUMPY}. The background model is estimated in energy and radial bins assuming the appropriate CTAO IRFs. We compute the excess counts over $10^4$ realization of the signal and background models. In all cases, we find that the event excess is not significant over the background; therefore, we compute ULs at 95\% CL on $\langle\sigma v\rangle$ or 95\% CL lower limits (LLs) on $\tau$ using the profiles of the likelihood ratio~\autoref{eq:lkl_ratio} and the Wilks theorem for the definition of coverage. All pipelines can be found in the Online Material~\cite{zenodo_dsph}. In the remainder of the section, we go into more details on the specificity on the analysis, while the results are reported in the next section.

\subsection{Signal spectra and morphology}\label{sec:morph}
In order to compute the sensitivity we need to input a spectral and spatial model for the signal. We therefore integrate the DM density profiles of \autoref{fig:profile_density_14} over increasing radial/angular distances up to the tidal radius defined by \autoref{eq:tidal_radius}. The integration is done with \texttt{CLUMPY} utilities, but it can also be done independently.  In \autoref{fig:dsph_profile_ann} we report the annihilating DM profiles and in \autoref{fig:dsph_profile_dec} the decaying DM profiles\footnote{The tables of the astrophysical factor profiles integrated over larger apertures, together with their uncertainties, can be found in the Online Material~\citep{zenodo_dsph}.}. In each figure, we report the median value and the 68\% containment region. We also report a benchmark value for the angular resolution of CTAO for graphical comparison.

From \autoref{fig:dsph_profile_ann}, \autoref{fig:dsph_profile_dec} and \autoref{fig:jfactor_2Dmaps_einasto} one can see that some dSphs, especially the classical, and namely dSphs DraI, Scl, UMi and partly Wil1 have a better defined astrophysical factor, whereas CBe, RetII, SgrII and UMaII show larger uncertainties that also depend on the integration angle. When comparing with the CTAO angular resolution, there is in addition a wide spread in the halo extension, with none of the sources appearing as point-like. In \autoref{tab:extension}, we investigate the fraction of signal encompassed at different integration angle. The fact that the targets are extended has implications on the sensitivity: larger regions of interest (RoIs), where the signal is searched, imply larger contribution from the irreducible background. Given the fact that the signal intensity is not flat but decreasing toward larger distance from the center (see \autoref{fig:dsph_profile_ann} and \autoref{fig:dsph_profile_dec}), it is likely that the outer rims of the targets will not contribute to the signal as much as the inner parts. 
In ON/OFF searches, where the background is estimated from a control region\footnote{We define as ``ON/OFF'' observations both the case where the telescope points at the true source position, and when the control region is in a slightly offset position \citep[false-source or \textit{Wobble} method;][]{Fomin:1994aj}.}, the exact determination of the RoI is relevant \citep[see, e.g.,][]{MAGIC:2020xry}. However, our analysis is based on the so-called \textit{template background} method, in which we model the background acceptance over the entire field of view.

The morphological and physical similarities among different dSph haloes are at the basis of modeling their properties; e.g., \citet{str08} already found that typical dSphs appear to be hosted at the core of DM haloes of approximately the same mass and intrinsic properties, thus leading to similar expected $\gamma$-ray luminosities and hence to the determination of distance-dependent scaling relations, discussed in \S\ref{sec:dm}. Our conclusion is thus that the DM haloes around dSphs are concentrated enough such that the two profiles are similar. For this reason, we will concentrate only in Einasto profiles in the remainder of this work. We remark again that while this is sound for those targets with less uncertain astrophysical factor, we assume this to be true also for the entire sample. A future detailed study of the differences between the cuspy and cored DM density profiles will greatly benefit from improved spectro-photometric data sets, that will be collected by next-generation surveys and facilities before the start of the actual CTAO observations (see a dedicated discussion in \autoref{sec:future}).

\begin{figure*}
    \centering
    \includegraphics[width=0.99\linewidth]{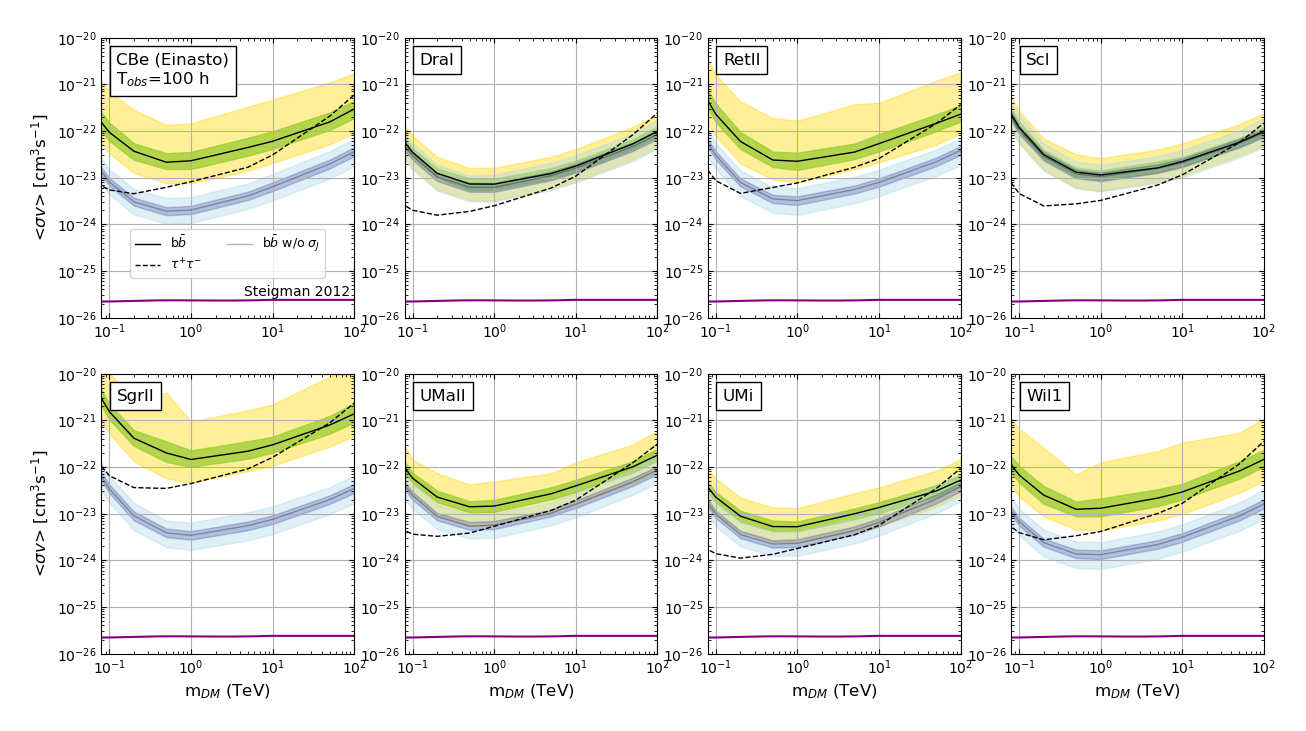}
    \caption{ULs on $\langle \sigma v \rangle$ for annihilating DM for the eight optimal targets, with astrophysical factors computed from the Einasto DM density profile. The median limits ({\it black solid and dashed lines}), along the corresponding 1$\sigma$ ({\it green shaded areas}) and 2$\sigma$ ({\it yellow shaded areas}) statistical uncertainties, are computed assuming 100-h observations of each source and DM annihilating into either the $b\bar{b}$ ({\it black solid lines)} or the $\tau^+\tau^-$ ({\it black dashed lines}) channel and including the uncertainties on $J_{\mathrm{ann}}$ as in \autoref{eq:lkl_combined_noise}. The same limits obtained by excluding that uncertainty (\textit{solid gray lines and blue/light-blue shaded areas}) are also shown. In all panels, the thermal relic value from \citet[{\it purple solid line}]{Steigman:2012nb} is indicated.}
    \label{fig:results_cusp}
\end{figure*}

In \autoref{fig:jfactor_2Dmaps_einasto} we report the 2D skymaps for the selected dSphs for the annihilating DM case and a Einasto profile. These are also computed with \texttt{CLUMPY} but can be generated independently. They are the 2D-representation of \autoref{fig:dsph_profile_ann} and are one of the morphological inputs for the subsequent analysis\footnote{In order to cast the \texttt{CLUMPY} skymaps into \texttt{gammapy}-readable 2D maps in FITS format, one can use the script \texttt{makeFitsImage.py} available in the Online Material~\citep{zenodo_dsph}.}. In the plots, the angular resolution is also shown. One can clearly see again how some dSphs are more extended than others. FITS files for these skymaps can also be found in the Online Material, together with those generated for the decaying DM scenario. 

\subsection{Computation of the background}
The principal source of background is due to misclassified events coming from primary cosmic rays, especially protons, electrons and helium nuclei. To some extent they contribute to an irreducible background, depending on the primary energy and direction. The template prepares a background model sampled in spatial and energy bins, and randomized according to a Poissonian distribution when generating an actual background instance accounting for integration time, energy, angular offset and bin-by-bin acceptance estimated via MC simulations. 

In the template method, for the $i-$the energy bin and $j-$th angular bin, the modeled counts can therefore be written as:
\begin{equation}\label{eq:signal_model}
    n_{ij}(\alpha_s,\alpha_b)=\alpha_s\,n^s_{ij}+\sum_{b}\alpha_b\,n^b_{ij}
\end{equation}
where $\alpha_s$ are the set of parameters influencing the signal count (those in \autoref{eqn:dm_flux} and those from the IRFs) and $\alpha_b$ are the set of parameters influencing the background counts. Once a background instance is generated, we proceed to model the signal for annihilating DM by considering $\langle\sigma v\rangle$ as the parameter of interest and treating all others as nuisance. Analogue consideration can be made for decaying DM replacing $\tau$ as parameter of interest.

In case of only Poissonian fluctuations on $n_{ij}$, the combined likelihood of having $n_{ij}$ counts in all energy and spatial bins assuming our model with $\langle\sigma v\rangle$ as parameter of interest and nuisance parameters $\nu$ can be written as:
\begin{equation}\label{eq:lkl_combined}
 \mathcal{L}(\langle\sigma v\rangle;\mathbf{\nu}|\mathbf{D}) = \prod_{i=1}^{N_E}\prod_{j=1}^{N_P} \frac{\mu_{ij}^{n_{ij}}e^{-\mu_{ij}}}{n_{ij}!}\, ,
\end{equation}
where $\mu_{ij}=\alpha_s\,\mu^{(s)}_{ij}+\sum_{b}\alpha_b\,\mu^{(b)}_{ij}$ is the Poissonian mean of the expected signal ($s$) and background ($b$) counts for each energy and spatial bin, and $\mathbf{D}$ is the simulated data set.

\begin{figure*}
    \centering
    \includegraphics[width=\textwidth]{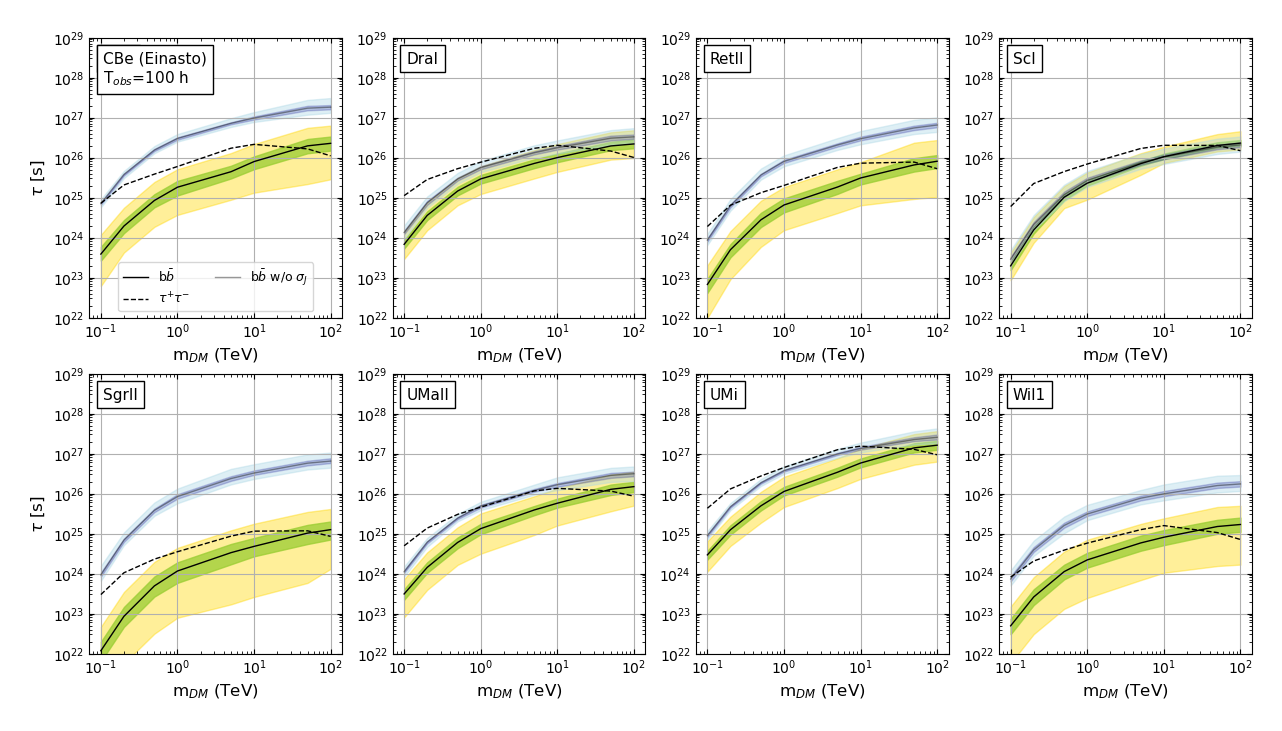}
    \caption{LLs on the particle lifetime for DM (Einasto DM density profile) decaying into the $b\bar{b}$ ({\it solid lines}) and $\tau^+\tau^-$ channels ({\it dashed lines}). In all plots, the statistical uncertainty bands at 1$\sigma$ ({\it green shaded areas}) and 2$\sigma$ ({\it yellow shaded areas}) obtained including the uncertainties on $J_{\mathrm{dec}}$ as in \autoref{eq:lkl_combined_noise} are reported on the $b\bar{b}$ limits. The same limits obtained by excluding that uncertainty (\textit{blue/light-blue shaded areas and solid gray line}) are also shown.}
    \label{fig:results_decay}
\end{figure*}

According to the Neyman-Pearson lemma~\citep{Neyman:1933}, the test statistics that rejects false hypotheses with higher power is the (inverse of the) likelihood ratio between the absolute maximum likelihood (MLE) estimation of the parameters of interest and nuisance combined $\hat{\vec{\alpha}};\hat{\mathcal{\nu}}$ and the MLE of the parameter of interest setting the nuisance parameters as those obtained before. The ratio of the $\ln$~likelihoods can be thus written as: 
\begin{equation}\label{eq:lkl_ratio}
-2\ln\lambda(\langle\sigma v\rangle|\mathbf{D})=\frac{\mathcal{L}(\langle\sigma v\rangle;\hat{\hat{\mathcal{\nu}}}|\mathbf{D})}{\mathcal{L}(\widehat{\langle\sigma v\rangle};\hat{\mathcal{\nu}}|\mathbf{D})}
\end{equation}
\autoref{eq:lkl_ratio} is distributed as a $\chi^2$ probability distribution with one degree of freedom corresponding to our parameter of interest, according to the \citet{Wilks:1938} theorem. If the target is not detected, we produce one-sided upper limits (ULs) at 95\% CL by solving the equation $-2 \ln{\lambda(\langle\sigma v\rangle)} = 4$, with $\langle\sigma v\rangle,\tau$ restricted to the physical region (i.e. positive). 

\subsection{Uncertainties on the astrophysical factor}
The most relevant contribution to systematic uncertainties stems from the computation of the astrophysical factor. As shown in \autoref{fig:dsph_profile_ann} and \autoref{fig:dsph_profile_dec}, the uncertainties in the profiles can be larger than 1 dex  especially in the case of ultra-faint dSphs; furthermore, the uncertainty depends on the integration angle, generally increasing with the aperture (an example of posterior distribution for different integration angles is shown in \autoref{fig:jfactor_PDF}). This means that, in principle, one should compute a different posterior distribution for each spatial bin of \autoref{eq:lkl_combined}, which may significantly complicate the computation. To overcome this issue, we modify the likelihood \autoref{eq:lkl_ratio} following the approach of \citep{CTA:DM_line}, i.e. including in \autoref{eq:lkl_combined} the distributions of astrophysical factor realizations $\mathcal{G}(\Tilde{J},\sigma_J)$ parametrized as a log-normal (see \autoref{app:jfactor_sys}):
\begin{equation}\label{eq:lkl_combined_noise}
 \mathcal{L}(\vec{\alpha};\mathbf{\nu}|\mathbf{D},J)= \mathcal{L}(\vec{\alpha};|n_{ij}(\mathbf{\nu}),J)   = \prod_{i=1}^{N_E}\prod_{j=1}^{N_P} \frac{\mu_{ij}^{n_{ij}}e^{-\mu_{ij}}}{n_{ij}!}\cdot \ln\mathcal{G}(\Tilde{J},\sigma_J)
\end{equation}
We then re-run our simulations, using \autoref{eq:lkl_combined_noise} with $10^4$ trials for each target. Throughout this work, we compare the results obtained with and without the extra term for the astrophysical factor uncertainties in \autoref{eq:lkl_combined_noise}, since this plays a fundamental role in both the characterization of the DM exclusion limits and the choice of optimal targets.

\subsection{Other systematic uncertainties}
The discussion of additional systematics is deferred to \autoref{app:systematics}. Here we provide a brief summary. Overall, all other systematics are largely subdominant with respect to the uncertainty in the astrophysical factor:
\begin{itemize}
 \item As discussed, we used a template-background approach. An alternative method, based on the so-called ON/OFF analysis, in which the background is estimated from a background control (OFF) region and applied to the signal region (ON) may produce different constraints. The change is around 10\% at 300 GeV and less then 2\% above 300 GeV, see upper left Fig. \autoref{fig:systematics}. 
 \item The use of \texttt{gammapy} rather than alternative reconstruction tools, such as the public software \texttt{ctools}~\citep{ctools} or the Asimov dataset-based sensitivity code \texttt{swordfish} \citep{Edwards:2017kqw,Edwards:2017mnf} provide differences of less than 10\% throughout the DM mass range.
\item The uncertainties on the IRFs, obtained by bracketing the IRFs, introduce a difference of less than 2\%; See \autoref{app:systematics} for more information.
\item In our analysis, we neglected the contribution of other sources of $\gamma$-rays, such as as those coming from the diffuse $\gamma$-ray background. The amount of such contribution depends on the Galactic latitude and is increasing toward the galactic center plane. At the latitude of the dSphs into consideration, this contribution is negligible.
\end{itemize}
Similar systematic uncertainties are found in other DM-oriented CTAO searches \citep{cta19,CTAConsortium:2023yak,CTA:DM_line}.

\begin{figure*}
    \centering
    \includegraphics[width=0.48\textwidth]{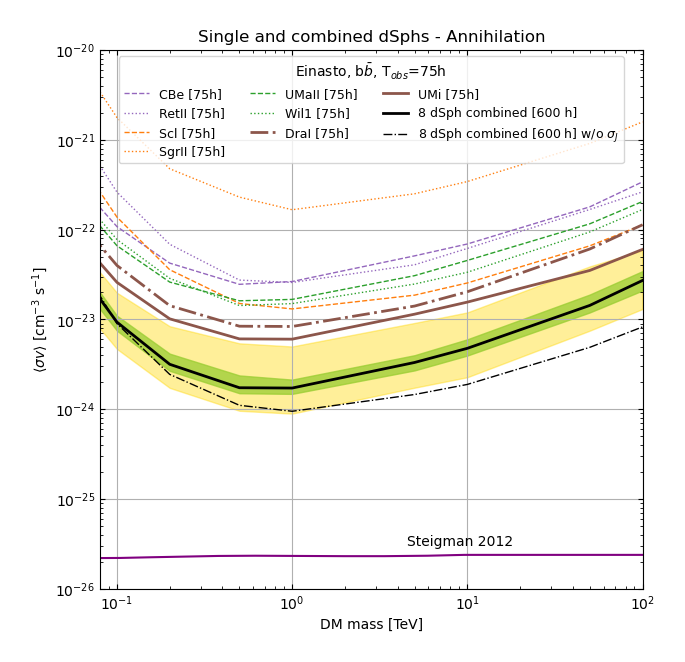}
    \includegraphics[width=0.48\textwidth]{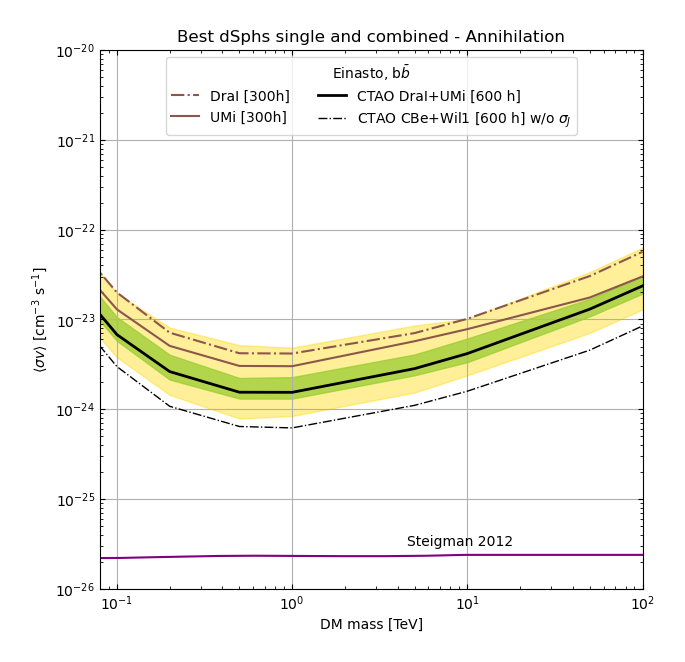}
    \includegraphics[width=0.48\textwidth]{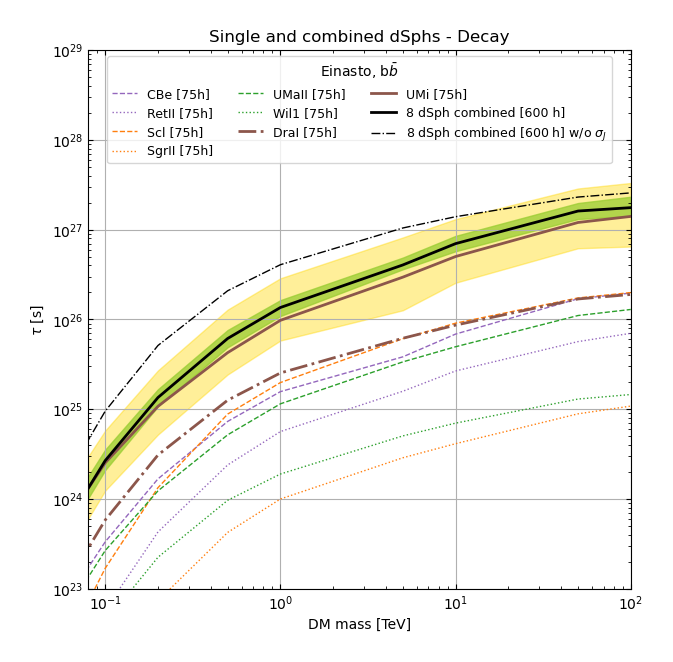}
    \includegraphics[width=0.48\textwidth]{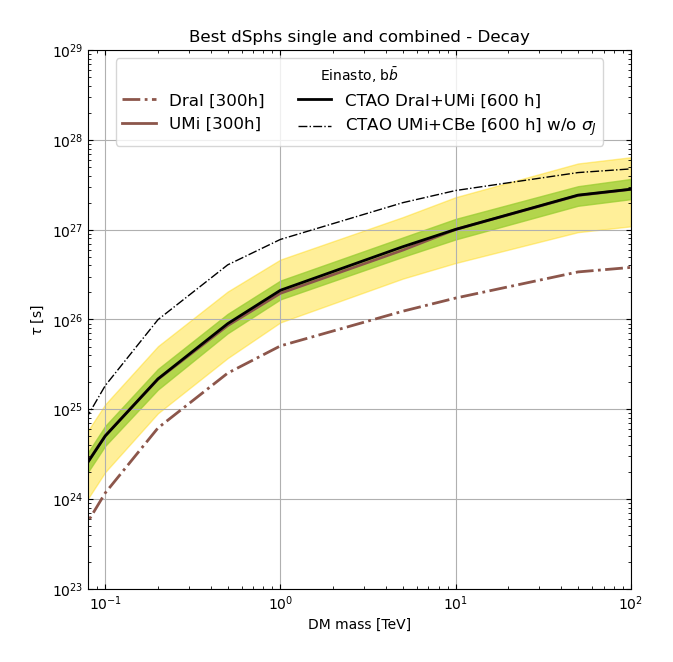}
        \caption{Constraints on the DM annihilation cross section and decay lifetime from combined likelihood analyses, both including ({\it black solid lines}) and excluding uncertainties in the dSph astrophysical factors ({\it black dot-dashed lines}). {\it Left panels:} combination of the limits from all of the optimal dSphs observed for 75 h each. {\it Right panels:} combination of the limits of the two best dSphs observed for 300 h each. In all panels, the uncertainties on the combined cross-section limit due to photon statistics at 1$\sigma$ ({\itshape green shaded area}) and 2$\sigma$ CL ({\itshape yellow shaded area}) are reported along with the thermal-relic limit \citep[{\it purple solid line};][]{Steigman:2012nb}}
    \label{fig:results_combined}
\end{figure*}


\begin{figure*}
    \centering
        \includegraphics[width=0.48\textwidth]{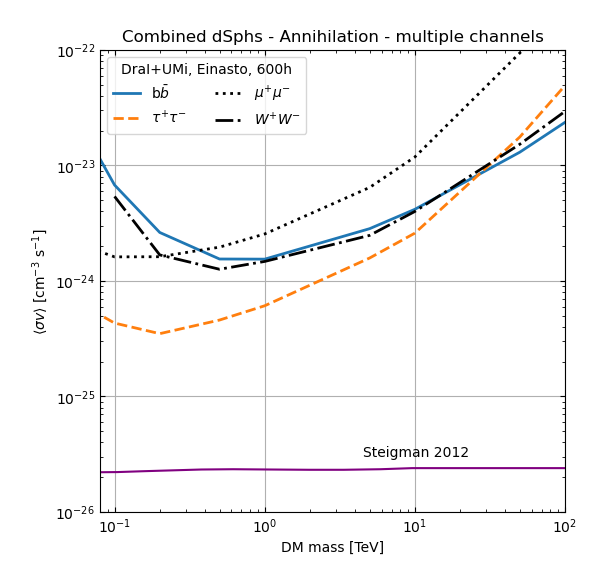}
        \includegraphics[width=0.48\textwidth]{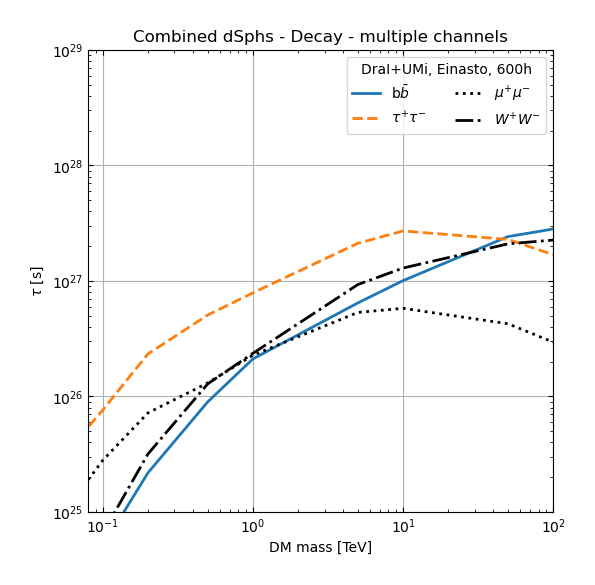}
    \caption{Constraints on the DM annihilation cross section and decay lifetime for other choices of the SM interaction channel. {\it Left panel:} ULs on DM annihilation with 600 h of combined observations of Wil1 and CBe (Einasto DM density profile); the thermal-relic cross section by \citet[{\it purple solid line}]{Steigman:2012nb} is indicated as reference. {\it Right panel:} LLs on the particle lifetime for DM decay with 600 h of combined observations of CBe and UMi (Einasto DM density profile).}
    \label{fig:results_otherspectra}
\end{figure*}

\section{Results}
\label{sec:results}
\textbf{In the case of} no detection for any combination of the DM channels and celestial targets presented above, we can compute the CTAO expected limits at 95\% CL to the DM parameters using the procedure described in \autoref{sec:methods}. To this end, we consider $\langle \sigma v\rangle$ and $\tau$ as free parameters in \autoref{eq:lkl_ratio} and maximize the likelihood of \autoref{eq:lkl_combined_noise} that includes the uncertainties on the astrophysical factor. We organize the results as follows: in \autoref{sec:result_cusp}, we present the ULs for DM annihilation considering cuspy DM density profiles for the optimal targets defined in our study (see \autoref{fig:results_cusp}); in \autoref{sec:results_decay}, we show the LLs on the particle lifetime for models of decaying DM (see \autoref{fig:results_decay}); in \autoref{sec:results_combined}, we discuss the CTAO sensitivity for the case of a combined likelihood analysis of multiple sources (see \autoref{fig:results_combined}). In all of these subsections, we consider the cases of DM annihilation and decay into the $b\bar{b}$ and $\tau^+\tau^-$ channels, taken as representative examples of a soft and a hard DM $\gamma$-ray spectrum, respectively (see \autoref{sec:dm-particle}). Furthermore, we show how the cross-section limits change when considering other annihilation channels such as $W^+W^-$ and $\mu^+\mu^-$ (see \autoref{fig:results_otherspectra}). These results are put into context and discussed in \autoref{sec:discussion}. We provide numerical values for all limits in the Online Material~\citep{zenodo_dsph}.

\subsection{Upper limits to the cross section for DM annihilation}
\label{sec:result_cusp}
In \autoref{fig:results_cusp}, we report the ULs on the velocity-averaged annihilation cross section for the eight optimal dSphs, assuming 100-h observations with the respective CTAO array (either CTAO-N for Northern sources, or CTAO-S for Southern targets) and annihilation in the $b\bar{b}$ or the $\tau^+\tau^-$ channels. For the $b\bar{b}$ channel, we also report the 1$\sigma$ and 2$\sigma$ statistical uncertainties along with the comparison of the limits with those that would have been obtained by neglecting the astrophysical factor uncertainty -- i.e. using \autoref{eq:lkl_combined} in place of \autoref{eq:lkl_combined_noise}. DraI and UMi provide the strongest constraints considering such an uncertainty, replaced by Wil1 and CBe if we only consider the median value of $J_{\mathrm{ann}}$. The reason for the more constraining nature of the leptonic limits with respect to the hadronic ones is due to the different spectral shape in this mass range (see \autoref{fig:spectra}).

\subsubsection{Impact of the uncertainties on the astrophysical factors} 
\autoref{fig:results_cusp} clearly displays the impact of the astrophysical factor uncertainty on the DM limits. The green/yellow shaded area refers to limits obtained accounting for such uncertainties, and the blue/light blue shaded area to those obtained considering the $J_{\rm ann}$ median values only. This choice has no or negligible impact for the classical dSphs DraI and Scl thanks to both the good knowledge of their stellar velocities and the availability of more member stars with respect to the ultra-faint dSphs. These latter targets exhibit a worsening of the limits by a factor of $\gtrsim$10: this finding has strong implications in terms of both sensitivity reach and especially target selection (see \autoref{sec:strategy}), demonstrating the need to collect high-quality data on dSphs for this kind of studies.

\subsubsection{Impact of cuspy and cored DM density profiles}
\label{sec:corecusp}
We have shown in \autoref{fig:profile_density_14} how cored DM density profiles are very similar to the cuspy ones, thus providing astrophysical factors that are comparable to each other within 1$\sigma$ uncertainties when integrated along the line of sight. We refer to \autoref{tab:dsph_jfactors} for a depiction of this fact; the limits to the DM parameters obtained for cored DM distribution would therefore be compatible with those shown in \autoref{fig:results_cusp}, as already found by \citet{ack15} with cored profiles implying changes in the predicted limits by a factor of at most $\sim$40\%. For this reason, we do not report them independently in this paper, always referring to the astrophysical factors obtained for the case of a cuspy DM profile in the following. However, in case of significant detection, \citet{Hiroshima:2019wvj} demonstrated that CTAO has the capabilities to discriminate between cuspy and cored profiles in dSphs.

\subsection{Lower limits on the DM lifetime}
\label{sec:results_decay}
The case of decaying DM requires a similar analysis, however: (i) the signal model having the particle DM lifetime $\tau$ as free parameter (\autoref{eqn:dm_flux});
(ii) the spectral photon yield only extending up to $m_{\mathrm{DM}}/2$ due to the energy budget of the process;
(iii) the astrophysical factor being the integral of the linear DM density rather than its square (see \autoref{eqn:jfac}) over the line of sight and the solid angle, shown in \autoref{fig:dsph_profile_dec} as a function of the integration angle $\alpha_{\rm int}$. The LLs on the DM particle lifetime, reaching values above $10^{27}$~s for the prototypical $b\bar{b}$ and $\tau^+\tau^-$ channels, are shown in \autoref{fig:results_decay}, with DraI and UMi again providing the strongest constraints taking into account their astrophysical factors uncertainties in place of CBe and UMi when neglecting them.

\subsection{Combined results for multiple targets}
\label{sec:results_combined}
Although the observational strategy of CTAO on dSphs is not ultimately defined, it was discussed e.g. in \citet{cta19} that an observing time of $\sim$100 h will be allocated on one or more dSphs. Considering that the amount of observing time allocated for the CTAO key science programs (KSPs) -- with the indirect searches of $\gamma$-ray signals from DM annihilation or decay in dSph haloes being one of them -- will considerably decrease with time in favor of guest programs, it is reasonable to propose that CTAO invests a total of $500 - 600$~h  shared between both sites for the observation of dSphs. To discuss the distribution of such a significant amount of time among the optimal dSphs, we present three scenarios:
({\it a}) a combination of the observations of all the eight optimal dSphs, observed for 75~h each, taking into account the respective astrophysical factor uncertainties;
({\it b}) a combination of 600-h observations of the two overall best targets (300 h/target) taking into account their astrophysical factor uncertainties;
({\it c}) a combination of 600-h observations of the two overall best targets (300 h/target) neglecting their astrophysical factor uncertainties (see below).

\subsubsection{Combined results for the annihilating DM scenario} 
For case ({\it a}), we combine the sources by profiling the expected uncertainties on the values of $J_{\rm ann}$ over their log-normal posterior distributions (see \autoref{eq:lkl_combined_noise}). For case ({\it b}), we instead consider the two targets providing the largest expected signal,  i.e. DraI and UMi. Finally, Wil1 and CBe are considered for case ({\it c}). We report the results in \autoref{fig:results_combined}.

Comparing strategies ({\it a}) and ({\it b}), we find that both allow to reach similar limits, with the combination of two deep observation marginally more sensitive. This is related to the fact that, while the targets with the largest astrophysical factors dominate the limits, the inclusion of the uncertainties in the likelihood maximization process plays a significant role. We remark that, neglecting the astrophysical factor uncertainties, the combined limits of all the optimal dSphs would be somewhat more constraining, and that the best targets would be CBe and Wil1, whose combined limits would be significantly more constraining (dot-dashed lines in the upper right panel of \autoref{fig:results_combined}).

Furthermore, we compute the cross-section ULs for the bosonic $W^+W^-$ channel and the leptonic $\mu^+\mu^-$ channel in addition to those for the $b\bar{b}$ and $\tau^+\tau^-$ channels. In \autoref{fig:results_otherspectra}, we compare these limits together by analyzing the combination of 300-h observation of Wil1 and CBe. This figure shows that the $b\bar{b}$ and $\tau^+\tau^-$ channels are indeed representative of several cases: in fact, the $b\bar{b}$ spectrum is similar to other hadronic or bosonic channels such as $W^+W^-$. The  $\tau^+\tau^-$ hard channel represents the more constraining case, with a lighter leptonic channel such as the $\mu^+\mu^-$ providing  weaker constraints but similar in shape. The $\tau^+\tau^-$ channel is the most constraining for both the annihilation and decay cases, except at very large masses (above $\sim$40~TeV) where the $b\bar{b}$ channel start to be dominant.

\begin{figure*}
    \centering
    \begin{minipage}{0.49\textwidth}
        \includegraphics[width=\textwidth]{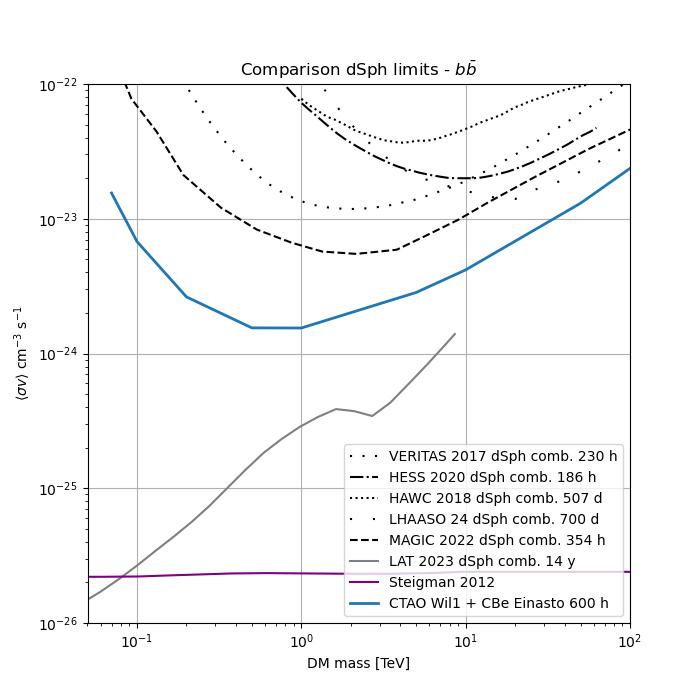}
    \end{minipage}
    \begin{minipage}{0.49\textwidth}
        \includegraphics[width=\textwidth]{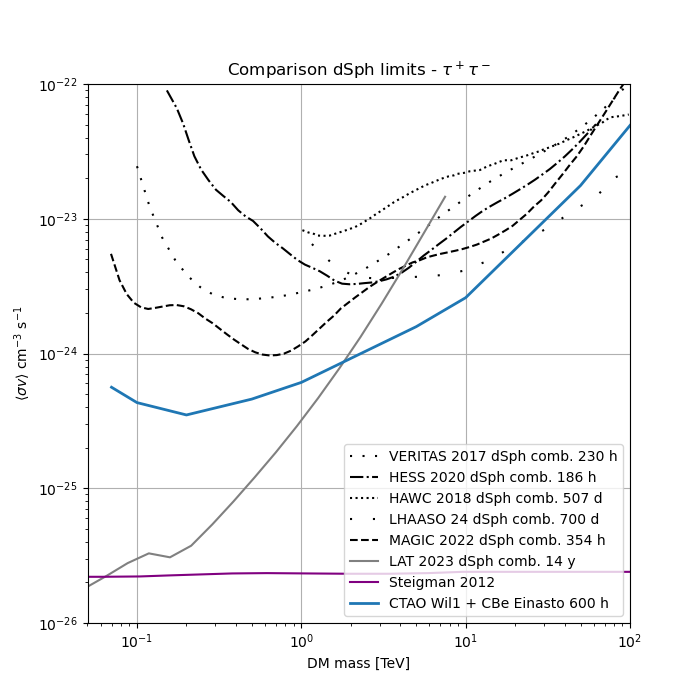}
    \end{minipage}
    \caption{ULs to the DM annihilation cross section with simulated CTAO observations of dSphs ({\it blue solid lines}) compared to the most constraining limits obtained with observations of dSphs by current $\gamma$-ray facilities -- MAGIC \citep{acc22}, HAWC \citep{HAWC:2017mfa}, HESS~\citep{HESS:2020zwn}, {\it Fermi}-LAT~\citep{Hoof:2018hyn}, VERITAS~\citep{VERITAS:2017tif}. The CTAO limits for dSphs are obtained with 600-h combined observations of Wil1 and CBe, assuming an Einasto DM profile. {\it Left panel:} limits for the $b\bar{b}$ channel. {\it Right panel:} limits for the $\tau^+\tau^-$ channel. In both panels, the value of the thermal-relic cross section by \citet[{\it purple solid line}]{Steigman:2012nb} is indicated for reference.}
    \label{fig:compare_dsph}
\end{figure*}

\subsubsection{Combined results for the decaying DM scenario} 
We also report LLs on the particle lifetime for DM decay channels that produce the $dN_\gamma/dE_\gamma$ photon yield term of the expected DM flux in \autoref{eqn:dm_flux}, obtained by repeating, for the case of decaying DM, the analysis presented above. Also in this case, the two best targets are DraI and UMi if we take into account the uncertainty on $J_{\mathrm{dec}}$ or CBe and UMi if we neglect it. We reach conclusions that are similar to those that hold annihilating case, with the difference that the limits are dominated by UMi alone. We show these results in \autoref{fig:results_combined}, also comparing them in \autoref{fig:results_otherspectra} for different decay channels.

\section{Discussion}
\label{sec:discussion}
We now discuss the results obtained in \autoref{sec:results}, and place them in the wider context. We discuss the prospects for discovery of new dSphs or for refinement of their astrophysical factors in the next decade (see \autoref{sec:future}). We compare our results compared to present and future $\gamma$-ray experiments and observatories in \autoref{sec:context}, and subsequently describe possible optimizations of the observational strategy in \autoref{sec:strategy}.

\subsection{Expected observations of dSphs in the next decade}
\label{sec:future}
Over several decades of efforts, it is plausible that only a fraction of all dSphs residing in the MW halo has been discovered so far. Estimating a maximal number of them requires a certain set of assumptions, based on both $N$-body simulations and theoretical arguments. If we follow \citet{Newton:2017xqg}, we expect for a MW-like galaxy $124^{+40}_{-27}$ (68\% confidence) satellite galaxies brighter than $M_V=0$ within 300~kpc of the Sun. The prediction is based on the {\it Aquarius} simulation, from the number of satellites detected by the SDSS and the Dark Energy Survey \citep[DES;][]{san16}. Of the expected 124, $46^{+12}_{-8}$ are ultra-faint dSphs ($-8 \leq M_V \leq -3)$ and $61^{+37}_{-23}$ are hyper-faint ($M_V \geq -3$). Roughly half the predicted number of ultra-faint dSphs has actually been discovered, and the observation of hyper-faint targets requires a survey $\sim$4 mag deeper than DES.

\begin{figure*}
    \centering
    \includegraphics[width=0.48\textwidth]{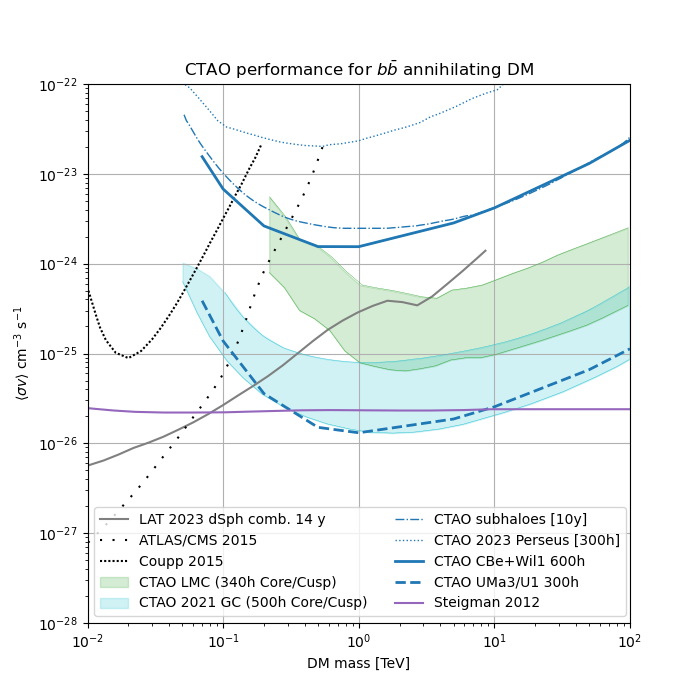}
    \includegraphics[width=0.48\textwidth]{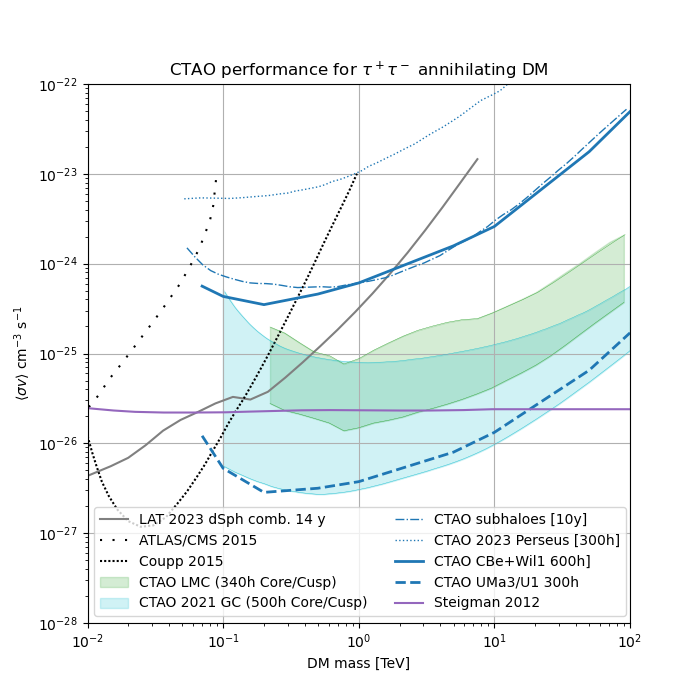}
    \caption{ULs to the DM annihilation cross section with simulated CTAO observations of dSphs ({\it blue solid lines}) compared to the most constraining limits obtained by current $\gamma$-ray facilities \citep[{\it Fermi}-LAT;][]{Hoof:2018hyn} and to CTAO prospects on other targets -- the LMC \citep{ach23}, the GC \citep{ach21}, the DM sub-haloes~\citep{Coronado-Blazquez:2021avx} and the Perseus cluster~\citep{CTAConsortium:2023yak}) -- and from direct detection experiments and accelerators~\citep{BAUER201516}. The CTAO limits for dSphs are obtained with 600-h combined observations of DraI and UMi, assuming an Einasto DM profile. {\it Left panel:} cross section limits for the $b\bar{b}$ channel. {\it Right panel:} limits for the $\tau^+\tau^-$ channel. In both panels, the value of the thermal-relic cross section by \citet[{\it purple solid line}]{Steigman:2012nb} is indicated. The projected limits for UMaIII/U1 discussed in the text ({\it blue dashed line}) are also shown.}
    \label{fig:compare_instrument}
\end{figure*}

In order to improve the modeling of the DM distribution, one needs accurate spectroscopic information to infer the velocity distribution, and ultimately the actual gravitational potential through the Jeans formalism. Also, if the number of identified stars is few tens, any prediction on the dSph astrophysical factor less uncertain than a factor of $10 - 100$ is hard to obtain \citep[see the cases of Seg1 and TriII as examples; see also figure 3 of][]{chi19}. To date, several observational projects have surveyed in great detail some fraction of the sky in search of dSphs, with a comparable number currently underway or planned:
\begin{itemize}
    \item DES \citep{san16}, that ended in 2019, was an international effort dedicated to map $\mathcal{O}(10^{11})$ galaxies, in order to unveil the nature of the so-called dark energy\footnote{See \url{https://www.darkenergysurvey.org/}.} \citep{cal09,ame10,lim11,kun12};
    \item {\it Gaia} \citep{omu00}, launched in 2013 and ended in 2025, is a satellite mission aimed at charting a three-dimensional map of the MW, with accurate positional and radial velocity measurements needed to produce a stereoscopic and kinematic census of $\mathcal{O}(10^9)$ stars in our Galaxy and throughout the Local Group\footnote{See \url{https://sci.esa.int/web/gaia}.};    
    \item the SDSS \citep{yor00} is the first facility providing multi-epoch optical spectroscopy across a large fraction of the sky, as well as now offering contiguous integral-field spectroscopic coverage of the MW and Local Volume galaxies\footnote{See \url{https://www.sdss.org}.}; in past years, its observations allowed the discovery the first ultra-faint dSphs;
    \item the Hyper Suprime-Cam Subaru Strategic Program \citep[HSC-SSP;][]{aih18} is a three-layered, multi-band ($grizy$ plus 4 narrow-band filters) imaging survey with the Hyper Suprime-Cam (HSC) on the 8.2-m Subaru Telescope that took data until 2020\footnote{See \url{https://hsc.mtk.nao.ac.jp/ssp/survey/}.};
    \item the Panoramic Survey Telescope and Rapid Response System 1 \citep[Pan-STARRS1;][]{fle20} is an optical and near-infrared survey that covered the entire sky north of declination $-30^\circ$, including the Galactic plane, until 2014\footnote{See \url{https://panstarrs.stsci.edu/}.}.
    \item the DESI survey \citep{dey19} seeks to map the large-scale structure of the Universe over a wide range of look-back times with the Dark Energy Spectroscopic Instrument \citep{fla14} mounted since 2019 on the 4-m Mayall Telescope at the Kitt Peak National Observatory, targeting $\sim$3$\,\times 10^7$ pre-selected galaxies across $\sim$1/3 of the night sky\footnote{See \url{https://www.desi.lbl.gov/}.};
    \item the {\it Euclid} mission \citep{Euclid:2024yrr} detects main sequence stars up to Galactocentric radii of $\sim$100 kpc, providing details for dSphs lying in the outer MW halo\footnote{See \url{https://www.esa.int/Science_Exploration/Space_Science/Euclid}.};
    \item the WEAVE Project \citep{jin23} is a survey plan aimed at exploiting the capabilities of the WEAVE (WHT Enhanced Area Velocity Explorer) multi-object survey spectrograph for the 4.2-m William Herschel Telescope (WHT) at the Observatorio del Roque de los Muchachos (La Palma, Canary Islands), that allows astronomers to take optical spectra of up to $\sim$1000 targets over a $2^\circ$ FoV in a single exposure with a resolving power of up to 20,000\footnote{See \url{http://www.ing.iac.es/weave/index.html}.};
    \item the planned ESA “Analysis of Resolved Remnants of Accreted galaxies as a Key Instrument for Halo Surveys” (ARRAKIHS) mission will image 50 square degrees of the sky per year down to an unprecedented ultra-low surface brightness simultaneously achieved in two visible bands, providing important insights on the populations of ultra-faint dSphs~\citep{Arrakhis}.
\end{itemize}

To catalog the entire population of dSphs, these surveys are accompanied by follow-ups of DES, in particular the Survey of the Magellanic Stellar History\footnote{See \url{https://datalab.noirlab.edu/smash/smash.php}.} \citep[SMASH;][]{nid17} since 2012 and the Magellanic Satellites Survey \citep[MagLiteS;][]{drl17} since 2017 that are dedicated to surveying ample sky regions around the Magellanic Clouds. In addition, future instruments tailored for this specific task will become operational; such efforts include the recently inaugurated ``Vera C. Rubin'' Observatory\footnote{See \url{https://www.lsst.org/} and \url{https://rubinobservatory.org/}.} \citep{ive19}, and 4MOST (4-metre Multi-Object Spectroscopic Telescope), a wide-field, high-multiplex, fibre-fed, optical spectroscopic survey facility to be mounted on ESO's 4-m-class telescope VISTA\footnote{See \url{https://www.4most.eu/cms/home/}.} \citep{dej19}, for both of which the first light is expected in 2026.

Spectroscopic measurements are already within reach of facilities such as the Deep Imaging Multi-Object Spectrograph \citep[DeIMOS\footnote{See \url{https://www2.keck.hawaii.edu/inst/deimos/}.};][]{fab03}, the Michigan/Magellan Fiber System \citep[M2FS\footnote{See \url{https://ui.adsabs.harvard.edu/abs/2012SPIE.8446E..4YM/abstract}.};][]{mat12} and GIRAFFE\footnote{See \url{https://www.eso.org/sci/facilities/paranal/instruments/flames/inst/Giraffe.html}.} \citep{roy14}, all currently operational since $\gtrsim$10 years. In the future, the European Extremely Large Telescope \citep[E-ELT\footnote{See \url{https://elt.eso.org/}.};][]{gil07}, currently under construction at the Cerro Armazones (Chile) site -- close to CTAO-S -- will also sample dSph member stars with unprecedented sensitivity.

A complementary approach to evaluate the future prospects consists in performing analytical or semi-analytical estimates of the number of MW satellites, with the aim of predicting the abundance of potentially highly DM-dominated dSphs to be discovered in the future. For example, \citet{Ando:2019rvr} compute the statistics of objects discovered by LSST by adopting models of MW halo substructures and phenomenological prescriptions connecting sub-haloes to satellite galaxies. In this way, they find that $\sim$1 target with $\log{J_{\rm ann}(<0.5^\circ)} \geq 19.4$ is expected 5\% of the time, consistent with our computation of the astrophysical factor for CBe. More optimistically, \citet{Coronado-Blazquez:2021avx} pointed out, based on predictions from the DM-only {\itshape Via Lactea II} simulations \citep[see e.g.][for results from hydrodynamical simulations]{kel19}, that CTAO will likely detect $\gtrsim$5 to $\gtrsim$10 targets with $\log{J_{\rm ann}}($<$0.5^\circ) \gtrsim 19$, and $\sim$1 with $\log{J_{\rm ann}}($<$0.5^\circ) \gtrsim 20$, considering the cumulated exposure over the sky.

Such results further highlight that, with a wide range of spectro-photometric instruments already available and more performing facilities planned for the near future, dedicated observational campaigns for the discovery of dSphs and more detailed investigation of the stellar content of those presently known should become operational before the advent of CTA. The expected results from observations carried out with such facilities will impact for both the discovery of new nearby, low-luminosity dSphs that had remained undetected so far \citep[e.g.,][]{drl20} and are possibly hosted inside massive DM haloes, and for the better determination of the kinematic properties of (ultra-faint) sources with sparsely populated stellar samples (see \autoref{tab:dsph_parameters}), for which an improvement in the determination of $J_{\rm ann}$ and $J_{\rm dec}$ is anticipated\footnote{Preliminary tests show that the Jeans analysis on the dSph kinematics produces uncertainties that scale with $N^{-1/2}$ (i.e. according to a Poissonian statistics), where $N$ is the number of confirmed member stars.}.

\begin{figure*}
    \centering
    \includegraphics[width=0.48\textwidth]{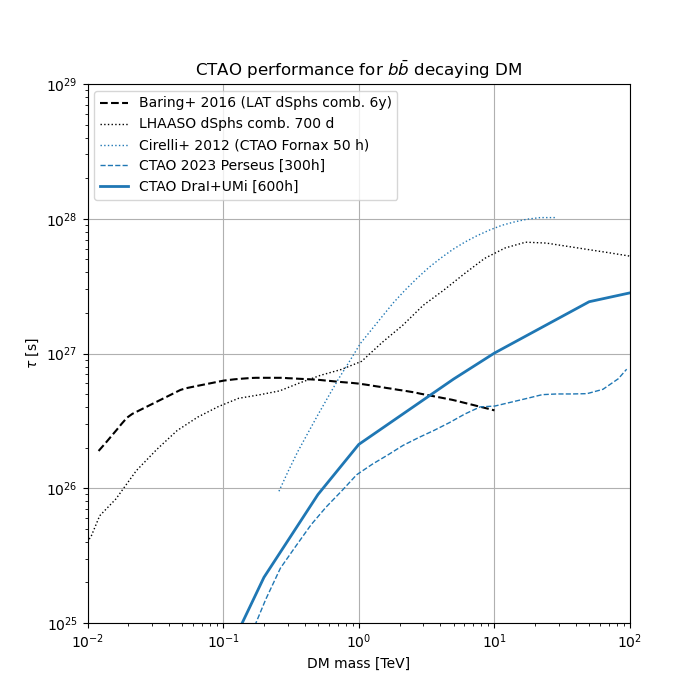}
    \includegraphics[width=0.48\textwidth]{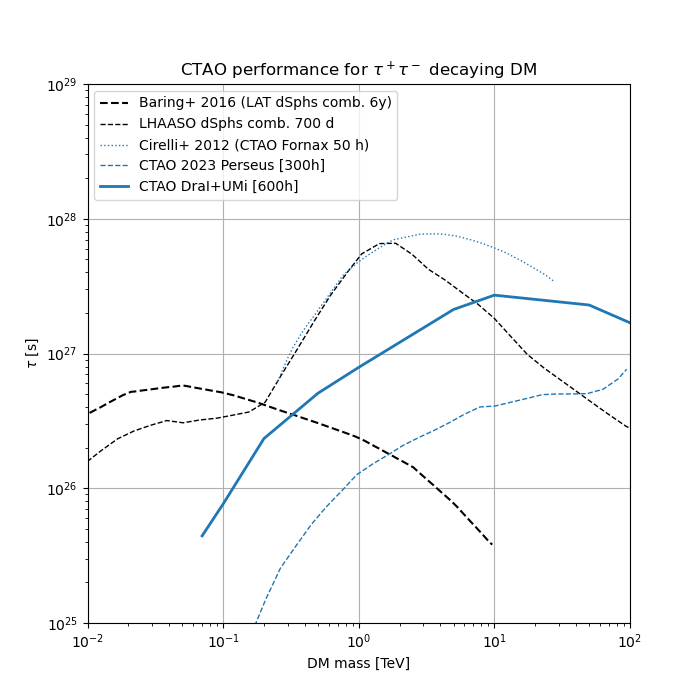}
    \caption{LLs on the DM decay lifetime for the $b\bar{b}$ channel with simulated CTAO observations of dSphs ({\it blue solid lines}) compared to the most constraining limits obtained by current $\gamma$-ray facilities -- {\it Fermi}-LAT \citep{Baring:2015sza} and LHAASO~\citep{LHAASO:2024upb} -- and the CTAO prospects for the Fornax~\citep{Cirelli:2012ut} and the Perseus clusters~\citep{CTAConsortium:2023yak}. The CTAO limits for dSphs are obtained with 600-h combined observations of CBe and UMi, assuming an Einasto DM profile. {\it Left panel:} cross section limits for the $b\bar{b}$ channel. {\it Right panel:} limits for the $\tau^+\tau^-$ channel.}
    \label{fig:compare_instrument_decay}
\end{figure*}

\subsection{`Champion' dSphs: the case of UMaIII}
\label{sec:uma3}
As discussed above, $N$-body simulations \citep[e.g.,][]{Diemand:2008in} predict $\mathcal{O}(1)$ DM sub-haloes with $J_{\rm ann} \gtrsim 10^{20}$ GeV$^2$ cm$^{-5}$ in the MW halo \citep[see][]{Coronado-Blazquez:2021avx}. If such highly DM-dominated sub-haloes have formed dSphs, they would constitute optimal targets for CTAO. In the following we argue that, if the observational data are confirmed, the recent discovery of the dSph U1/Ursa Major III \citep[UMaIII hereafter;][]{Smith:2024} sparks interest, since this object could represent such a `champion'. \citet{Smith:2024} report that UMaIII would be the closest ultra-faint dSph ($d_\odot \sim 10$~kpc), and would possess a velocity dispersion of member stars yielding the highest value of $J_{\rm ann}$ over all known dSphs\footnote{The UMaIII astrophysical factor from \citet{Crnogorcevic:2023ijs} is $\sim$500~times higher than that of Scl, the first discovered dSph.}.

The validity of this prediction is under debate: when in fact the star with the highest impact on the kinematic modeling of this system is removed from the analysis, the velocity spread reduces, completely vanishing when removing the second one \citet[see also][]{Fisher:2025smx}. However, by adopting the $J_{\rm ann}$-to-$d_\odot$ relation (\autoref{eq:J-distance2}), \citet{Crnogorcevic:2023ijs} find $J_{\rm ann} \sim 10^{20.9}$ GeV$^2$ cm$^{-5}$ for UMaIII, with an uncertainty of 0.1~dex only: this allows them to set an extremely robust constraint on the annihilation of DM with {\it Fermi}-LAT data~\citep{Crnogorcevic:2023ijs}. We recall that this is not the first time that a cluster of stars is wrongly identified as the most DM-dominated target, having this already happened with Seg1 \citep{sim11,bon15c} and more recently with TriII~\citep{gen16,kir17}.

Given that `champion' dSphs are predicted by $N$-body simulations and could possibly be correctly identified by the time in which CTAO will have started its operations, we investigate in the following the prospects of using targets like UMaIII in indirect DM searches. However, at present no robust spectro-photometric data are available from dedicated observing campaign of this source: therefore, we cannot include UMaIII in our analysis, since it would be excluded by the data quality selection. To overcome this issue, we thus limit ourselves to scale the expected limits for the dSphs analyzed so far to the value of $J_{\rm ann}$ found by \citet{Smith:2024} and \citet{Crnogorcevic:2023ijs}.

\subsection{Comparison with other limits -- Annihilating DM}
\label{sec:context}
We now first proceed to compare our limits with those obtained with dSph observations from other instruments. We show this comparison in \autoref{fig:compare_dsph} for the $b\bar{b}$ and the $\tau^+\tau^-$ channels; for conciseness, rather than providing an exhaustive census of all the limits available in the literature, we select only the more recent or more constraining results. To this aim, we use:
\begin{itemize}
    \item data from the combination of 4 dSphs observed with MAGIC between 2011 and 2019 \citep{acc22}; this dataset is composed of 52.1 h on DraI, 49.5 h on CBe, 94.8 h on UMaII and 157.9 h on Seg1, for a total of 354.3-h exposure time\footnote{For this dataset, the astrophysical factors used for that work are compatible with those used here, except for Seg1.};
    \item H.E.S.S. data \citep{HESS:2020zwn} including RetII (18.3 h), TucII (16.4 h), TucIII (39.0 h), TucIV (39.2 h) and GruII (29 h), for a total of $\sim$186 h; 
    \item VERITAS limits \citep{VERITAS:2017tif} obtained from $\sim$230-h observations of 5 dSphs (Seg1, UMi, DraI, BooI and Wil1);
    \item HAWC results \citep{HAWC:2017mfa} based on the combined results from 15 dSphs (BooI, CVnI/II, CBe, DraI, Her, LeoI/II/IV, Seg1, Sex, UMaI/II, UMi and TriII) in 507 days of data;
    \item {\it Fermi}-LAT results \citep{mcd23} obtained with 30 dSphs (the so-called `Measured' sample) observed for 14 years\footnote{Wil1 is removed from this analysis due to the claim of strong evidence for tidal disruption and/or non-equilibrium kinematics \citep{wil11}.}.
\end{itemize}
At the time of CTAO, interesting limits are also expected from annihilating DM in dSphs~\citep{Cembranos:2019noa} from the Square Kilometer Array \citep[SKA;][]{Braun:2015} and its precursors~\citep{Kar:2019hnj}.

Looking at \autoref{fig:compare_dsph}, we see that the CTAO limits obtained with the combination of the two best targets at hand -- DraI and UMi -- for a duration of observations of 600~h, would provide limits a factor of $\sim$10 better than any other current limits achieved by IACTs and shower front detectors. When comparing to {\it Fermi}-LAT \citep{mcd23}, the CTAO limits would be the most constraining above $\sim$10~TeV for the $b\bar{b}$ channel and above $\sim$1~TeV for the $\tau^+\tau^-$ case. We advise some caution in making a direct comparison, due to the fact that our limits are obtained profiling over the astrophysical factor uncertainties while some of those with which we compare are not. As discussed in \autoref{sec:results}, neglecting such uncertainties can indeed have a major impact on the final results.

In \autoref{fig:compare_instrument}, we also compare the CTAO sensitivity in the $b\bar{b}$ and $\tau^+\tau^-$ channels using dSphs with other relevant indirect DM limits, either obtained for other classes of astrophysical targets that are suitable for DM searches with CTAO or derived with different techniques\footnote{The search for narrow-line DM signatures with CTAO is reported in a separate work \citep{CTA:DM_line}.}. Specifically, we report limits obtained with ATLAS/CMS and COUPP by~\citet{BAUER201516} adapted to the $\langle \sigma v\rangle-m_{\rm \mathrm{DM}}$ plane\footnote{This choice is partially model-dependent.}, the 14-yr {\it Fermi}-LAT limits from \citet[called {\footnotesize MEASURED} limit there]{mcd23}, the limits predicted for CTAO observations of the GC \citep[called {\footnotesize SENS} limit there]{ach21}, those derived from observations of the Perseus galaxy cluster \citep[called {\footnotesize SENSMAX} limit there]{CTAConsortium:2023yak} and the Large Magellanic Cloud \citep[LMC;][called {\footnotesize NFWmean} limit there]{ach23}; we also add predictions for CTAO in search of DM annihilation signals from dark clumps \citep[called {\footnotesize EXPO} limit there]{Coronado-Blazquez:2021avx}. Also here, some caution should be adopted when comparing other CTAO predictions with those obtained in this work, since all of them have been derived for fixed values of the astrophysical factors without profiling over the relative uncertainties.

 \autoref{fig:compare_instrument} reveals that dSph observations with CTAO provide the most constraining limits in the multi-TeV WIMP mass range. The limits obtained with dSphs are significantly more constraining than those obtained for the Perseus cluster and dark clumps, demonstrating the relevance of using such a class of objects as targets for indirect DM searches. The dSph limits are instead less constraining than those predicted for both the GC and the LMC; however, we remark that such sources are subject to larger uncertainties on the DM modeling than dSphs, given the still poor modeling of the interplay between baryons and DM in the MW halo. Also, the assumption of a cored DM density profile in place of a cuspy one alters the predicted bounds up to a factor of $\sim$20$\div$50 \citep{ach21}. For the LMC observation, the DM signal is expected to be significantly extended ($\sim$10$^\circ$) and contaminated by the astrophysical foreground emission \citep{ach23}. For these reasons, especially in the case of an established DM signal, dSphs would still be attractive targets to corroborate and further study this type of $\gamma$-ray emission.

\subsection{Comparison with other limits -- Decaying DM}
As done for the annihilation case, we show the comparison for the decaying DM case in \autoref{fig:compare_instrument_decay}, adopting the $b\bar{b}$ and the $\tau^+\tau^-$ as representative channels. We exclude UMaIII from these plots due to its compactness, which -- similarly to Scl -- would lead to a low integrated signal yield for DM decay. We compare our limits with a selection of recent and constraining results:
\begin{itemize}
    \item \citet{Cirelli:2012ut}, who analyzed the Fornax galaxy cluster as observed by H.E.S.S. and forecast limits for CTAO by adopting an educated guess of 50-h observations;
    \item \citet{Baring:2015sza}, who analyzed six years of \textit{Fermi}-LAT data from 16 dSphs with a stacked analysis. This limits compare well with other \textit{Fermi}-LAT based limits \citep[e.g.,][]{Ando:2015qda,Blanco:2018esa} obtained from the analysis of the diffuse extragalactic $\gamma$-ray background emission;
    \item CTAO prospects on the observation of the Perseus galaxy cluster for 300~h \citep{CTAConsortium:2023yak};
    \item \citet{LHAASO:2024upb}, reporting limits obtained in 700 days of observations on a combination of dSphs\footnote{This combination includes Seg1.}.
\end{itemize}

\autoref{fig:compare_instrument_decay} shows that the CTAO limits for dSphs in the $\tau^+\tau^-$ channel are competitive above 10~TeV with respect to the current ones. Interestingly, the limits obtained in this work are orders of magnitude more constraining than those obtained when observing the Perseus galaxy cluster \citep{CTAConsortium:2023yak}: this is due to the larger distance of this source compared to our `champion' dSphs \citep[redshift $z \sim 0.02$;][]{bil18}. A more suited galaxy cluster for decaying DM searches is Fornax \citep[$z \sim 0.005$;][]{fir08}; however, the limits reported by \citet{Cirelli:2012ut} are mostly based on the adopted educated guesses.

\subsection{Observational strategy}
\label{sec:strategy}
The results obtained above allow us to discuss the observational strategy to be proposed for CTAO toward the class of dSphs. Considering the competition with other astrophysical targets and the preliminary scheme of the KSPs~\citep{cta19}, we estimated in \autoref{sec:results_combined} that -- although the formal schedule of observations is not fully defined yet and an ample amount of time will be devoted to guest programs, thus leading to a potential overestimation of the allocated observing time -- CTAO could plausibly provide to $\sim$600~h on such sources. Having also shown how the constraints on the DM parameter space are strongly dominated by those objects with the highest astrophysical factors, we point to a strategy in which deep exposures of few `champion' dSphs are performed. However, the choice of the best dSphs is strongly affected by uncertainties related to the astronomical knowledge of their stellar content and motion (see \autoref{app:astro}). As discussed in \autoref{sec:future}, future astronomical campaigns on dSphs will be extremely relevant to obtain more accurate determinations of the DM density profile for both known dSphs and newly discovered candidates. An alternative strategy could be instead that of observing multiple dSphs to reduce the risk of uncertain astrophysical factors.

\section{Summary and conclusions}
\label{sec:conc}
In this work, we have presented the prospects for the detectability of $\gamma$-rays from processes of DM annihilation and decay in dSph haloes with CTA, assuming DM to be composed of non-SM particles belonging to the WIMP class. 

\subsection{Summary of results} 
Our main results can be summarized as follows:
\begin{enumerate}
    \item We have consistently selected the best targets according to their DM content, starting from a complete list (see \autoref{tab:dsph_firstcut} and \autoref{tab:dsph_excluded}) and narrowing down to the optimal dSphs in terms of observability, expected signal strength and uncertainty on their astrophysical factors. Collecting all the most updated and complete samples of spectro-photometric data of such dSphs, we have made use of the public \texttt{CLUMPY} code to compute the DM distribution in the selected dSph haloes in a common framework of data treatment for both classical and ultra-faint targets (see \autoref{sec:dm-astro} and \autoref{app:astro}), and have compared our astrophysical factors with the results currently available in the literature (see \autoref{fig:dsph_comparison_01deg} and \autoref{fig:dsph_jfactor_05}). Our calculations of $J_{\rm ann}$ and $J_{\rm dec}$ are in agreement with what has been found in both single-target and ensemble analyses of the DM content in dSph haloes, with the exception of those known targets that have biased observational data or exhibit tidal disruption features (see \autoref{sec:dm-astro}) We have also provided updated calculations of the corresponding astrophysical factors $J_{\rm ann}$ and $J_{\rm dec}$ (also known as $J$- and $D$-factors). This is of direct relevance for DM searches from dSphs also well outside the context of CTAO.
       \item The uncertainties on the astrophysical factors due to the poor knowledge of stellar content and motion (extensively discussed in \autoref{app:astro}) depend on the target, and may amount to more than an order of magnitude in the worst cases. This can in principle significantly bias our estimates, thus demanding deep surveys to adequately sample the stellar content of this class of targets for both the already known ones and those newly discovered (see \autoref{sec:future}).
        \item We have presented the signal model and analysis pipeline (see \autoref{sec:methods}), based on the public \texttt{gammapy} software -- the official CTAO analysis code -- and discussed the contribution of systematic uncertainties (see \autoref{app:systematics}). These amount to at most 10\%, and are subdominant with respect to the uncertainties on the astrophysical factor. 
    \item We have produced limits for annihilating DM both in individual targets (see \autoref{fig:results_cusp}) and from a combined analysis (see \autoref{sec:results_combined}) down to $\langle\sigma v\rangle \sim 5\times10^{-25}$ cm$^3$ s$^{-1}$, as well as for the case of decaying DM up to $\tau\sim 10^{28}$~s (see \autoref{fig:results_decay}), for combined observations of the best dSphs.
    \item The derivation of limits for annihilating or decaying DM with the inclusion of the astrophysical factor uncertainties plays an important role, since it may affect both the achieved sensitivity in the DM parameter space -- a worsening of the derived limits by a factor $\gtrsim$10 (see \autoref{sec:result_cusp}) -- and the optimal target selection. We recommend profiling the likelihood over the astrophysical factor uncertainties in all future works to improve the accuracy in assessing the effective reach of the indirect DM searches with the Cherenkov technique.
    \item Limits at high DM masses obtained for dSphs with CTAO are up to one order of magnitude better than the corresponding limits obtained with current instruments (see \autoref{fig:compare_dsph}). These limits are more constraining than those predicted for CTAO toward the Perseus galaxy cluster thanks to the dSph proximity and lack of significant background emission, but less than those derived from simulated observations of the GC or the LMC due to the intrinsically larger content of DM of such targets; however, the latter two are subject to large systematic uncertainties due to their DM modeling (see \autoref{fig:compare_instrument}).
    \item In case new DM-dominated dSphs will be discovered as predicted by numerical simulations (see \autoref{sec:future}), or the most optimistic astrophysical factors for objects like UMaIII will be confirmed, the CTAO reach with even a single target for annihilating DM will fall well below the thermal-relic limit in reasonable observing times (see \autoref{fig:compare_instrument}).
\end{enumerate}

\subsection{Conclusions} 
In this work we argued that dSphs are a valid class of targets for indirect DM searches in VHE $\gamma$-rays, due to (i) the absence of a significant background contamination than in closer targets such as the GC and the LMC, thus providing independent constraints and corroborating evidence in case of signal hints at those targets, and (ii) stronger expected signals with respect to more distant sources such as the galaxy clusters. An optimal exploitation of dSphs for DM searches will, however, rely in future deep spectro-photometric surveys on current (or yet-to-be discovered) dSphs aimed at reducing the uncertainties in the DM modeling, in order to obtain the best set of dSphs to be pointed at in the CTAO era. In such conditions, the best observational strategy will be likely based on deep pointings of few optimal targets rather than shorter snapshots of many objects.

\section*{Author contributions}
This publication is the result of a collaborative effort within the CTAO Consortium. The project was conceptualized and supervised by M.D. and F.G.S., who also defined the methodology and led the draft preparation. F.G.S. carried out the formal analysis of the astrophysical factors using \texttt{CLUMPY}, while G.R.F. performed the dark matter limits analysis using \texttt{gammapy} and \texttt{ctools}. A.M. contributed to the validation of the results and the preparation of the manuscript. K.D. Nakashima (FAU-Erlangen) contributed in earlier phases by computing the diffuse $\gamma$-ray background and cross-checking the analysis. The initial manuscript drafts were reviewed by A. Chen (Witwatersrand University), M.-{\'A}. S{\'a}nchez-Conde (UAM-IFT), and T. Bringmann (Oslo University) on behalf of CTAO-SAPO. The rest of the authors in one way or the other contributed to the development of the CTAO analysis tools and IRFs and participated in the review of the manuscript through internal discussions and approval processes coordinated by the Dark Matter and Exotic Physics Working Group (DMEP-WG) leaders and the CTAO Speakers and Publications Office (CTAO-SAPO). 

\section*{Acknowledgements}
We thank the anonymous referee for their helpful comments. We also thank J. Palacio (MPIK), K. D. Nakashima (FAU-Erlangen) and G. D'Amico (UiB and INAF-OAPd) for earlier contributions and discussions about this project. FGS acknowledges financial support from the PRIN MIUR project ``ASTRI/CTA Data Challenge'' (PI: P. Caraveo), contract 298/2017, and from the INAF Mini-Grant 2023 ``MADaMA -- Measuring the Amount of Dark Matter in Astrophysical targets''.

We gratefully acknowledge financial support from the following agencies and organizations: State Committee of Science of Armenia, Armenia;
The Australian Research Council, Astronomy Australia Ltd, The University of Adelaide, Australian National University, Monash University, The University of New South Wales, The University of Sydney, Western Sydney University, Australia; Federal Ministry of Education, Science and Research, and Innsbruck University, Austria;
Conselho Nacional de Desenvolvimento Cient\'{\i}fico e Tecnol\'{o}gico (CNPq), Funda\c{c}\~{a}o de Amparo \`{a} Pesquisa do Estado do Rio de Janeiro (FAPERJ), Funda\c{c}\~{a}o de Amparo \`{a} Pesquisa do Estado de S\~{a}o Paulo (FAPESP), Funda\c{c}\~{a}o de Apoio \`{a} Ci\^encia, Tecnologia e Inova\c{c}\~{a}o do Paran\'a - Funda\c{c}\~{a}o Arauc\'aria, Ministry of Science, Technology, Innovations and Communications (MCTIC), Brasil;
Ministry of Education and Science, National RI Roadmap Project DO1-153/28.08.2018, Bulgaria; 
The Natural Sciences and Engineering Research Council of Canada and the Canadian Space Agency, Canada; 
ANID PIA/APOYO AFB230003, ANID-Chile Basal grant FB 210003, N\'ucleo Milenio TITANs (NCN19-058), FONDECYT-Chile grants 1201582, 1210131, 1230345, and 1240904; 
Croatian Science Foundation, Rudjer Boskovic Institute, University of Osijek, University of Rijeka, University of Split, Faculty of Electrical Engineering, Mechanical Engineering and Naval Architecture, University of Zagreb, Faculty of Electrical Engineering and Computing, Croatia;
Ministry of Education, Youth and Sports, MEYS  LM2018105, LM2023047, EU/MEYS CZ.02.1.01/0.0/0.0/16\_013/0001403, CZ.02.1.01/0.0/0.0/18\_046/0016007,  CZ.02.1.01/0.0/0.0/16\_019/0000754 and CZ.02.01.01/00/22\_008/0004632, Czech Republic; 
Academy of Finland (grant nr.317636 and 320045), Finland;
Ministry of Higher Education and Research, CNRS-INSU and CNRS-IN2P3, CEA-Irfu, ANR, Regional Council Ile de France, Labex ENIGMASS, OCEVU, OSUG2020 and P2IO, France; 
The German Ministry for Education and Research (BMBF), the Max Planck Society, the German Research Foundation (DFG, with Collaborative Research Centres 876 \& 1491), and the Helmholtz Association, Germany; 
Department of Atomic Energy, Department of Science and Technology, India; 
Istituto Nazionale di Astrofisica (INAF), Istituto Nazionale di Fisica Nucleare (INFN), MIUR, Istituto Nazionale di Astrofisica (INAF-OABRERA) Grant Fondazione Cariplo/Regione Lombardia ID 2014-1980/RST\_ERC, Italy; 
ICRR, University of Tokyo, JSPS, MEXT, Japan; 
Netherlands Research School for Astronomy (NOVA), Netherlands Organization for Scientific Research (NWO), Netherlands; 
University of Oslo, Norway; 
Ministry of Science and Higher Education, DIR/WK/2017/12, the National Centre for Research and Development and the National Science Centre, UMO-2016/22/M/ST9/00583, Poland; 
Slovenian Research and Innovation Agency, grants P1-0031, I0-E018, J1-60014, Slovenia; 
South African Department of Science and Technology and National Research Foundation through the South African Gamma-Ray Astronomy Programme, South Africa; 
The Spanish groups acknowledge funds from ``ERDF A way of making Europe", the Spanish Ministry of Science, Innovation and Universities, and the Spanish Research State Agency (AEI) via MCIN/AEI/10.13039/501100011033, grants CNS2023-144504, PDC2023-145839-I00, PID2022-137810NB-C22, PID2022-139117NB-C41/C42/C43/C44, PID2022-136828NB-C41/C42, PID2022-138172NB-C41/C42/C43, PID2021-124581OB-I00, PID2021-125331NB-I00, and budget lines 28.06.000X.411.01 and 28.06.000X.711.04 of PGE 2023, 2024 and 2025; the ``Centro de Excelencia Severo Ochoa" program through grants no. CEX2019-000920-S, CEX2020-001007-S, CEX2021-001131-S; the ``Unidad de Excelencia Mar\'ia de Maeztu" program through grants no. CEX2019-000918-M and CEX2020-001058-M; the ``Ram\'on y Cajal" program through grant RYC2021-032991-I; and the ``Juan de la Cierva" program through grants no. JDC2022-048916-I and JDC2022-049705-I. They also acknowledge the projects with refs. PR47/21 TAU and TEC-2024/TEC-182, both funded by the Comunidad de Madrid regional government. Funds were also granted by the ``Consejer\'ia de Universidad, Investigaci\'on e Innovaci\'on" of the regional government of Andaluc\'ia (Refs.~AST22\_00001\_9 and AST22\_0001\_16) and ``Plan Andaluz de Investigaci\'on, Desarrollo e Innovaci\'on" (Ref. FQM-322); and by the ``Programa Operativo de Crecimiento Inteligente" FEDER 2014-2020 (Refs.~ESFRI-2017-IAC-12 and ESFRI-2020-01-IAC-12) and Spanish Ministry of Science, Innovation and Universities, 15\% co-financed by ``Consejer\'ia de Econom\'ia, Industria, Comercio y Conocimiento" of the Gobierno de Canarias regional government. The Generalitat de Catalunya regional government is also gratefully acknowledged via its ``CERCA" program and grants 2021SGR00426 and 2021SGR00679. Spanish groups were also kindly supported by European Union funds via the Horizon Europe Research and innovation programme under Grant Agreement no. 101131928; NextGenerationEU, grants no. PRTR-C17.I1, CT19/23-INVM-109, and the ‘MicroStars’ ERC, ref. 101076533. This research used computing and storage resources provided by the Port d'Informaci\'o Cient\'ifica (PIC) data center;
Swedish Research Council, Royal Physiographic Society of Lund, Royal Swedish Academy of Sciences, The Swedish National Infrastructure for Computing (SNIC) at Lunarc (Lund), Sweden; 
State Secretariat for Education, Research and Innovation (SERI) and Swiss National Science Foundation (SNSF), Switzerland; The National Research Foundation of Ukraine (project 2023.03/0149 and 2023.05/0024), Ukraine;
Durham University, Leverhulme Trust, Liverpool University, University of Leicester, University of Oxford, Royal Society, Science and Technology Facilities Council, UK; 
U.S. National Science Foundation, U.S. Department of Energy, Argonne National Laboratory, Barnard College, University of California, University of Chicago, Columbia University, Georgia Institute of Technology, Institute for Nuclear and Particle Astrophysics (INPAC-MRPI program), Iowa State University, the Smithsonian Institution, V.V.D. is funded by NSF grant AST-1911061, Washington University McDonnell Center for the Space Sciences, The University of Wisconsin and the Wisconsin Alumni Research Foundation, USA.

The research leading to these results has received funding from the European Union's Seventh Framework Programme (FP7/2007-2013) under grant agreements No~262053 and No~317446.
This project is receiving funding from the European Union's Horizon 2020 research and innovation programs under agreement No~676134.

\section*{Data Availability}

This research has made use of {\ttfamily gammapy}\footnote{See \url{https://www.gammapy.org}.} \citep{gammapy}, a com\-munity-developed core Python package for TeV $\gamma$-ray astronomy, and of the CTAO instrument response functions (version {\ttfamily prod5-v0.1}) provided by the CTA Consortium and Observatory\footnote{See \url{https://www.cta-observatory.org/science/ctao-performance/} for more details.} \citep{cta_irf}. Most of the analysis is stored as publicly available notebooks in the Online Material~\citep{zenodo_dsph}. {\ttfamily CLUMPY} is licensed under the GNU General Public License (GPLv2).



\bibliographystyle{mnras}
\bibliography{biblio} 




\appendix

\begin{table*}
\caption{Same as \autoref{tab:dsph_firstcut}, but for those targets that do not satisfy our first selection cut (in alphabetical order).}
\label{tab:dsph_excluded}
\begin{center}
\resizebox{.99\textwidth}{!}{
\begin{tabular}{llllllllll}
\hline
\hline
Name & Abbr. & Type & R.A. & dec. & Distance  & ZA$_{\rm culm}$ & ZA$_{\rm culm}$ & Month & Ref.\\
 & & & (hh\,mm\,ss) & (dd\,mm\,ss) & (kpc) & N (deg) & S (deg) & & \\
\hline
Andromeda XVIII & AndXVIII & uft & 00 02 14.5 & +45 05 20 & $1330 \pm 104$ & 16.3 & 69.7 & Sep & 1,2\\
Antlia II & AntII & uft & 09 35 33.7 & $-$36 46 12 & $132 \pm 6$ & 65.8 & 12.5 & Feb & 3\\
Aquarius II & AqrII & uft & 22 33 55.1 & $-$09 19 48 & $108 \pm 3$ & 38.3 & 15.8 & Aug & 4\\
Bo{\"o}tes IV & Bo{\"o}IV & uft & 15 34 45.5 & +43 43 48 & $209 \pm 20$ & 14.2 & 67.5 & May & 5\\
Canes Venatici I & CVnI & uft & 13 28 03.5 & +33 33 21 & $216 \pm 8$ & 4.8 & 58.2 & Apr & 1,6\\
Canes Venatici II & CVnII & uft & 12 57 10.0 & +34 19 15 & $159 \pm 8$ & 5.6 & 58.9 & Apr & 1,6\\
Carina I & CarI & cls & 06 41 36.7 & $-$50 57 58 & $106 \pm 1$ & 79.7 & 26.3 & Dec & 1,7\\
Centaurus I & CenI & uft & 12 38 21.5 & $-$40 54 00 & $116 \pm 2$ & 70.0 & 16.7 & Apr & 8\\
Cetus I & CetI & uft & 00 26 11.0 & $-$11 02 40 & $748 \pm 31$ & 39.8 & 13.6 & Sep & 1,9\\
Cetus III & CetIII & uft & 02 05 19.3 & $-$04 16 12 & $251 \pm 19$ & 32.5 & 20.8 & Oct & 10\\
Columba I & ColI & uft & 05 31 26.4 & $-$28 01 48 & $182 \pm 18$ & 56.8 & 3.4 & Dec & 11\\
Crater II & CrtII & uft & 11 49 14.5 & $-$18 24 36 & $118 \pm 1$ & 47.5 & 5.8 & Mar & 12\\
Eridanus II & EriII & uft & 03 44 21.5 & $-$43 31 48 & $330 \pm 16$ & 72.3 & 18.9 & Nov & 13\\
Fornax & For & cls & 02 39 59.3 & $-$34 26 57 & $146 \pm 1$ & 63.2 & 9.8 & Oct & 1,7\\
Grus I & GruI & uft & 22 56 42.4 & $-$50 09 48 & $120 \pm 17$ & 78.9 & 25.5 & Sep & 14\\
Hercules & Her & uft & 16 31 02.0 & +12 47 30 & $137 \pm 11$ & 16.0 & 37.4 & May & 1,15\\
Hydra II & HyaII & uft & 12 21 42.1 & $-$31 59 07 & $134 \pm 10$ & 60.7 & 7.4 & Mar & 16\\
Indus II & IndII & uft & 20 38 52.8 & $-$46 09 36 & $214 \pm 16$ & 74.9 & 21.5 & Aug & 11\\
Leo I & LeoI & cls & 10 08 28.1 & +12 18 23 & $272 \pm 10$ & 16.5 & 36.9 & Feb & 1,17\\
Leo II & LeoII & cls & 11 13 28.8 & +22 09 06 & $240 \pm 9$ & 6.6 & 46.8 & Mar & 1,18\\
Leo IV & LeoIV & uft & 11 32 57.0 & $-$00 32 00 & $151 \pm 4$ & 29.3 & 24.1 & Mar & 1,19\\
Leo V & LeoV & uft & 11 31 09.6 & +02 13 12 & $169 \pm 5$ & 26.5 & 26.9 & Mar & 1,19\\
Leo T & LeoT & uft & 09 34 53.4 & +17 03 05 & $377 \pm 28$ & 11.7 & 41.7 & Feb & 1,20\\
Pegasus III & PegIII & uft & 22 24 25.2 & +05 24 36 & $215 \pm 12$ & 22.5 & 40.0 & Aug & 21\\
Phoenix I & PheI & uft & 01 51 06.3 & $-$44 26 41 & $427 \pm 31$ & 73.2 & 19.8 & Oct & 1,20\\
Pictor I & PicI & uft & 04 43 48.0 & $-$50 16 48 & $126 \pm 24$ & 79.0 & 25.7 & Nov & 13\\
Pisces II & PscII & uft & 22 58 31.0 & +05 57 09 & $182 \pm 13$ & 22.8 & 30.6 & Sep & 1,22\\
Tucana I & TucI & uft & 22 41 49.6 & $-$64 25 10 & $855 \pm 35$ & --- & 39.8 & Sep & 1,9\\
Ursa Major I & UMaI & uft & 10 34 52.8 & +51 55 12 & $105 \pm 2$ & 23.2 & 76.6 & Mar & 1,23\\
\hline
\end{tabular}
}
\end{center}
\scriptsize{
References: $^1$\citet{McConnachie:2012vd}, $^2$\citet{mak17}, $^{3}$\citet{tor19}, $^{4}$\citet{tor16}, $^{5}$\citet{hom19}, $^6$\citet{oka12}, $^7$\citet{kar15}, $^{8}$\citet{mau20}, $^9$\citet{dam13}, $^{10}$\citet{hom18}, $^{11}$\citet{drl15}, $^{12}$\citet{tor16b}, $^{13}$\citet{bec15}, $^{14}$\citet{kop15}, $^{15}$\citet{gar18}, $^{16}$\citet{mar15b}, $^{17}$\citet{ste14}, $^{18}$\citet{tul13}, $^{19}$\citet{med18}, $^{20}$\citet{rip14}, $^{21}$\citet{kim16}, $^{22}$\citet{san12}, $^{23}$\citet{bro12}.
}
\end{table*}

\section{Derivation of the astrophysical factors for the CTAO optimal dSphs}
\label{app:astro}

Hereafter we describe the details of the input data sets and the operational set-up of the MCMC Jeans analysis that has been performed with the \texttt{CLUMPY} software in order to make the target selection and derive the astrophysical factors of the optimal dSphs for the DM searches with CTA. This appendix completes the information reported in \autoref{sec:dm-astro}.

\subsection{Original source selection and discarded targets}
In \autoref{sec:dm-astro} we have presented our target selection based on distance criteria and quality of stellar information. Out of a starting sample of 64 dSphs candidates, we have narrowed down the initial list first to 35 surviving candidates, reported in \autoref{tab:dsph_firstcut} and with a second selection to 14 plausible targets for both hemispheres, reported in \autoref{tab:dsph_parameters_2}. For completeness, we also report in \autoref{tab:dsph_excluded} the list of 29 targets that were excluded according to the first selection cut.

\subsection{\texttt{CLUMPY} set-up for the computation of the astrophysical factors}
We follow the prescriptions by \citet{bon15c} to derive the astrophysical factors of dSph haloes in a self-consistent and conservative way. In \autoref{sec:dm-astro}, we have discussed the retrieval of the stellar information for each target from literature data of the 35 targets that survived our first selection cut. 

\subsubsection{Brightness profiles} 
For these selected targets, we attempted a fit the brightness density $n^*(r)$ of each galaxy with a 3D Zhao-Hernquist profile \citep{her90,zha96}:
\begin{equation}\label{eqn:nfw}
    n^*(r) = \frac{n^*_s}{\left(
    \frac{r}{r^*_s}
    \right)^{\gamma^*}\left[
    1 + \left(
    \frac{r}{r^*_s}
    \right)^{\alpha^*}
    \right]^{\frac{\beta^*-\gamma^*}{\alpha^*}}}\, ,
\end{equation}
projected onto the corresponding circularized\footnote{$r_{\rm circ} = r\sqrt{1-\epsilon}$, with $\epsilon$ the dSph eccentricity.} 2D surface brightness profile $\Sigma(r)$. The spatial distribution of the baryonic content over the dSph volume inferred in this way is used in \texttt{CLUMPY} as a proxy for the DM spatial distribution to solve the Jeans equation~\autoref{eqn:jeans}.

\begin{figure*}
\centering
\includegraphics[width=0.48\linewidth]{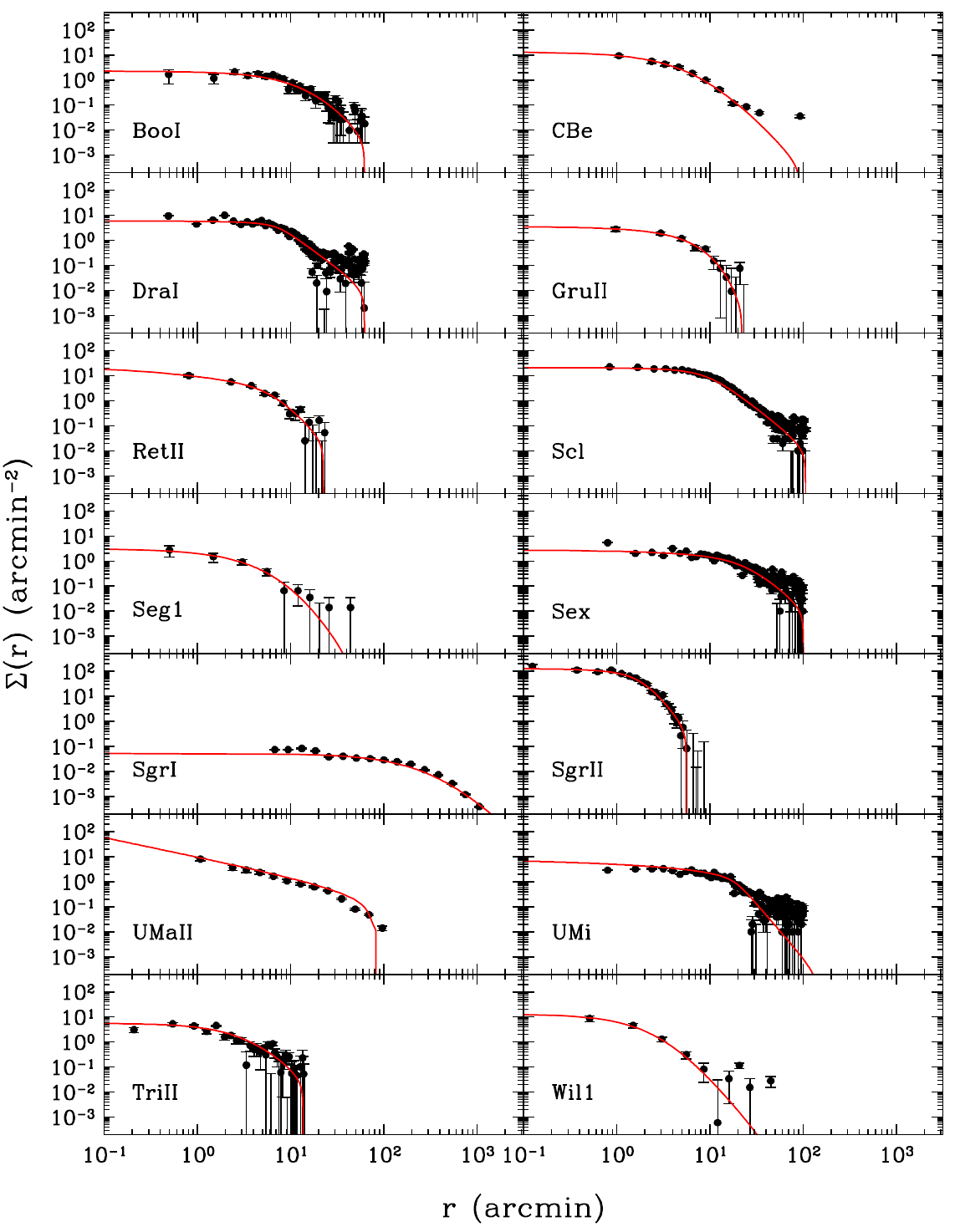}
\includegraphics[width=0.48\linewidth]{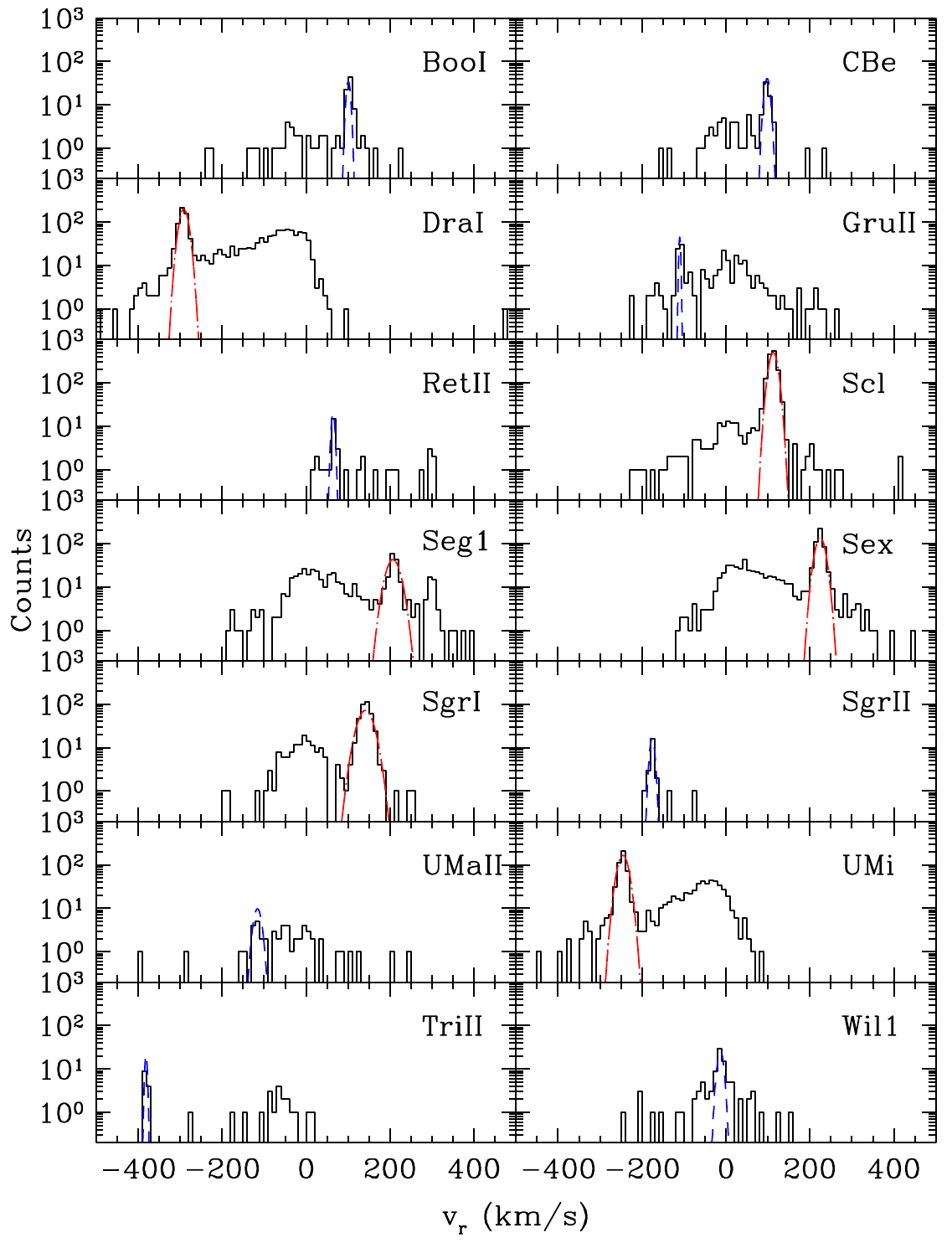}
\caption{{\it Left panels:} Best-fit integrated brightness profiles $\Sigma(r)$ of the dSphs that pass the selection on the astrometric data availability as a function of the object's projected (2D) radial coordinate $r$ from the dSph centroid. In each panel, the projected 3D profile resulting from the fit ({\itshape red line}) is shown superimposed on the corresponding background-subtracted data set ({\itshape black dots}). {\it Right panels:} Distributions of radial velocities for the most updated stellar samples available for each optimal dSph ({\itshape black solid histograms}; see \autoref{tab:dsph_parameters} for the references). For each dSph, a Gaussian is shown to represent the velocity distribution of confirmed member stars obtained either from the application of the EM algorithm by \citet[{\itshape red dot-dashed lines}]{wal09} or from the binary classification extracted from the relevant literature ({\itshape blue dashed lines}; see \autoref{tab:dsph_parameters} for the references).}
\label{fig:dsph_brightness_profile}
\end{figure*}

We performed this fit using the IDL package \texttt{MPFIT} \citep{mar09}, requesting that the final profiles give finite numbers of stars -- and hence finite dSph brightnesses -- once integrated up to infinity. This in turn translates to constraints on the inner slopes $\gamma^* < 3$ and outer slopes $\beta^* > 3$. Such choices forced us to remove Segue 2, Tucana II, Tucana III and Tucana IV from the sample of optimal targets, since no binned surface brightness data were at present available for these dSphs, and Bo{\"o}tes II, for which no profile with finite luminosity can be obtained\footnote{This is hinting to a severe tidal disruption of the Bo{\"o}tes II baryonic content, due to its potential association with the Sagittarius stellar stream \citep{cha22}, although an inaccurate measurement of the surface brightness due to its intrinsic faintness \citep{mar08} is equally possible.}. 

\subsubsection{Stellar kinematics} 
For the dSph stellar kinematics, we were forced to remove from the analysis those candidates that did not have adequately populated stellar samples to obtain a significant MCMC fit, i.e. the targets with number of confirmed member stars $N_{\rm mem} < N_{\rm par}$ ($N_{\rm par}$ being the number of free parameters in the fitting procedure, which is set to 7: 3 for the DM density profile, and 4 for the velocity anisotropy in our framework). This further constraint led us to exclude Bo{\"o}tes III, Carina II, Carina III, Laevens 3, Cetus II, Eridanus III, Horologium I, Horologium II, Indus I, Pictor I, Phoenix II, Tucana V and Virgo I, since such targets had no spectroscopic measurements or spectroscopic samples containing $\leq$5 member candidates \citep{koc09,kop15b}. 

To take into account non-radial components of the stellar velocities -- that hint to a partial, although not dominant, rotational support of the dSph structure -- we adopted in \texttt{CLUMPY} the most conservative priors by \citet{bon15a} for the treatment of the velocity anisotropy, using the Baes \& van Hese profile \citep{bae07}:
\begin{equation}\label{eqn:velanis}
    \beta_{\rm ani}(r) = \frac{\beta_0 + \beta_\infty (r/r_a)^\eta}{1 + (r/r_a)^\eta}\, ,
\end{equation}
with 4 free parameters (central anisotropy $\beta_0$, asymptotic anisotropy $\beta_\infty$, anisotropy scale radius $r_a$ and sharpness index $\eta$). 

\subsubsection{Stellar association} 

\begin{table*}
\begin{center}
\caption{Parameters of the 3D brightness profiles (see \autoref{fig:dsph_brightness_profile}) for the selected dSphs.}
\label{tab:dsph_parameters}
\begin{small}
\begin{tabular}{lc|cccccccc}
\hline
\hline
Name & Site & $M_V$ & $\epsilon$ & $\rho_{\rm s}^*$ & $r_{\rm s}^*$ & $\alpha^*$ & $\beta^*$ & $\gamma^*$ & Ref. \\
 & & (mag) & & ($10^5$ L$_\odot$ kpc$^{-3}$) & (kpc) & & &\\
\hline
Bo{\"o}I & N & $-6.3 \pm 0.2$ & $0.39 \pm 0.06$ & $1.14 \pm 0.21$ & $0.461 \pm 0.021$ & 1.1 & 7.7 & 0.0 & 1,2 \\
CBe & N & $-4.1 \pm 0.5$ & $0.38 \pm 0.14$ & $1.08 \pm 0.50$ & $0.0740 \pm 0.0035$ & 1.1 & 5.4 & 0.0 & 1,3 \\
DraI & N & $-8.8 \pm 0.3$ & $0.31 \pm 0.02$ & $4.5 \pm 1.3$ & $0.1473 \pm 0.0079$ & 6.8 & 3.8 & 0.0 & 1,4 \\
GruII & S & $-3.9 \pm 0.2$ & $\sim$0.2 & $1.58 \pm 0.29$ & $0.166 \pm 0.016$ & 1.3 & 7.6 & 0.0 & 5 \\
RetII & S & $-3.6 \pm 0.2$ & $0.6 \pm 0.2$ & $2.04 \pm 0.19$ & $0.0408 \pm 0.0026$ & 3.5 & 4.7 & 1.1 & 6 \\
Scl & S & $-11.1 \pm 0.5$ & $0.32 \pm 0.03$ & $23 \pm 11$ & $0.2100 \pm 0.0050$ & 3.2 & 4.0 & 0.6 & 1,4 \\
Seg1 & N & $-1.5 \pm 0.8$ & $0.48 \pm 0.13$ & $1.21 \pm 0.89$ & $0.0739 \pm 0.0064$ & 1.1 & 9.2 & 0.2 & 1,7 \\
Sex & S & $-9.3 \pm 0.5$ & $0.35 \pm 0.05$ & $0.56 \pm 0.26$ & $0.493 \pm 0.018$ & 2.7 & 4.0 & 0.6 & 1,4 \\
SgrI & S & $-13.5 \pm 0.3$ & $0.64 \pm 0.02$ & $0.277 \pm 0.076$ & $1.869 \pm 0.060$ & 1.1 & 4.9 & 0.0 & 1,8 \\
SgrII & S & $-5.2 \pm 0.4$ & $\sim$0.2 & $42.9 \pm 3.9$ & $0.0371 \pm 0.0028$ & 3.5 & 5.7 & 0.1 & 9,10 \\
TriII & N & $-1.8 \pm 0.5$ & $\sim$0.2 & $7.3 \pm 3.4$ & $0.0342 \pm 0.0023$ & 1.2 & 5.3 & 0.0 & 11 \\
UMaII & N & $-4.2 \pm 0.6$ & $0.63 \pm 0.05$ & $49.4 \pm 27.3$ & $0.0265 \pm 0.0015$ & 0.1 & 2.1 & 2.0 & 1,3 \\
UMi & N & $-8.8 \pm 0.5$ & $0.56 \pm 0.05$ & $21.7 \pm 10.0$ & $0.336 \pm 0.010$ & 4.0 & 7.3 & 0.7 & 1,4 \\
Wil1 & N & $-2.7 \pm 0.8$ & $0.47 \pm 0.08$ & $4.4 \pm 3.3$ & $0.0251 \pm 0.0046$ & 1.2 & 5.9 & 0.0 & 1,7 \\
\hline\hline
\end{tabular}
\end{small}
\end{center}
{\footnotesize References: 
$^1$\citet{McConnachie:2012vd}; $^2$\citet{bel06}; $^3$\citet{mun10}; $^4$\citet{irw95}; $^5$\citet{drl15}; $^6$\citet{bec15}; $^7$\citet{mar08}; $^8$\citet{maj03}; $^9$\citet{lae15a}; $^{10}$\citet{mut18}; $^{11}$\citet{lae15b}; $^{12}$\citet{kop11}; $^{13}$\citet{sim07}; $^{14}$\citet{wal15}; $^{15}$\citet{sim20}; $^{16}$\citet{wal15b}; $^{17}$\citet{wal09b}; $^{18}$\citet{sim11}; $^{19}$\citet{iba97}; $^{20}$\citet{lon20}; $^{21}$\citet{kir17}; $^{22}$\citet{spe18}; $^{23}$\citet{wil11}.}
\end{table*}

To select only fiducial member stars for each target, we processed the stellar kinematics data sets of the classical dSphs, plus those of Segue 1 and Sagittarius I, through the expectation-maximization (EM) algorithm by \citet{wal09}. We also applied additional static cuts on the spectroscopic stellar parameters in the case of Draco I \citep{wal15} and Segue 1 \citep{bon16b}. This procedure was feasible only for those objects that had sufficiently populated ($N \gtrsim 100$) stellar samples; for the remaining ultra-faint dSphs, we adopted binary (0/1) memberships already available in the literature. It is to be noted that no spatial cuts were applied to the data that were processed with the EM algorithm: in fact, since for all targets we aim at computing precise astrophysical factor profiles up to large angular distances from the dSph centroid in order to exploit the large FoV of the CTAO telescopes (see \autoref{sec:dm-astro}), we did not preliminarily remove any candidate member star based on their coordinates. This is clearly biasing the $J_{\rm ann}$ and $J_{\rm dec}$ calculation for some sources, like Sagittarius I, that show significant non-equilibrium features; nevertheless, we maintain them in our analysed sample flagging them adequately (see \autoref{tab:dsph_jfactors}).

Due to the relatively low number of members ($<$50) for many of the analyzed dSphs, we chose the \texttt{CLUMPY} unbinned analysis method \citep{bon15b,bon15c}, which runs over the velocity data of the single member stars. After this selection, we were left with a sample of 14 targets; for them, the brightness profiles and the distribution of stellar velocities are reported in \autoref{fig:dsph_brightness_profile}. We report the number of member stars identified in this way for such dSphs in Tab. \ref{tab:dsph_parameters}. Note that the scale number density $n^*_s$ in Eq. \ref{eqn:nfw} has been converted to a brightness density $\rho^*_s$ scaling the volume integral of the stellar number density profile to the dSph $V$-band absolute magnitude.

\subsubsection{DM density profiles} 
To compute the DM density profiles, we chose both a cuspy Einasto profile \citep{ein65} with 3 free parameters (DM scale density $\rho_s$, DM scale radius $r_s$ and DM sharpness index $\alpha$), and a cored Burkert profile \citep{bur95} with only scale density and radius as free parameters (see \autoref{eq:dm_profiles}). Such choices imply MCMC fits with a total of 7 and 6 free parameters, respectively. The priors adopted for such free parameters are the same listed in \citet{bon15a}. On such parameter sets, we run 200 independent MC chains of $10^5$ realizations each for every target. From the resulting posterior distributions of profile parameters, we derive their average best-fit values, reporting them in \autoref{tab:einasto_parameters} and \autoref{tab:burkert_parameters} along with the corresponding uncertainties at 68\% CL.

\subsubsection{Tidal radii}\label{app:rtid}
We then derived the distribution of tidal radii $R_{\rm tid}$ for each dSph from the output profiles as made by \citet{bon15c}, iteratively solving the tidal equation \citep{Springel:2008cc,mol15}:
\begin{equation}\label{eq:tidal_radius}
    R_{\rm tid} = \left[
    \dfrac{M_{\rm dSph}}{\left(
    2 - \dfrac{d\ln{M_{\rm MW}}}{d\ln{r}}\bigg\rvert_{d_{\rm GC}}
    \right) \cdot M_{\rm MW}\left(
    <d_{\rm GC}
    \right)}
    \right]^{1/3} \times d_{\rm GC}\, ,
\end{equation}
where $M_{\rm dSph}$ is the total dSph DM mass, $M_{\rm MW}$ is the MW mass \citep[here assumed to be $10^{12}$ M$_\odot$;][]{wat19} and $d_{\rm GC}$ is the dSph Galactocentric distance \citep{McConnachie:2012vd}. We also report the average values of $R_{\rm tid}$, along with the corresponding uncertainties at 68\% CL, in \autoref{tab:einasto_parameters} and \autoref{tab:burkert_parameters} for the two adopted DM density profiles, respectively.

\subsubsection{Boost to the astrophysical factors from DM substructures}
\label{sec:boost} 
$N$-body simulations and the cold, collisionless nature of WIMPs predict that MW-like DM haloes form by hierarchical clustering of smaller substructures, that are generally referred to as DM sub-haloes \citep{Kuhlen:2012ft,Zavala:2019gpq}. The effect of substructures on DM searches has been subject to strong debates in the past \citep[e.g.,][]{pin11,san11,gao12,san14,mol17,and19}. Depending on the target, the sub-halo contribution can in fact strongly increase the astrophysical factor in the annihilation case where the signal yield depends on the integration of a squared DM density from extremely compact and dense sub-haloes (see \autoref{eqn:jfac}). Such a contribution is modeled according to the sub-halo number density, the sub-halo radial distribution within the host halo, and the so-called concentration parameter. Such functions are obtained using data from $N$-body simulations such as {\itshape Via Lactea II} and {\itshape Aquarius} \citep{Diemand:2008in,Springel:2008cc,agu23}. When integrating all of the sub-halo contributions, one gets a boost factor $\mathcal{B}$ on the astrophysical factor such as:
\begin{equation}\label{eq:jboost}
    J'(<\alpha)=\;\left[
    1+\mathcal{B}(<\alpha)
    \right]J(<\alpha)\, ,
\end{equation}
where $J'$ is the astrophysical factor integrated over a certain aperture angle $\alpha$, taking into account both the sub-halo mass distribution and the concentration-to-mass relation. The net signal enhancement produced by substructures can thus be factorized with $\mathcal{B}$, as shown in \autoref{eq:jboost}.

In the annihilation case, $\mathcal{B}$ is relevant for large and massive objects, such as galaxy clusters, and less relevant for the dSphs. This is also due to the fact that dSphs are affected by tidal stripping on their outer rims, which truncates the dSph DM density profile at large radii removing in this way a significant amount of substructures that are mainly located in these regions \citep{mol17}. Depending on the dSph, the boost factor can be of the order of at most $\mathcal{B} \sim 0.3$ \citep[see e.g. figure 7 of][]{mol17}; for this reason, we decide to arbitrarily set $\mathcal{B}=0$ for our discussion. This choice is further motivated by the intrinsic uncertainties on $J_{\rm ann}$ (or $J_{\rm dec}$) of the main halo that dominate over this factor; in addition, since the value of $\mathcal{B}$ linearly affects our results, it is easily modifiable for future analyses with different assumptions on the boost factor. Finally, for the case of decaying DM, the boost factor is even more negligible, because $J_{\mathrm{dec}}$ integrates the linear DM density (see \autoref{eqn:jfac}).

\subsubsection{Astrophysical factors} 
We finally computed the profiles of $J_{\rm ann}(<\alpha_{\rm int})$ and $J_{\rm dec}(<\alpha_{\rm int})$ as functions of the integration angle $\alpha_{\rm int}$ from the dSph center by running the \texttt{CLUMPY} executable over the posterior distributions of the DM profile parameters. From such profiles, we extracted the full radial dependence of $J_{\rm ann}$ and $J_{\rm dec}$, as well as their values at the typical reference values of $\alpha_{\rm int} = 0.1^\circ$ -- close to the average CTAO angular resolution (see \autoref{tab:cta_arrays}) -- and $0.5^\circ$.

\begin{table}
    \centering
    \caption{Average best-fit Einasto profile parameters for the sample of optimal dSphs.}
    \label{tab:einasto_parameters}
    \resizebox{\columnwidth}{!}{
    \begin{tabular}{lccccc}
        \hline
        \hline
        Name & $\rho_s$ ($10^7$ M$_{\odot}$ kpc$^{-3}$) & $r_s$ (kpc) & $\alpha$ & $R_{\rm tid}$ (kpc) & $\chi^2/n_{\rm d.o.f.}$ \\
        \hline
        CBe   & $4.3 \pm 2.8$ & $6.1^{+4.1}_{-5.7}$ & $0.60^{+0.32}_{-0.24}$ & $6.3^{+19.0}_{-4.3}$ & $76.3/52$ \\
        Dral  & $1.3^{+1.6}_{-1.0}$ & $0.91^{+0.89}_{-0.33}$ & $0.25^{+0.23}_{-0.10}$ & $4.83^{+1.16}_{-0.83}$ & $642.9/462$ \\
        RetII & $4.2^{+6.7}_{-3.3}$ & $0.47^{+3.26}_{-0.33}$ & $0.63^{+0.23}_{-0.31}$ & $1.7^{+4.5}_{-1.0}$ & $21.5/11$ \\
        Scl   & $6.1^{+3.4}_{-2.6}$ & $0.31^{+0.12}_{-0.10}$ & $0.54^{+0.30}_{-0.31}$ & $2.95^{+0.55}_{-0.30}$ & $1501.4/1113$ \\
        SgrII & $9.4^{+20.9}_{-9.2}$ & $0.49^{+4.80}_{-0.41}$ & $0.45^{+0.34}_{-0.25}$ & $3.7^{+13.9}_{-2.7}$ & $27.7/14$ \\
        UMalI & $1.7^{+13.6}_{-1.4}$ & $0.96^{+1.68}_{-0.80}$ & $0.56^{+0.35}_{-0.36}$ & $2.1^{+1.7}_{-1.0}$ & $30.9/13$ \\
        UMi   & $0.80^{+0.92}_{-0.51}$ & $3.8^{+3.3}_{-1.5}$ & $0.53^{+0.25}_{-0.20}$ & $14.7^{+6.6}_{-4.1}$ & $682.4/460$ \\
        Wil1  & $23^{+82}_{-21}$ & $0.12^{+2.25}_{-0.10}$ & $0.39^{+0.38}_{-0.20}$ & $1.20^{+4.08}_{-0.51}$ & $52.1/37$ \\
        \hline
    \end{tabular}
    }
\end{table}

\subsection{Comparison with the results from the literature}
\label{app:litcomp}
In general, the determination of astrophysical factors for dSph haloes is affected by several uncertainties and systematics. If not recognized and appropriately removed or mitigated, such spurious contributions may significantly alter the analysis of the dSph stellar kinematics, leading to a wrong estimate of the DM content. The major sources of uncertainties and systematics affecting $J_{\rm ann}$ and $J_{\rm dec}$ values are:

\begin{itemize}
    \item {\bfseries Difficulty/impossibility to obtain tangential components of the member star velocities} -- The possibility to neglect the rotational support in dSph dynamics, quantified by the dispersion of stellar proper motions, is key to considering such objects as DM dominated; in fact, the presence of a non-negligible stellar-velocity tangential component may significantly alter the distribution of measured radial velocities, thus artificially increasing an intrinsically low DM amount. However, since the typical proper motion of a dSph is roughly of $0.2 - 0.5$ mas yr$^{-1}$ and the inner radial velocity dispersion is of the order of 10 km s$^{-1}$, the proper-motion dispersion of the dSph member stars is of the order of 0.01 mas yr$^{-1}$; for bright stars ($G < 15$), this amount is already at the limit of current and future stellar surveys, such as the {\it Gaia} second \citep[DR2;][]{gai18,gai20a,gai20b} and early third data release\footnote{See \url{https://www.cosmos.esa.int/web/gaia/earlydr3}.} (EDR3).
    \item {\bfseries Stellar velocity dispersion dominated by tidal forces} -- Another source of alteration of the dSph stellar velocity dispersion comes from the risk that the analyzed dwarf galaxy does not reside inside a gravitationally undisturbed DM mini-halo of primordial origin, but is rather a remnant of a bigger object that has been tidally disrupted by close encounters with the MW. In the first case, the measured radial velocity dispersion $\sigma_r$ is \citep{bin08}:
    \begin{equation}\label{eqn:dmvel}
        \sigma_r^{\rm (DM)} = \sqrt{\frac{GM_{\rm DM}}{3R_{\rm tid}}}\, ,
    \end{equation}
    whereas in the second case one gets:
    \begin{equation}\label{eqn:tidvel}
        \sigma_r^{\rm (tid)} = \sqrt{\frac{2GM_{\rm MW}R_{\rm tid}}{3d_{\rm GC}^2}}\, .
    \end{equation}
    It is therefore clear that, lacking hints of ongoing tidal interaction in the target, a large $\sigma_r$ can potentially lead to its wrong attribution to an extreme DM domination; a clear example is offered by the SgrI dSph, which exhibits a rather large -- although uncertain -- value of $J_{\rm ann}$ \citep[but see][]{ven24}. The current deep stellar surveys may help to identify tidally disrupted sources by detecting the stellar streams produced by the gravitational encounters of the dSph with the MW \citep[e.g.,][]{drl15,mut18}, thus allowing the reanalysis (or exclusion) of targets located within such features; this might likely be the case of SgrII, which presumably lies inside the trailing arm of the Sagittarius stream \citep{lae15a}.

    \begin{table}
    \centering
    \caption{Average best-fit Burkert profile parameters for the sample of optimal dSphs.}
    \label{tab:burkert_parameters}
    \resizebox{\columnwidth}{!}{
    \begin{tabular}{lcccc}
        \hline
        \hline
        Name & $\rho_s$ ($10^7$ M$_{\odot}$ kpc$^{-3}$) & $r_s$ (kpc) & $R_{\rm tid}$ (kpc) & $\chi^2/n_{\rm d.o.f.}$ \\
        \hline
        CBe   & $57^{+22}_{-27}$   & $1.7^{+3.7}_{-1.6}$   & $19^{+55}_{-16}$   & $76.3/53$ \\
        Dral  & $38^{+24}_{-16}$   & $0.32^{+0.15}_{-0.10}$   & $4.30^{+0.86}_{-0.54}$   & $643.5/463$ \\
        RetII & $38^{+55}_{-21}$   & $0.28^{+5.32}_{-0.22}$   & $5.8^{+19.3}_{-5.1}$   & $21.6/12$ \\
        Scl   & $40.8^{+8.4}_{-8.9}$   & $0.247^{+0.048}_{-0.026}$   & $3.71^{+0.30}_{-0.18}$   & $1501.6/1114$ \\
        SgrII & $260^{+140}_{-170}$   & $0.36^{+2.52}_{-0.31}$   & $4.2^{+36.4}_{-2.8}$   & $27.9/15$ \\
        UMalI & $16^{+181}_{-13}$   & $0.43^{+3.60}_{-0.34}$   & $2.2^{+6.5}_{-1.0}$   & $31.0/14$ \\
        UMi   & $16.3^{+2.6}_{-2.1}$   & $1.62^{+1.01}_{-0.47}$   & $15.3^{+8.6}_{-3.9}$   & $682.5/461$ \\
        Wil1  & $290^{+440}_{-230}$   & $0.070^{+2.770}_{-0.038}$   & $1.35^{+26.35}_{-0.48}$   & $52.5/38$ \\
        \hline
    \end{tabular}
    }
\end{table}

\begin{table*}
\caption{Astrophysical factors for DM annihilation and decay of the 14 dSphs remaining after the 2nd selection criterion, integrated up to $0.1^\circ$ and $0.5^\circ$ for both the Einasto and Burkert DM density profiles (see text for details), along with the corresponding uncertainties at 68\% CL. For profiles yielding unconstrained values, ULs are given at a 95\% CL. In the table, all the astrophysical factors for DM annihilation ($J_{\rm ann}$) in logarithmic GeV$^2$ cm$^{-5}$ and all those for DM decay ($J_{\rm dec}$) in logarithmic GeV cm$^{-2}$.}
\label{tab:dsph_jfactors}
\begin{center}
\resizebox{\textwidth}{!}{
\begin{tabular}{l|cccc|cccc}
\hline
\hline
\multicolumn{1}{l}{} & \multicolumn{4}{c}{E{\scriptsize INASTO}} & \multicolumn{4}{c}{B{\scriptsize URKERT}}\\
\cline{2-5}
\cline{6-9}
  & $\log{J_{\rm ann}}$ & $\log{J_{\rm ann}}$ & $\log{J_{\rm dec}}$ & $\log{J_{\rm dec}}$ & $\log{J_{\rm ann}}$ & $\log{J_{\rm ann}}$ & $\log{J_{\rm dec}}$ & $\log{J_{\rm dec}}$\\
Name & $(<0.1^\circ)$ & $(<0.5^\circ)$ & $(<0.1^\circ)$ & $(<0.5^\circ)$ & $(<0.1^\circ)$ & $(<0.5^\circ)$ & $(<0.1^\circ)$ & $(<0.5^\circ)$\\
\hline
Bo{\"o}I & $17.8^{+0.4}_{-0.3}$ & $18.5^{+0.7}_{-0.5}$ & $17.2^{+0.4}_{-0.2}$ & $18.4^{+0.6}_{-0.4}$ & $18.0^{+0.6}_{-0.6}$ & $19.3^{+0.7}_{-0.6}$ & $17.6^{+0.5}_{-0.4}$ & $19.0^{+0.6}_{-0.5}$\\
CBe & $18.7^{+0.4}_{-0.5}$ & $19.6^{+0.8}_{-0.7}$ & $17.6^{+0.6}_{-0.3}$ & $18.9^{+0.8}_{-0.5}$ & $18.9^{+0.8}_{-0.4}$ & $20.3^{+0.8}_{-1.1}$ & $18.2^{+0.5}_{-0.8}$ & $19.6^{+0.5}_{-1.1}$\\
DraI & $18.3^{+0.3}_{-0.2}$ & $18.7^{+0.1}_{-0.1}$ & $17.3^{+0.1}_{-0.1}$ & $18.3^{+0.1}_{-0.1}$ & $18.1^{+0.2}_{-0.2}$ & $18.7^{+0.1}_{-0.1}$ & $17.3^{+0.1}_{-0.1}$ & $18.3^{+0.1}_{-0.1}$\\
GruII & $14.9^{+1.0}_{-1.3}$ & $15.5^{+1.1}_{-2.0}$ & $15.5^{+0.7}_{-1.8}$ & $16.2^{+1.2}_{-3.1}$ & $\lesssim$15.6 & $\lesssim$18.0 & $\lesssim$16.3 & $\lesssim$18.0\\
RetII & $18.3^{+0.3}_{-0.3}$ & $19.0^{+0.8}_{-0.5}$ & $17.3^{+0.4}_{-0.3}$ & $18.5^{+0.6}_{-0.7}$ & $18.6^{+0.2}_{-0.7}$ & $20.1^{+0.1}_{-1.5}$ & $18.2^{+0.1}_{-1.1}$ & $18.9^{+0.7}_{-1.2}$\\
Scl & $18.2^{+0.3}_{-0.2}$ & $18.4^{+0.2}_{-0.1}$ & $17.2^{+0.1}_{-0.1}$ & $17.9^{+0.1}_{-0.1}$ & $17.9^{+0.1}_{-0.1}$ & $18.3^{+0.1}_{-0.1}$ & $17.2^{+0.1}_{-0.1}$ & $18.0^{+0.1}_{-0.1}$\\
Seg1 & $16.2^{+1.7}_{-2.4}$ & $16.9^{+1.6}_{-3.4}$ & $16.2^{+1.2}_{-1.0}$ & $17.3^{+1.5}_{-1.8}$ & $\lesssim$18.0 & $\lesssim$21.1 & $\lesssim$17.5 & $\lesssim$21.1\\
Sex & $17.3^{+0.6}_{-0.8}$ & $18.0^{+0.3}_{-0.3}$ & $17.1^{+0.1}_{-0.1}$ & $18.3^{+0.1}_{-0.1}$ & $16.6^{+0.3}_{-0.1}$ & $17.9^{+0.1}_{-0.1}$ & $17.0^{+0.1}_{-0.1}$ & $18.3^{+0.1}_{-0.1}$\\
$^\ddag$SgrI & $18.2^{+1.0}_{-2.1}$ & $18.9^{+0.7}_{-1.4}$ & $17.1^{+0.3}_{-0.7}$ & $18.3^{+0.3}_{-0.6}$ & $\lesssim$16.2 & $\lesssim$18.0 & $\lesssim$16.5 & $\lesssim$18.0\\
SgrII & $18.6^{+1.0}_{-0.8}$ & $18.9^{+1.7}_{-1.0}$ & $17.4^{+0.8}_{-0.7}$ & $18.3^{+1.1}_{-1.1}$ & $19.2^{+0.9}_{-0.6}$ & $19.6^{+1.7}_{-1.1}$ & $17.8^{+0.8}_{-0.6}$ & $18.7^{+1.3}_{-1.1}$\\
TriII & $16.0^{+2.2}_{-3.7}$ & $16.6^{+2.4}_{-4.5}$ & $16.1^{+1.3}_{-2.7}$ & $16.9^{+1.6}_{-3.9}$ & $\lesssim$18.6 & $\lesssim$21.6 & $\lesssim$17.8 & $\lesssim$21.6\\
$^\dag$UMaII & $18.1^{+0.7}_{-0.7}$ & $18.9^{+0.5}_{-0.4}$ & $17.3^{+0.3}_{-0.2}$ & $18.4^{+0.3}_{-0.3}$ & $17.8^{+1.2}_{-0.8}$ & $18.9^{+0.5}_{-0.6}$ & $17.3^{+0.2}_{-0.3}$ & $18.4^{+0.4}_{-0.4}$\\
UMi & $18.2^{+0.1}_{-0.1}$ & $19.3^{+0.2}_{-0.2}$ & $17.6^{+0.1}_{-0.1}$ & $18.9^{+0.2}_{-0.1}$ & $18.2^{+0.1}_{-0.1}$ & $19.4^{+0.2}_{-0.1}$ & $17.7^{+0.2}_{-0.1}$ & $19.0^{+0.2}_{-0.1}$\\
$^\dag$Wil1 & $18.9^{+0.4}_{-0.4}$ & $19.2^{+0.7}_{-0.5}$ & $17.3^{+0.4}_{-0.3}$ & $18.0^{+1.0}_{-0.5}$ & $19.0^{+0.7}_{-0.4}$ & $19.3^{+1.5}_{-0.6}$ & $17.4^{+1.0}_{-0.3}$ & $18.1^{+1.7}_{-0.5}$\\
\hline
\end{tabular}
}
\end{center}
$^\dag$\footnotesize{Kinematically altered targets.\\}
$^\ddag$\footnotesize{Kinematically altered targets with non-negligible $\gamma$-ray background emission.}
\end{table*}

    \item {\bfseries Contamination of member samples by foreground stars} -- Since the dSphs are often viewed in projection only, with no or very few hints about surrounding stellar structures, foreground population of stars with age and metallicity that balance the distance difference with respect to the dSph cannot be distinguished through photometric measurements. Therefore, the measurement of spectroscopic velocities is crucial in order to fully disentangle the dSph stellar population by foreground contamination. The erroneous inclusion of such spurious populations may deeply alter the calculation of correct dSph astrophysical factors: this is the case of Seg1, which had a $J_{\rm ann} \gtrsim 10^{19}$ GeV$^2$ cm$^{-5}$ \citep[e.g.,][]{ack14,ger15} until the discovery by \citet{sim11} of high-velocity foreground stars ($v \sim 300$ km s$^{-1}$) superimposed on the dSph structure, which led \citet{bon15a} and \citet{bon16b} to the revision of its DM content ($J_{\rm ann} \ll 10^{18}$ GeV$^2$ cm$^{-5}$).
    \item {\bfseries Missing consideration of the dSph triaxiality} -- Undisturbed dSphs are often treated as if they possessed a quasi-spheroidal symmetry, which allows the application of the spherical Jeans analysis with a negligible approximation on the real dSph geometry. However, such an assumption is in contrast with both (i) the observed non-spherical distribution of luminous matter in both classical and ultra-faint dSphs \citep{irw95,McConnachie:2012vd}, and (ii) the prediction of non-spherical DM haloes in $\Lambda$-CDM models of structure formation \citep{hay16}. Axisymmetric studies of the mass distribution in dSphs actually predict triaxial dark haloes \citep{hay12,hay15}; based on this evidence, \citet{hay16} adopted such models to derive the values of $J_{\rm ann}$ and $J_{\rm dec}$ for dSphs. They found that the median values of such quantities computed in this way are in general lower than those obtained with the assumption of spherical haloes, and with larger error bars; an increase of $\sim$0.4 dex in the uncertainties at 68\% probability is also evidenced by \citet{bon15c} when taking into account the DM halo triaxiality.
\end{itemize}

\begin{table}
    \centering
    \caption{Ranking of the optimal dSphs, based on their relative values of $J_{\rm ann}$ (Einasto profile) integrated up to the reported angular sizes. The angle $\alpha_{\rm tid} = \arctan{(R_{\rm tid}/d_\odot)}$ corresponds to the projection of the tidal radius $R_{\rm tid}$ (see \autoref{tab:dsph_parameters_2}) at the dSph distance $d_\odot$. Our final ranking is the one reported in the $0.1^\circ$ column.}
    \label{tab:ranking}
    \begin{tabular}{lcccc}
    \hline
    \hline
        Name & $0.1^\circ$ & $0.5^\circ$ & $1^\circ$ & $\alpha_{\rm tid}$\\
    \hline
        BooI & $10^{\rm th}$ & $8^{\rm th}$ & $8^{\rm th}$ & $8^{\rm th}$\\
        CBe & $2^{\rm nd}$ & $1^{\rm st}$ & $1^{\rm st}$ & $2^{\rm nd}$\\
        DraI & $4^{\rm th}$ & $7^{\rm th}$ & $9^{\rm th}$ & $10^{\rm th}$\\
        RetII & $5^{\rm th}$ & $4^{\rm th}$ & $4^{\rm th}$ & $3^{\rm rd}$\\
        Scl & $6^{\rm th}$ & $10^{\rm th}$ & $10^{\rm th}$ & $11^{\rm th}$\\
        Sex & $11^{\rm th}$ & $11^{\rm th}$ & $11^{\rm th}$ & $9^{\rm th}$\\
        SgrI & $9^{\rm th}$ & $9^{\rm th}$ & $7^{\rm th}$ & $7^{\rm th}$\\
        SgrII & $3^{\rm rd}$ & $5^{\rm th}$ & $6^{\rm th}$ & $6^{\rm th}$\\
        UMaII & $8^{\rm th}$ & $6^{\rm th}$ & $5^{\rm th}$ & $5^{\rm th}$\\
        UMi & $7^{\rm th}$ & $2^{\rm nd}$ & $2^{\rm nd}$ & $1^{\rm st}$\\
        Wil1 & $1^{\rm st}$ & $3^{\rm rd}$ & $3^{\rm rd}$ & $4^{\rm th}$\\
    \hline
    \end{tabular}
\end{table}

\begin{figure*}
\centering
\includegraphics[scale=0.6,angle=-90]{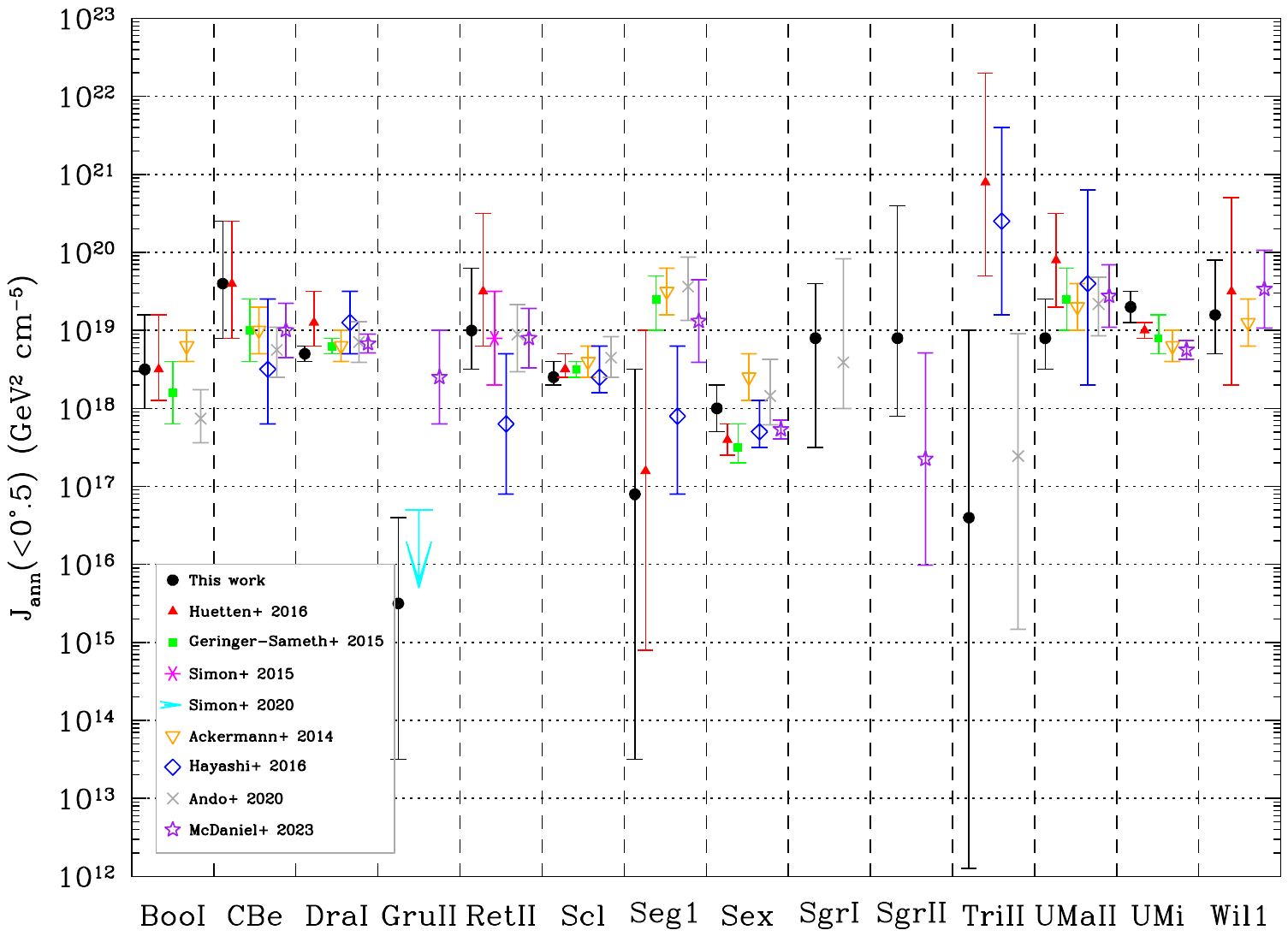}
\caption{Same as in \autoref{fig:dsph_comparison_01deg}, but within $0.5^\circ$ of integration \citep[see legend]{ack14,ger15,sim15,hay16,hut16,and20,sim20,mcd23}.}
\label{fig:dsph_jfactor_05}
\end{figure*}

Considering such potential issues, we check the consistency of our set of astrophysical factors for the selected optimal dSphs by comparing them to the values available in the literature for both $\alpha_{\rm int} = 0.1^\circ$ and $\alpha_{\rm int} = 0.5^\circ$. To this aim, we collect all the relevant estimates of $J_{\rm ann}$ at such integration angles obtained from the analysis of the dSph kinematics \citep{ack14,bon15b,bon15c,ger15,sim15,gen16,hay16,sim20,and20}. For sources lacking an estimate of $J_{\rm ann}(<0.1^\circ)$ but having values of $J_{\rm ann}$ derived at larger integration angles (e.g., $0^\circ.2$ or $0.5^\circ$), such as GruII \citep{sim20} and RetII \citep{sim15}, we extrapolate them to $\alpha_{\rm int} = 0.1^\circ$ based on such values. We show the comparison between our values of $J_{\rm ann}(<0.1^\circ)$ and those from the literature in \autoref{fig:dsph_comparison_01deg}, as well as between our $J_{\rm ann}(<0.5^\circ)$ estimates and the literature in \autoref{fig:dsph_jfactor_05}. Such plots reveals that, prior to the analysis by \citet{bon15b}, the astrophysical factor of Seg1 was overestimated by a factor of $>$100 due to the inclusion in its member sample of the spurious stellar population with $\langle v_r \rangle \sim 300$ km s$^{-1}$ (see \autoref{fig:dsph_brightness_profile}). An even more severe overestimation by $>$4 orders of magnitude was made for TriII due to poorly determined kinematics of its member stars \citep{kir17}. The need for selecting clean kinematic samples in gravitationally undisturbed objects to obtain a reliable measurement of the DM amount in a dSph halo is again well exemplified by the case of SgrI, which would be classified as a DM-dominated source ($J_{\rm ann} \gtrsim 10^{18}$ GeV$^2$ cm$^{-5}$) if the gravitational disturbance due to its proximity to the dense Galactic bulge were not known.

\subsection{Ranking for source extension}
\label{sec:ranking_extension}
As a final consistency check on our dSph selection, we report in \autoref{tab:ranking} the positional ranking of every source of interest, based on their relative $J_{\rm ann}$ within a specified integration angle. As can be inferred by this table, once excluding SgrI due to its well-known altered dynamical status (see \autoref{app:litcomp}), the only target that significantly challenges our selection of optimal sources is BooI; such an object has in fact a larger $J_{\rm ann}$ with respect to Scl for integration angles up to $0.5^\circ$, and also with respect to DraI when integrating up to $1^\circ$. However, this does not happen at the prioritized integration angle of $0.1^\circ$, and only affects the bottom of our ranking -- with the position of the best 6 sources (CBe, RetII, SgrII, UMaII, UMi and Wil1) not being significantly altered. Since the first ranked objects at any integration angle will likely be the only ones observed by CTAO during the first part of its scientific operations, we leave unaltered our selection of optimal targets presented above.

\begin{figure*}
    \centering
    \begin{minipage}{0.49\textwidth}
        \includegraphics[width=\textwidth]{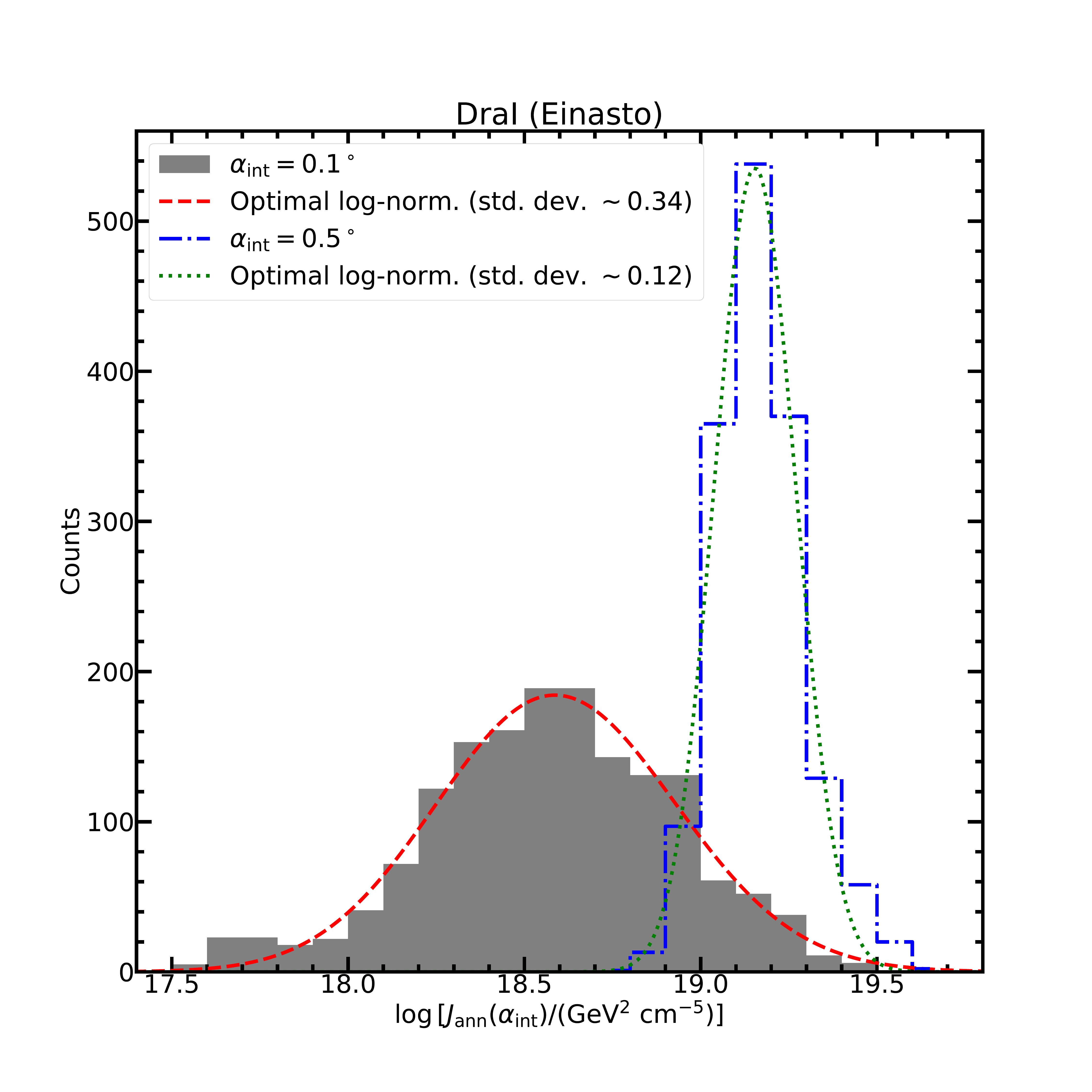}
    \end{minipage}
    \begin{minipage}{0.49\textwidth}
        \includegraphics[width=\textwidth]{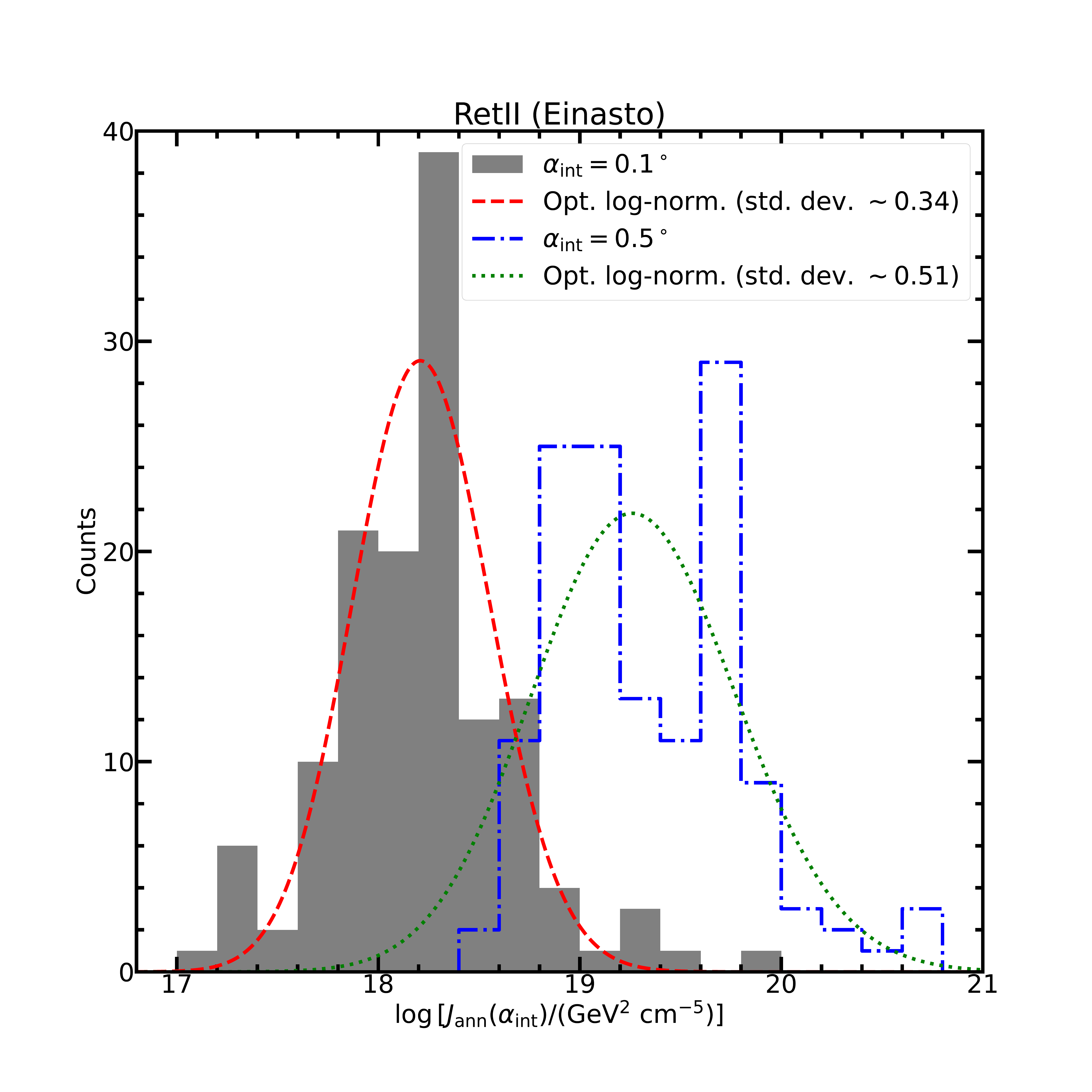}
    \end{minipage}
    \caption{Examples of posterior distributions for the annihilation astrophysical factors of DraI ({\itshape left panel}) and RetII ({\itshape right panel}) using the Einasto DM density profile. In both panels, histograms are computed at integration angles of $0.1^\circ$ ({\itshape gray areas}) and $0.5^\circ$ ({\itshape blue dot-dashed lines}). The optimal log-normal curves ({\itshape red dashed and green dotted lines}) are shown superimposed to the relative histogram, along with the corresponding standard deviations ({\itshape see legend}).}
    \label{fig:jfactor_PDF}
\end{figure*}

\section{Studies on sources of systematic uncertainties}
\label{app:systematics}
In this appendix, we evaluate the impact of some analysis choices on our results. In detail, in \autoref{app:extension} we discuss the effect of the dSph extension; in \autoref{app:skyregion} we compare the results obtained with different apertures of the integration region; in \autoref{app:onoff_template} we describe the performance of the ON/OFF method, in comparison with the template-background one; in \autoref{app:irf_sys} we discuss the effect of IRF systematics; finally, in \autoref{app:ctools} we compare our results with those obtained with other publicly available tools like \texttt{ctools} and \texttt{Swordfish}.

\subsection{Statistical uncertainties on the astrophysical factor}
\label{app:jfactor_sys}
Our calculation of the astrophysical factors with the \texttt{CLUMPY} software (see \autoref{sec:dm}) allows us to compute the posterior distributions of the values of $J_{\rm ann}$ and $J_{\rm dec}$ at arbitrary values of the integration angle $\alpha_{\rm int}$ (see \autoref{app:astro}). However, as shown in \autoref{fig:dsph_profile_ann} and \autoref{fig:dsph_profile_dec} and described in \autoref{app:astro}, such an analysis results in relevant uncertainties affecting the values of $J_{\rm ann}$ and $J_{\rm dec}$, especially at large integration angles. An accurate estimate of such systematics is influenced by several factors, mostly related to the quality of the photometric and spectroscopic data available for each target, as well as the working hypothesis on the DM distribution around the source (anisotropy, triaxiality, etc. as discussed in \autoref{app:astro}).

As an example, in \autoref{fig:jfactor_PDF} we report the histograms of such posterior distributions for the DraI and RetII DM haloes -- modeled with the Einasto DM density profile -- at two different integration angles ($0.1^\circ$ and $0.5^\circ$). We note that the distributions appear log-normal, with extended tails towards both small and large values of the astrophysical factor. In addition, the general dependence of such distributions on the integration angle is opposite for classical (with large stellar statistics) and ultra-faint dSphs (with $\lesssim$50 member stars), with the former tending to narrow around a saturation value of the astrophysical factor for increasing $\alpha_{\rm int}$ and the latter becoming more spread out at large integration angles. Therefore, we conclude that there is no unique reference for a statistical distribution of the uncertainties on the astrophysical factors, making their inclusion in \autoref{eq:lkl_combined_noise} non-trivial.

In this work, in line with what is done in \citet{CTA:DM_line}, we use a log-normal distribution as our benchmark for the statistical uncertainties on the astrophysical factors, with a width equal to the average of the upper and lower 1$\sigma$ uncertainties computed at an integration angle of $0.5^\circ$ (see \autoref{fig:jfactor_PDF}). We use the distributions parametrized in this way for each dSph in \autoref{eq:lkl_combined_noise}; a full treatment of these uncertainties in a 2D approach could be explored in a future publication.

\subsection{Effect of tidal stripping on the DM density profile}
\label{app:stripping}
\begin{figure*}
    \centering
    \includegraphics[width=\textwidth]{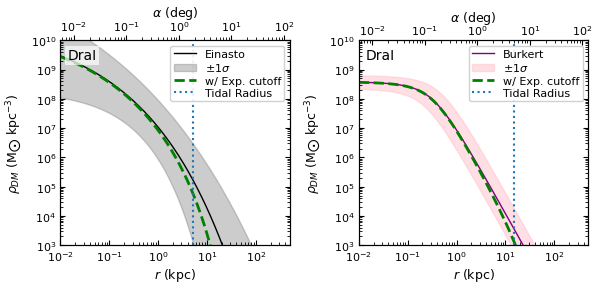}
    \caption{Impact of the exponential cut-off produced by tidal interactions on the DM density profiles for the representative case of DraI. In both panels, the median profiles modified by the functional form presented in \autoref{eq:expdec} \citep[{\it green dashed lines};][]{err21} are shown superimposed to the \texttt{CLUMPY} best-fit median profiles ({\it black and purple solid lines}) along with the corresponding uncertainties at 68\% CL ({\it grey and pink bands}). The respective tidal radii ({\it blue dotted lines}) are also indicated.}
    \label{fig:expdec}
\end{figure*}

We have detailed in \autoref{app:litcomp} how tidal forces can dominate the stellar velocity dispersions in dSphs. Even in case of DM-dominated sources as those analysed in this paper, tidal interactions can still strip their outer envelope of the more loosely bound DM particles, leading to cut-off DM density profiles with respect to the theoretical expectations \citep{hay04,kaz04,pen10}. We already account for a degree of tidal stripping in the dSph haloes by stopping the integration of $J_{\rm ann}$ and $J_{\rm dec}$ at the tidal angle $\alpha_{\rm tid}$ (see \autoref{sec:dm-astro} and \autoref{sec:ranking_extension}); here, we instead evaluate potential biases introduced with this technique with respect to a proper modeling of stripped halo profiles. Following \citet{err21}, we thus multiply the DM density profiles $\rho_{\rm Ein/Bur}(r)$ presented in \autoref{eq:dm_profiles} by an exponential cut-off:
\begin{equation}\label{eq:expdec}
    \rho_{\rm DM}(r) = \rho_{\rm Ein/Bur}(r) \times \frac{e^{-r/R_{\rm tid}}}{\left(
    1 + r_{\rm s}/R_{\rm tid}
    \right)^{0.3}},
\end{equation}
with $r_{\rm s}$ and $R_{\rm tid}$ selected according to the accounted profile. We show the impact of such a cut-off on the DM density profiles in \autoref{fig:expdec} for the representative case of DraI.

We then compare the integrals of $J_{\rm ann}$ truncated to $R_{\rm tid}$ with no exponential cut-off to the values obtained by integrating the entire DM density profiles defined in \autoref{eq:expdec}: we find that the latter are systematically lower than the former by $\sim$0.2 dex on average for the Einasto shape, and by $\sim$0.1 dex for the Burkert functional form. Lower astrophysical factors clearly affect the expected $\gamma$-ray signal, thus worsening the detection prospects; however, we note that a (systematic) $\lesssim$0.2-dex change in the value of $J_{\rm ann}$ or $J_{\rm dec}$ produces an effect that falls at most within the (statistical) 2$\sigma$ uncertainty of the limits on the DM parameters presented in \autoref{fig:results_cusp} to \autoref{fig:results_combined}. In addition, the magnitude of the signal loss becomes significant only when integrating up to large angular regions that approach $\alpha_{\rm tid}$. We therefore conclude that the impact of tidally induced cut-offs on the $J_{\rm ann}$ and $J_{\rm dec}$ values may be mitigated by adopting appropriate observing strategies -- e.g., focusing on the innermost regions of the dSph haloes or dividing the FoV into adequately defined RoIs (see \autoref{sec:morph}).

\subsection{Source extension}
\label{app:extension}
A target is considered extended if its size is larger than the telescope angular resolution. This quantity is approximately defined as the standard deviation of the telescope PSF, i.e. the distribution of photons in the focal plane from a perfectly par-axial beam. The PSF is normally modeled as a double Gaussian\footnote{\citet{DaVela:2018ire} demonstrated that a more accurate modeling of the PSF is through a King function, but this is not relevant for the purpose of this work.}. If the angular distribution of the detected events coincides with the PSF, and thus no extension can be determined, the target is considered to be point-like; in this condition, the ON/OFF analysis (see \autoref{app:onoff_template}) is the most accurate. Conversely, the template method is better suited for an extended source.

In general, the telescope acceptance changes as a function of the source position in the FoV -- the farther from the telescope optical axis, the worse the acceptance. Therefore, for the analysis of an extended source one has to convolve the signal with the corresponding acceptance, which normally possesses circular symmetry; in this way, `acceptance' rings can be defined. In the case of an unknown source extension, the search for signals depends on the definition of the RoI in the ON/OFF method. The optimal dSphs selected in this work are all mildly extended, as shown in \autoref{tab:extension} (see also \autoref{fig:dsph_profile_ann} and \autoref{fig:dsph_profile_dec}). Our results indicate that, together with some very compact objects like DraI for the DM annihilation scenario and Scl for the DM decay one, there are significantly extended targets such as UMi and CBe, confirming the advantage of a template background analysis for all of the selected sources. 

\subsection{Impact of the aperture of the acceptance region}
\label{app:skyregion}
\begin{figure*}
    \centering
    \includegraphics[width=0.49\linewidth]{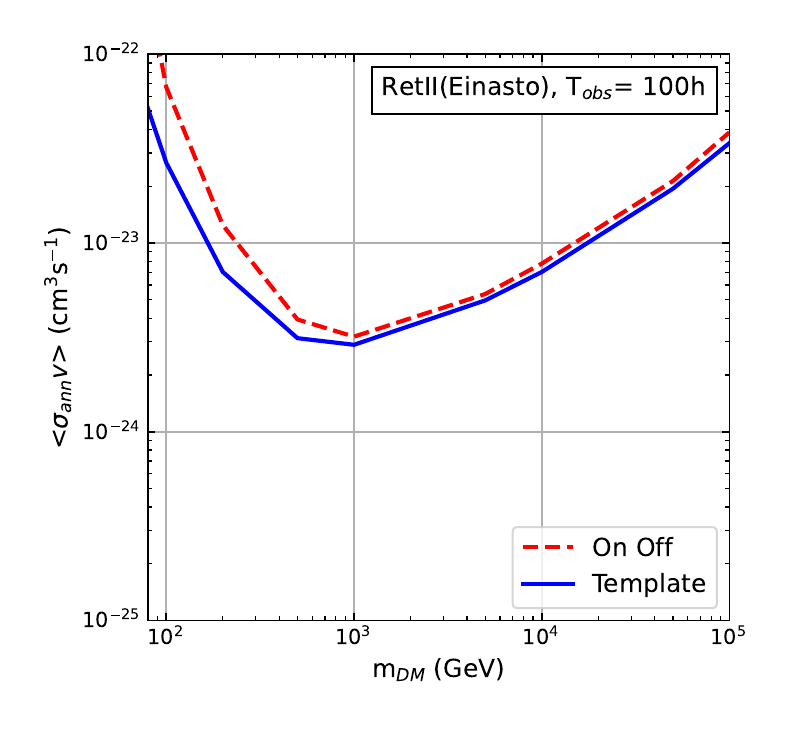}
    \includegraphics[width=0.49\linewidth]{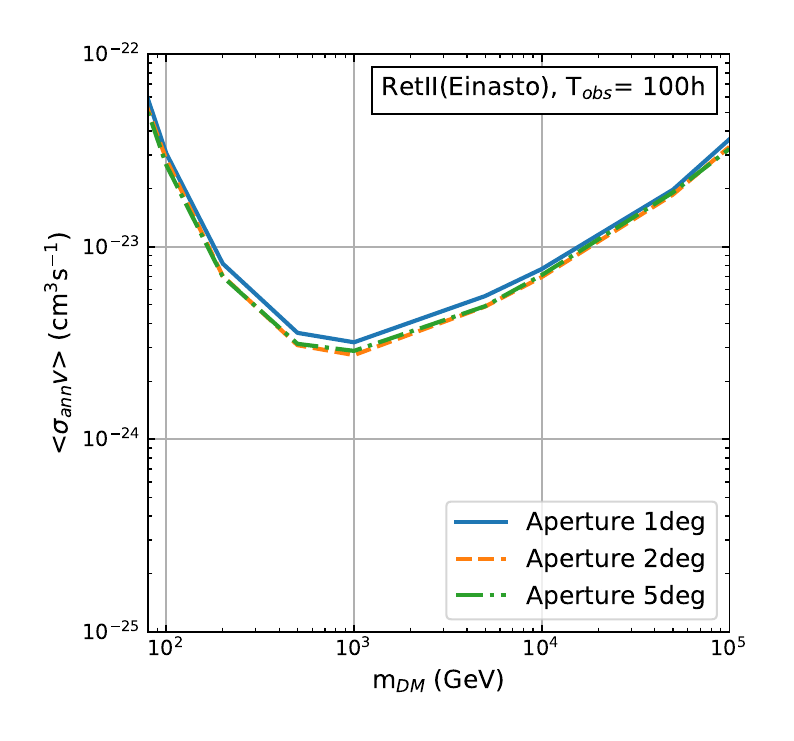}
        \includegraphics[width=0.49\linewidth]{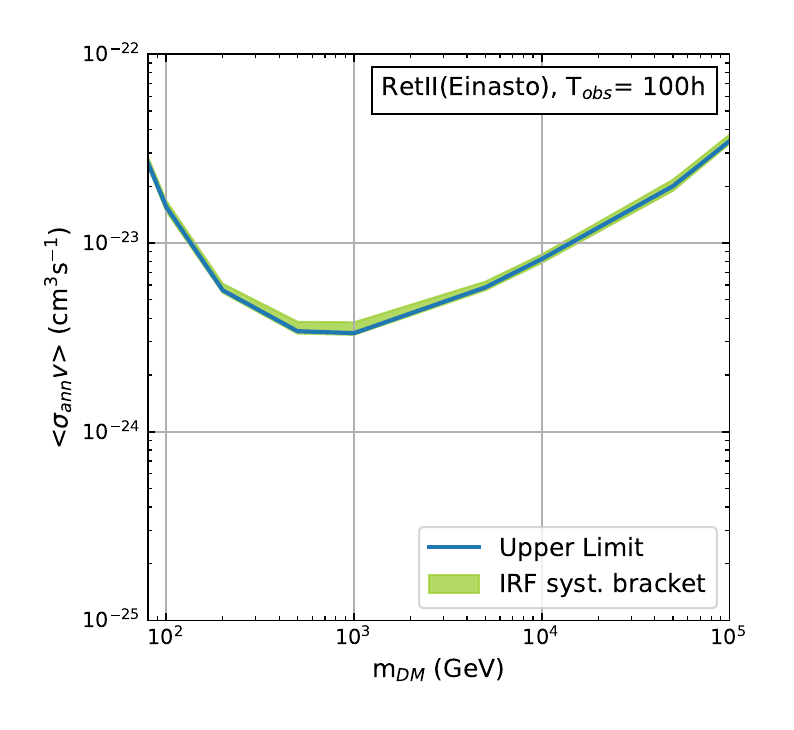}
            \includegraphics[width=0.49\linewidth]{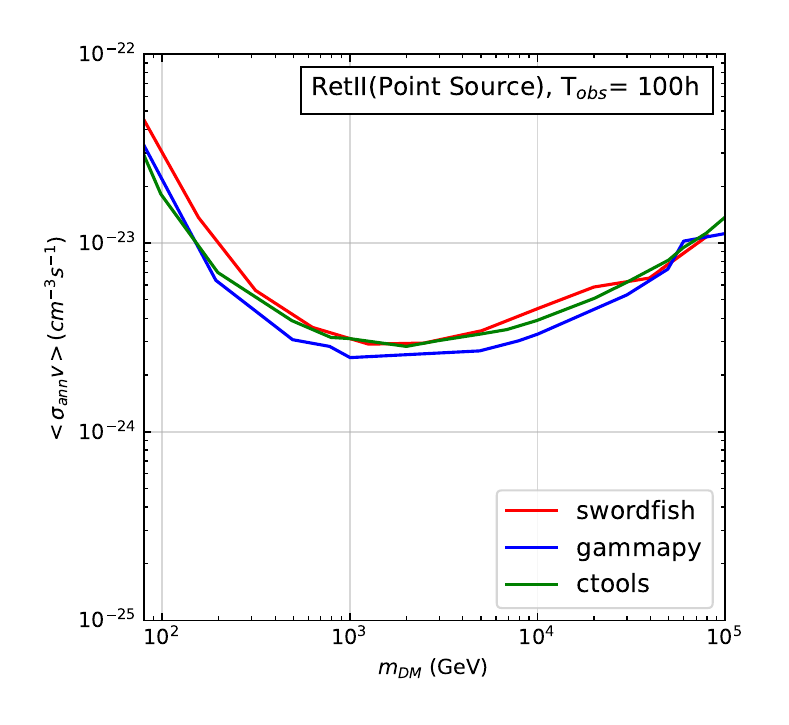}
 \caption{Checks on systematic uncertainties potentially affecting the derivation of particle DM parameters. {\it Top left panel:} comparison of the cross-section limits obtained using a template-based method ({\it blue solid line}) and an ON/OFF analysis ({\it orange dashed line}). {\it Top right panel:} comparison of the limits for different FoV sizes -- $1 \times 1$ deg$^2$ ({\it blue solid line}), $2 \times 2$ deg$^2$ ({\it orange dashed line}) and $5 \times 5$ deg$^2$ ({\it green dot-dashed line}). {\it Bottom left panel:} limits derived taking into account the IRF systematics ({\it green shaded area}). {\it Bottom right panel:} Comparison of limits computed with different codes for Cherenkov $\gamma$-ray analysis -- \texttt{gammapy} ({\it orange line}), \texttt{ctools} ({\it green line}) and \texttt{Swordfish} ({\it blue line}). In all panels, the ULs are computed assuming 100-h observations of the RetII dSph (Einasto astrophysical factor) for DM annihilating in the $b\bar{b}$ channel.}
    \label{fig:systematics}
\end{figure*}

In \autoref{app:onoff_template} we mentioned the use of a FoV of $2\times2$~deg$^2$ for the computation of event counts; to evaluate the impact that different choices of the FoV extensions have on our results, we produce likewise analyses on sky regions of $1\times1$~deg$^2$ and $5\times5$~deg$^2$, respectively. In \autoref{fig:systematics} (top right), we present the comparison between such choices; one can see that different FoV sizes have no or negligible effect on the derivation of the DM limits. Therefore, we conclude that the choice of a different aperture size with respect to our benchmark does not significantly impact our results.

\subsection{Comparison of ON/OFF and template analysis}
\label{app:onoff_template}
In \autoref{sec:methods} we stated our choice to use the so-called template background method for our analysis. Within this framework, the background counts are estimated through a model over the entire field of view. This model is currently generated through the analysis of MC simulations reproducing the expected behavior of the CTAO arrays, and cannot be validated before actual data from the instruments will be available. The alternative method for the background estimate is the commonly used ON/OFF method, in which background counts in an ON region -- i.e. the region containing the source emitting $\gamma$-rays -- are computed from a separate (OFF) background control region where no signal is expected. Normally, several OFF regions are identified in the FoV to increase the precision of the background intensity and shape measurements.

Since the background is measured rather than modeled, this technique has the advantage of being more accurate than the template method; however, it starts to be difficult to apply when the source is moderately to strongly spatially extended. The dSphs with moderate extension are borderline, and thus an ON/OFF analysis is still possible for such targets; therefore, we tested the two methods on the sample dSph RetII. In \autoref{fig:systematics} (top left), we compare the template and ON/OFF analyses of 100-h observations of a DM Einasto distribution in the RetII halo, assuming WIMPs of $m_\mathrm{DM} = 2.5$ TeV annihilating into the $b\bar{b}$ channel. The template analysis is computed over an aperture of $2 \times 2$ deg$^2$, whereas the ON/OFF analysis is computed over 5 background control regions of 0.11$^\circ$ each.

One can see that the two methods agree well between 300~GeV and 2~TeV, i.e. in the region with the largest expected instrumental sensitivity; the template method is better performing over the ON/OFF analysis below 300~GeV and above 2~TeV. Such a behavior is expected, since at low energies the amount of residual background is large and the background classification is more complex -- thus allowing the background model to improve the accuracy and precision of the analysis, and ultimately the sensitivity -- and at high energies the statistic of events is lower -- thus a more precise background estimate produces a higher signal-to-noise ratio. The actual difference in performance between the two analysis methods will be possible to be accurately evaluated only when real CTAO data will be collected, but we conclude that the choice of one analysis procedure over the other does not greatly impact our main results.

\subsection{Systematics on the instrument response functions (IRFs)}
\label{app:irf_sys}
An accurate reconstruction of signal and background counts would require the use of IRFs for the specific sky direction and atmospheric conditions, that are not available at the moment. For this reason, we do not directly include the systematic uncertainties of the IRFs into the likelihood formalism (see \autoref{eq:lkl_combined}); instead, we evaluate their impact by bracketing IRFs. This approach follows the procedure developed by the Fermi-LAT Collaboration \citep{Ackermann_2012}. We have biased the energy reconstruction by 6$\%$, the effective area by 5$\%$ and the background rate by 0.8$\%$, corresponding to $\pm$1$\sigma$ estimates matching the projected systematic uncertainties for CTA. Although potentially inaccurate, such an approximation allows us to bracket the maximum systematic uncertainties in this pre-operation prediction. We present the results in the bottom left panel of \autoref{fig:systematics}, evidencing how systematic errors from IRFs are well below the Poisson statistic fluctuations.


In addition, regarding the background counts, we estimate that they can be determined with systematics below 1\% \citep{cta_irf}, therefore becoming potentially relevant for observations longer than 100~h. This term should also be included in \autoref{eq:lkl_combined}; however, considering that it is subdominant with respect to other instrumental systematics uncertainties discussed here and the uncertainties on the astrophysical factors, we neglect it for the purposes of this work.

\subsection{Comparison of results obtained with other codes}
\label{app:ctools}
Finally, we compare in \autoref{fig:systematics} (bottom right) the DM limits obtained with \texttt{gammapy} \citep{gammapy}, for the template method, with those obtained with the independent open-source analysis pipeline \texttt{ctools}, based on the well-known {\it Fermi}-LAT tools and also valid to be used with CTAO data \citep{ctools}. We also make predictions of the null hypothesis using the Asimov dataset generated with the open-source \texttt{Swordfish} \citep{Edwards:2017mnf,Edwards:2017kqw}, used in other CTAO projects \citep{ach21}. In the bottom right panel of \autoref{fig:systematics}, we show how -- also taking into account the uncertainty on the DM limits produced by the IRF systematics -- the results obtained with such additional software are in very good agreement with those presented in this work.

\clearpage
\section*{Affiliations}
\begin{scriptsize}
\begin{enumerate}[label=$^{\arabic*}$,ref=\arabic*,leftmargin=1.5em,labelsep=0.25em,labelwidth=1.25em]
\item Department of Physics, Tokai University, 4-1-1, Kita-Kaname, Hiratsuka, Kanagawa 259-1292, Japan\label{AFFIL::JapanUTokai}
\item Institute for Cosmic Ray Research, University of Tokyo, 5-1-5, Kashiwa-no-ha, Kashiwa, Chiba 277-8582, Japan\label{AFFIL::JapanUTokyoICRR}
\item ETH Z\"urich, Institute for Particle Physics and Astrophysics, Otto-Stern-Weg 5, 8093 Z\"urich, Switzerland\label{AFFIL::SwitzerlandETHZurich}
\item INFN and Universit\`a degli Studi di Siena, Dipartimento di Scienze Fisiche, della Terra e dell'Ambiente (DSFTA), Sezione di Fisica, Via Roma 56, 53100 Siena, Italy\label{AFFIL::ItalyUSienaandINFN}
\item Universit\'e Paris-Saclay, Universit\'e Paris Cit\'e, CEA, CNRS, AIM, F-91191 Gif-sur-Yvette Cedex, France\label{AFFIL::FranceCEAIRFUDAp}
\item FSLAC IRL 2009, CNRS/IAC, La Laguna, Tenerife, Spain\label{AFFIL::SpainFSLACIRLCNRSIAC}
\item University of Alabama, Tuscaloosa, Department of Physics and Astronomy, Gallalee Hall, Box 870324 Tuscaloosa, AL 35487-0324, USA\label{AFFIL::USAUAlabamaTuscaloosa}
\item Universit\'e C\^ote d'Azur, Observatoire de la C\^ote d'Azur, CNRS, Laboratoire Lagrange, France\label{AFFIL::FranceOCotedAzur}
\item Laboratoire Leprince-Ringuet, CNRS/IN2P3, \'Ecole polytechnique, Institut Polytechnique de Paris, 91120 Palaiseau, France\label{AFFIL::FranceLLREcolePolytechnique}
\item Departament de F{\'\i}sica Qu\`antica i Astrof{\'\i}sica, Institut de Ci\`encies del Cosmos, Universitat de Barcelona, IEEC-UB, Mart{\'\i} i Franqu\`es, 1, 08028, Barcelona, Spain\label{AFFIL::SpainICCUB}
\item Instituto de Astrof{\'\i}sica de Andaluc{\'\i}a-CSIC, Glorieta de la Astronom{\'\i}a s/n, 18008, Granada, Spain\label{AFFIL::SpainIAACSIC}
\item Institute for Computational Cosmology and Department of Physics, Durham University, South Road, Durham DH1 3LE, United Kingdom\label{AFFIL::UnitedKingdomICCUDurham}
\item Pontificia Universidad Cat\'olica de Chile, Av. Libertador Bernardo O'Higgins 340, Santiago, Chile\label{AFFIL::ChileUPontificiaCatolicadeChile}
\item Universidad Nacional Aut\'onoma de M\'exico, Delegaci\'on Coyoac\'an, 04510 Ciudad de M\'exico, Mexico\label{AFFIL::MexicoUNAMMexico}
\item D\'epartement de physique nucl\'eaire et corpusculaire, University de Gen\`eve,  Facult\'e de Sciences, 1205 Gen\`eve, Switzerland\label{AFFIL::SwitzerlandUGenevaDPNC}
\item Sorbonne Universit\'e, CNRS/IN2P3, Laboratoire de Physique Nucl\'eaire et de Hautes Energies, LPNHE, 4 place Jussieu, 75005 Paris, France\label{AFFIL::FranceLPNHEUSorbonne}
\item LUX, Observatoire de Paris, Universit\'e PSL, Sorbonne Universit\'e, CNRS, 5 place Jules Janssen, 92190, Meudon, France\label{AFFIL::FranceObservatoiredeParis}
\item INAF - Osservatorio Astrofisico di Arcetri, Largo E. Fermi, 5 - 50125 Firenze, Italy\label{AFFIL::ItalyOArcetri}
\item INFN Sezione di Perugia and Universit\`a degli Studi di Perugia, Via A. Pascoli, 06123 Perugia, Italy\label{AFFIL::ItalyUPerugiaandINFN}
\item INFN Sezione di Napoli, Via Cintia, ed. G, 80126 Napoli, Italy\label{AFFIL::ItalyINFNNapoli}
\item Universit\`a degli Studi di Napoli {\textquotedblleft}Federico II{\textquotedblright} - Dipartimento di Fisica {\textquotedblleft}E. Pancini{\textquotedblright}, Complesso Universitario di Monte Sant'Angelo, Via Cintia - 80126 Napoli, Italy\label{AFFIL::ItalyUNapoli}
\item INAF - Osservatorio Astronomico di Roma, Via di Frascati 33, 00078, Monteporzio Catone, Italy\label{AFFIL::ItalyORoma}
\item International Institute of Physics, Universidade Federal do Rio Grande do Norte, 59078-970, Natal, RN, Brasil\label{AFFIL::BrazilURioGrandedoNorteIIP}
\item Departamento de F{\'\i}sica, Universidade Federal do Rio Grande do Norte, 59078-970, Natal, RN, Brasil\label{AFFIL::BrazilURioGrandedoNortePhys}
\item INFN Sezione di Padova, Via Marzolo 8, 35131 Padova, Italy\label{AFFIL::ItalyINFNPadova}
\item Instituto de F{\'\i}sica Te\'orica UAM/CSIC and Departamento de F{\'\i}sica Te\'orica, Universidad Aut\'onoma de Madrid, c/ Nicol\'as Cabrera 13-15, Campus de Cantoblanco UAM, 28049 Madrid, Spain\label{AFFIL::SpainIFTUAMCSIC}
\item Instituto de F{\'\i}sica - Universidade de S\~ao Paulo, Rua do Mat\~ao Travessa R Nr.187 CEP 05508-090  Cidade Universit\'aria, S\~ao Paulo, Brazil\label{AFFIL::BrazilIFUSaoPaulo}
\item Department of Physics, Chemistry \& Material Science, University of Namibia, Private Bag 13301, Windhoek, Namibia\label{AFFIL::NamibiaUNamibia}
\item Centre for Space Research, North-West University, Potchefstroom, 2520, South Africa\label{AFFIL::SouthAfricaNWU}
\item School of Physics and Astronomy, Monash University, Melbourne, Victoria 3800, Australia\label{AFFIL::AustraliaUMonash}
\item IPARCOS-UCM, Instituto de F{\'\i}sica de Part{\'\i}culas y del Cosmos, and EMFTEL Department, Universidad Complutense de Madrid, E-28040 Madrid, Spain\label{AFFIL::SpainUCMAltasEnergias}
\item Faculty of Science and Technology, Universidad del Azuay, Cuenca, Ecuador.\label{AFFIL::EcuadorUAzuay}
\item Ruhr University Bochum, Faculty of Physics and Astronomy, Astronomical Institute (AIRUB), Universit\"atsstra{\ss}e 150, 44801 Bochum, Germany\label{AFFIL::GermanyUBochumPhysAst}
\item Instituto de Astrof{\'\i}sica de Canarias and Departamento de Astrof{\'\i}sica, Universidad de La Laguna, La Laguna, Tenerife, Spain\label{AFFIL::SpainIAC}
\item Universit\`a degli Studi di Trento, Via Calepina, 14, 38122 Trento, Italy\label{AFFIL::ItalyUTrento}
\item INFN Sezione di Bari and Universit\`a degli Studi di Bari, via Orabona 4, 70124 Bari, Italy\label{AFFIL::ItalyUandINFNBari}
\item Friedrich-Alexander-Universit\"at Erlangen-N\"urnberg, Erlangen Centre for Astroparticle Physics, Nikolaus-Fiebiger-Str. 2, 91058 Erlangen, Germany\label{AFFIL::GermanyUErlangenECAP}
\item INFN Sezione di Padova and Universit\`a degli Studi di Padova, Via Marzolo 8, 35131 Padova, Italy\label{AFFIL::ItalyUPadovaandINFN}
\item Institut f\"ur Physik \& Astronomie, Universit\"at Potsdam, Karl-Liebknecht-Strasse 24/25, 14476 Potsdam, Germany\label{AFFIL::GermanyUPotsdam}
\item University of the Witwatersrand, 1 Jan Smuts Avenue, Braamfontein, 2000 Johannesburg, South Africa\label{AFFIL::SouthAfricaUWitwatersrand}
\item Institut f\"ur Theoretische Physik, Lehrstuhl IV: Plasma-Astroteilchenphysik, Ruhr-Universit\"at Bochum, Universit\"atsstra{\ss}e 150, 44801 Bochum, Germany\label{AFFIL::GermanyUBochum}
\item Center for Astrophysics | Harvard \& Smithsonian, 60 Garden St, Cambridge, MA 02138, USA\label{AFFIL::USACfAHarvardSmithsonian}
\item Department of Physics, Humboldt University Berlin, Newtonstr. 15, 12489 Berlin, Germany\label{AFFIL::GermanyUBerlin}
\item Deutsches Elektronen-Synchrotron, Platanenallee 6, 15738 Zeuthen, Germany\label{AFFIL::GermanyDESY}
\item CIEMAT, Avda. Complutense 40, 28040 Madrid, Spain\label{AFFIL::SpainCIEMAT}
\item Max-Planck-Institut f\"ur Physik, Boltzmannstr. 8, 85748 Garching, Germany\label{AFFIL::GermanyMPP}
\item Pidstryhach Institute for Applied Problems in Mechanics and Mathematics NASU, 3B Naukova Street, Lviv, 79060, Ukraine\label{AFFIL::UkraineIAPMMLviv}
\item Center for Astrophysics and Cosmology (CAC), University of Nova Gorica, Nova Gorica, Slovenia\label{AFFIL::SloveniaUNovaGoricaCAC}
\item Univ. Savoie Mont Blanc, CNRS, Laboratoire d'Annecy de Physique des Particules - IN2P3, 74000 Annecy, France\label{AFFIL::FranceLAPPUSavoieMontBlanc}
\item ASI - Space Science Data Center, Via del Politecnico s.n.c., 00133, Rome, Italy\label{AFFIL::ASISpaceScienceDataCenter}
\item Politecnico di Bari, via Orabona 4, 70124 Bari, Italy\label{AFFIL::ItalyPolitecnicoBari}
\item INFN Sezione di Bari, via Orabona 4, 70126 Bari, Italy\label{AFFIL::ItalyINFNBari}
\item Institut de Fisica d'Altes Energies (IFAE), The Barcelona Institute of Science and Technology, Campus UAB, 08193 Bellaterra (Barcelona), Spain\label{AFFIL::SpainIFAEBIST}
\item FZU - Institute of Physics of the Czech Academy of Sciences, Na Slovance 1999/2, 182 00 Praha 8, Czech Republic\label{AFFIL::CzechRepublicFZU}
\item INAF - Osservatorio Astronomico di Brera, Via Brera 28, 20121 Milano, Italy\label{AFFIL::ItalyOBrera}
\item INFN Sezione di Pisa, Edificio C {\textendash} Polo Fibonacci, Largo Bruno Pontecorvo 3, 56127 Pisa\label{AFFIL::ItalyINFNPisa}
\item University School for Advanced Studies IUSS Pavia, Palazzo del Broletto, Piazza della Vittoria 15, 27100 Pavia, Italy\label{AFFIL::ItalyIUSSPaviaINAF}
\item University of Zagreb, Faculty of electrical engineering and computing, Unska 3, 10000 Zagreb, Croatia\label{AFFIL::CroatiaUZagreb}
\item University of Oslo, Department of Physics, Sem Saelandsvei 24 - PO Box 1048 Blindern, N-0316 Oslo, Norway\label{AFFIL::NorwayUOslo}
\item INAF - Osservatorio di Astrofisica e Scienza dello spazio di Bologna, Via Piero Gobetti 93/3, 40129  Bologna, Italy\label{AFFIL::ItalyOASBologna}
\item Dublin City University, Glasnevin, Dublin 9, Ireland\label{AFFIL::IrelandDCU}
\item Dublin Institute for Advanced Studies, 31 Fitzwilliam Place, Dublin 2, Ireland\label{AFFIL::IrelandDIAS}
\item Joseph-von-Fraunhofer-Str. 25, 44227 Dortmund, Germany\label{AFFIL::LamarrInstituteGermany}
\item Armagh Observatory and Planetarium, College Hill, Armagh BT61 9DB, United Kingdom\label{AFFIL::UnitedKingdomArmaghObservatoryandPlanetarium}
\item School of Physics, University of New South Wales, Sydney NSW 2052, Australia\label{AFFIL::AustraliaUNewSouthWales}
\item Cherenkov Telescope Array Observatory, Saupfercheckweg 1, 69117 Heidelberg, Germany\label{AFFIL::GermanyCTAOHeidelberg}
\item Unitat de F{\'\i}sica de les Radiacions, Departament de F{\'\i}sica, and CERES-IEEC, Universitat Aut\`onoma de Barcelona, Edifici C3, Campus UAB, 08193 Bellaterra, Spain\label{AFFIL::SpainUABandCERESIEEC}
\item Department of Physics, Faculty of Science, Kasetsart University, 50 Ngam Wong Wan Rd., Lat Yao, Chatuchak, Bangkok, 10900, Thailand\label{AFFIL::ThailandUKasetsart}
\item National Astronomical Research Institute of Thailand, 191 Huay Kaew Rd., Suthep, Muang, Chiang Mai, 50200, Thailand\label{AFFIL::ThailandNARIT}
\item INAF - Osservatorio Astronomico di Capodimonte, Via Salita Moiariello 16, 80131 Napoli, Italy\label{AFFIL::ItalyOCapodimonte}
\item Universidade Cidade de S\~ao Paulo, N\'ucleo de Astrof{\'\i}sica, R. Galv\~ao Bueno 868, Liberdade, S\~ao Paulo, SP, 01506-000, Brazil\label{AFFIL::BrazilUCidadeSPaulo}
\item Dep. of Physics, Sapienza, University of Roma, Piazzale A. Moro 5, 00185, Roma, Italy \label{AFFIL::ItalyURomaSapienza}
\item INAF - Istituto di Astrofisica Spaziale e Fisica Cosmica di Milano, Via A. Corti 12, 20133 Milano, Italy\label{AFFIL::ItalyIASFMilano}
\item CCTVal, Universidad T\'ecnica Federico Santa Mar{\'\i}a, Avenida Espa\~na 1680, Valpara{\'\i}so, Chile\label{AFFIL::ChileUTecnicaFedericoSantaMaria}
\item Aix Marseille Univ, CNRS/IN2P3, CPPM, Marseille, France\label{AFFIL::FranceCPPMUAixMarseille}
\item Universit\'e Paris Cit\'e, CNRS, Astroparticule et Cosmologie, F-75013 Paris, France\label{AFFIL::FranceAPCUParisCite}
\item Centre for Advanced Instrumentation, Department of Physics, Durham University, South Road, Durham, DH1 3LE, United Kingdom\label{AFFIL::UnitedKingdomUDurham}
\item Department of Physics and Astronomy, University of California, Los Angeles, CA 90095, USA\label{AFFIL::USAUCLA}
\item INFN Sezione di Torino, Via P. Giuria 1, 10125 Torino, Italy\label{AFFIL::ItalyINFNTorino}
\item Dipartimento di Fisica - Universit\`a degli Studi di Torino, Via Pietro Giuria 1 - 10125 Torino, Italy\label{AFFIL::ItalyUTorino}
\item Dipartimento di Fisica e Chimica {\textquotedblleft}E. Segr\`e{\textquotedblright}, Universit\`a degli Studi di Palermo, Via Archirafi 36, 90123, Palermo, Italy\label{AFFIL::ItalyUPalermo}
\item INFN Sezione di Catania, Via S. Sofia 64, 95123 Catania, Italy\label{AFFIL::ItalyINFNCatania}
\item Universidade Federal Do Paran\'a - Setor Palotina, Departamento de Engenharias e Exatas, Rua Pioneiro, 2153, Jardim Dallas, CEP: 85950-000 Palotina, Paran\'a, Brazil\label{AFFIL::BrazilUFPR}
\item IRFU, CEA, Universit\'e Paris-Saclay, B\^at 141, 91191 Gif-sur-Yvette, France\label{AFFIL::FranceCEAIRFUDPhP}
\item INAF - Osservatorio Astrofisico di Catania, Via S. Sofia, 78, 95123 Catania, Italy\label{AFFIL::ItalyOCatania}
\item University of Oxford, Department of Physics, Clarendon Laboratory, Parks Road, Oxford, OX1 3PU, United Kingdom\label{AFFIL::UnitedKingdomUOxford}
\item Universidad de Valpara{\'\i}so, Blanco 951, Valparaiso, Chile\label{AFFIL::ChileUdeValparaiso}
\item University of Wisconsin, Madison, 500 Lincoln Drive, Madison, WI, 53706, USA\label{AFFIL::USAUWisconsin}
\item INAF - Istituto di Radioastronomia, Via Gobetti 101, 40129 Bologna, Italy\label{AFFIL::ItalyRadioastronomiaINAF}
\item INAF - Istituto Nazionale di Astrofisica, Viale del Parco Mellini 84, 00136 Rome, Italy\label{AFFIL::ItalyINAF}
\item Instituto de Astronomia, Geof{\'\i}sica e Ci\^encias Atmosf\'ericas - Universidade de S\~ao Paulo, Cidade Universit\'aria, R. do Mat\~ao, 1226, CEP 05508-090, S\~ao Paulo, SP, Brazil\label{AFFIL::BrazilIAGUSaoPaulo}
\item INFN Sezione di Trieste and Universit\`a degli Studi di Udine, Via delle Scienze 208, 33100 Udine, Italy\label{AFFIL::ItalyUUdineandINFNTrieste}
\item Instituto de F{\'\i}sica de S\~ao Carlos, Universidade de S\~ao Paulo, Av. Trabalhador S\~ao-carlense, 400 - CEP 13566-590, S\~ao Carlos, SP, Brazil\label{AFFIL::BrazilIFSCUSaoPaulo}
\item Universidad de Alcal\'a - Space \& Astroparticle group, Facultad de Ciencias, Campus Universitario Ctra. Madrid-Barcelona, Km. 33.600 28871 Alcal\'a de Henares (Madrid), Spain\label{AFFIL::SpainUAlcala}
\item Max-Planck-Institut f\"ur Kernphysik, Saupfercheckweg 1, 69117 Heidelberg, Germany\label{AFFIL::GermanyMPIK}
\item Institut f\"ur Astronomie und Astrophysik, Universit\"at T\"ubingen, Sand 1, 72076 T\"ubingen, Germany\label{AFFIL::GermanyIAAT}
\item Department of Physics and Technology, University of Bergen, Museplass 1, 5007 Bergen, Norway\label{AFFIL::NorwayUBergen}
\item Department of Astronomy and Astrophysics, University of Chicago, 5640 S Ellis Ave, Chicago, Illinois, 60637, USA\label{AFFIL::USAUChicagoDAA}
\item Astroparticle Physics, Department of Physics, TU Dortmund University, Otto-Hahn-Str. 4a, 44227 Dortmund, Germany\label{AFFIL::GermanyUDortmundTU}
\item Escuela de Ingenier{\'\i}a El\'ectrica, Facultad de Ingenier{\'\i}a, Pontificia Universidad Cat\'olica de Valpara{\'\i}so, Avenida Brasil 2147, Valpara{\'\i}so, Chile\label{AFFIL::ChileEscIngElec}
\item Santa Cruz Institute for Particle Physics and Department of Physics, University of California, Santa Cruz, 1156 High Street, Santa Cruz, CA 95064, USA\label{AFFIL::USASCIPP}
\item Escola de Artes, Ci\^encias e Humanidades, Universidade de S\~ao Paulo, Rua Arlindo Bettio, CEP 03828-000, 1000 S\~ao Paulo, Brazil\label{AFFIL::BrazilEACHUSaoPaulo}
\item Astronomical Observatory of Taras Shevchenko National University of Kyiv, 3 Observatorna Street, Kyiv, 04053, Ukraine\label{AFFIL::UkraineAstObsofUKyiv}
\item Department of Physics and Astronomy, University of Utah, Salt Lake City, UT 84112-0830, USA\label{AFFIL::USAUUtah}
\item The University of Manitoba, Dept of Physics and Astronomy, Winnipeg, Manitoba R3T 2N2, Canada\label{AFFIL::CanadaUManitoba}
\item RIKEN, Institute of Physical and Chemical Research, 2-1 Hirosawa, Wako, Saitama, 351-0198, Japan\label{AFFIL::JapanRIKEN}
\item INFN Sezione di Roma La Sapienza, P.le Aldo Moro, 2 - 00185 Roma, Italy\label{AFFIL::ItalyINFNRomaLaSapienza}
\item Western Sydney University, Locked Bag 1797, Penrith, NSW 2751, Australia\label{AFFIL::AustraliaUWesternSydney}
\item INAF - Istituto di Astrofisica e Planetologia Spaziali (IAPS), Via del Fosso del Cavaliere 100, 00133 Roma, Italy\label{AFFIL::ItalyIAPS}
\item Physics Program, Graduate School of Advanced Science and Engineering, Hiroshima University, 739-8526 Hiroshima, Japan\label{AFFIL::JapanUHiroshima}
\item Department of Physics, Nagoya University, Chikusa-ku, Nagoya, 464-8602, Japan\label{AFFIL::JapanUNagoya}
\item Department of Information Technology, Escuela Polit\'ecnica Superior, Universidad San Pablo-CEU, CEU Universities, Campus Montepr{\'\i}ncipe, Boadilla del Monte, Madrid 28668, Spain\label{AFFIL::SpainUniversidadSanPabloCEU}
\item INFN Sezione di Roma Tor Vergata, Via della Ricerca Scientifica 1, 00133 Rome, Italy\label{AFFIL::ItalyINFNRomaTorVergata}
\item Alikhanyan National Science Laboratory, Yerevan Physics Institute, 2 Alikhanyan Brothers St., 0036, Yerevan, Armenia\label{AFFIL::ArmeniaNSLAlikhanyan}
\item Centro Brasileiro de Pesquisas F{\'\i}sicas, Rua Xavier Sigaud 150, RJ 22290-180, Rio de Janeiro, Brazil\label{AFFIL::BrazilCBPF}
\item INAF - Osservatorio Astronomico di Palermo {\textquotedblleft}G.S. Vaiana{\textquotedblright}, Piazza del Parlamento 1, 90134 Palermo, Italy\label{AFFIL::ItalyOPalermo}
\item Universidad Andr\'es Bello, Av. Fern\'andez Concha 700, Las Condes, Santiago, Chile\label{AFFIL::ChileUAndresBello}
\item N\'ucleo de Astrof{\'\i}sica e Cosmologia (Cosmo-ufes) \& Departamento de F{\'\i}sica, Universidade Federal do Esp{\'\i}rito Santo (UFES), Av. Fernando Ferrari, 514. 29065-910. Vit\'oria-ES, Brazil\label{AFFIL::BrazilUFES}
\item Astrophysics Research Center of the Open University (ARCO), The Open University of Israel, P.O. Box 808, Ra{\textquoteright}anana 4353701, Israel\label{AFFIL::IsraelOpenUniversityofIsrael}
\item Department of Physics, The George Washington University, Washington, DC 20052, USA\label{AFFIL::USAGWUWashingtonDC}
\item CTAO, Via Piero Gobetti 93/3, 40129 Bologna, Italy\label{AFFIL::ItalyCTAOBologna}
\item CTAO, Science Data Management Centre (SDMC), Platanenallee 6, 15738 Zeuthen, Germany\label{AFFIL::GermanyCTAOSDMC}
\item Institute of Space Sciences (ICE, CSIC), and Institut d'Estudis Espacials de Catalunya (IEEC), and Instituci\'o Catalana de Recerca I Estudis Avan\c{c}ats (ICREA), Campus UAB, Carrer de Can Magrans, s/n 08193 Cerdanyola del Vall\'es, Spain\label{AFFIL::SpainICECSIC}
\item General Education Center, Yamanashi-Gakuin University, Kofu, Yamanashi 400-8575, Japan\label{AFFIL::JapanUYamanashiGakuin}
\item School of Physics, Chemistry and Earth Sciences, University of Adelaide, Adelaide SA 5005, Australia\label{AFFIL::AustraliaUAdelaide}
\item Sendai College, National Institute of Technology, 4-16-1 Ayashi-Chuo, Aoba-ku, Sendai city, Miyagi 989-3128, Japan\label{AFFIL::JapanNITSendaiHirose}
\item Department of Physics, Faculty of Engineering Science, Yokohama National University, Yokohama 240{\textendash}8501, Japan\label{AFFIL::JapanUYokohamaNational}
\item Palack\'y University Olomouc, Faculty of Science, Joint Laboratory of Optics of Palack\'y University and Institute of Physics of the Czech Academy of Sciences, 17. listopadu 1192/12, 779 00 Olomouc, Czech Republic\label{AFFIL::CzechRepublicUOlomouc}
\item Josip Juraj Strossmayer University of Osijek, Trg Ljudevita Gaja 6, 31000 Osijek, Croatia\label{AFFIL::CroatiaUOsijek}
\item CETEMPS Dipartimento di Scienze Fisiche e Chimiche, Universit\`a degli Studi dell{\textquoteright}Aquila and GSGC-LNGS-INFN, Via Vetoio 1, L{\textquoteright}Aquila, 67100, Italy\label{AFFIL::ItalyCETEMPSandUandINFNAquila}
\item Research Center for Advanced Particle Physics, Kyushu University, 744 Motooka, Nishi-ku, Fukuoka 819-0395, Japan\label{AFFIL::JapanUKyushu}
\item Chiba University, 1-33, Yayoicho, Inage-ku, Chiba-shi, Chiba, 263-8522 Japan\label{AFFIL::JapanUChiba}
\item Department of Earth and Space Science, Graduate School of Science, Osaka University, Toyonaka 560-0043, Japan\label{AFFIL::JapanUOsaka}
\item Astronomical Observatory, Jagiellonian University, ul. Orla 171, 30-244 Cracow, Poland\label{AFFIL::PolandUJagiellonian}
\item Landessternwarte, Zentrum f\"ur Astronomie  der Universit\"at Heidelberg, K\"onigstuhl 12, 69117 Heidelberg, Germany\label{AFFIL::GermanyLSW}
\item IRAP, Universit\'e de Toulouse, CNRS, CNES, UPS, 9 avenue Colonel Roche, 31028 Toulouse, Cedex 4, France\label{AFFIL::FranceIRAPUToulouse}
\item Astronomical Institute of the Czech Academy of Sciences, Bocni II 1401 - 14100 Prague, Czech Republic\label{AFFIL::CzechRepublicASU}
\item Faculty of Science and Engineering, Waseda University, Shinjuku, Tokyo 169-8555, Japan\label{AFFIL::JapanUWaseda}
\item School of Physics, Aristotle University, Thessaloniki, 54124 Thessaloniki, Greece\label{AFFIL::GreeceUThessaloniki}
\item Universit\"at Innsbruck, Institut f\"ur Astro- und Teilchenphysik, Technikerstr. 25/8, 6020 Innsbruck, Austria\label{AFFIL::AustriaUInnsbruck}
\item Nicolaus Copernicus Astronomical Center, Polish Academy of Sciences, ul. Bartycka 18, 00-716 Warsaw, Poland\label{AFFIL::PolandNicolausCopernicusAstronomicalCenter}
\item National Astronomical Observatory of Japan (NAOJ), Division of Science, 2-21-1, Osawa, Mitaka, Tokyo 181-8588, Japan\label{AFFIL::JapanNAOJ}
\item Institute of Particle and Nuclear Studies,  KEK (High Energy Accelerator Research Organization), 1-1 Oho, Tsukuba, 305-0801, Japan\label{AFFIL::JapanKEK}
\item INAF - Istituto di Astrofisica Spaziale e Fisica Cosmica di Palermo, Via U. La Malfa 153, 90146 Palermo, Italy\label{AFFIL::ItalyIASFPalermo}
\item School of Physics and Astronomy, University of Leicester, Leicester, LE1 7RH, United Kingdom\label{AFFIL::UnitedKingdomULeicester}
\item Universit\`a degli studi di Catania, Dipartimento di Fisica e Astronomia {\textquotedblleft}Ettore Majorana{\textquotedblright}, Via S. Sofia 64, 95123 Catania, Italy\label{AFFIL::ItalyUCatania}
\item Anton Pannekoek Institute/GRAPPA, University of Amsterdam, Science Park 904 1098 XH Amsterdam, The Netherlands\label{AFFIL::NetherlandsUAmsterdam}
\item Department of Physics and Astronomy, University of Turku, Finland, FI-20014 University of Turku, Finland\label{AFFIL::FinlandUTurku}
\item INFN Sezione di Trieste and Universit\`a degli Studi di Trieste, Via Valerio 2 I, 34127 Trieste, Italy\label{AFFIL::ItalyUandINFNTrieste}
\item Escuela Polit\'ecnica Superior de Ja\'en, Universidad de Ja\'en, Campus Las Lagunillas s/n, Edif. A3, 23071 Ja\'en, Spain\label{AFFIL::SpainUJaen}
\item Department of Astronomy, University of Geneva, Chemin d'Ecogia 16, CH-1290 Versoix, Switzerland\label{AFFIL::SwitzerlandUGenevaISDC}
\item Saha Institute of Nuclear Physics, A CI of Homi Bhabha National Institute, Kolkata 700064, West Bengal, India\label{AFFIL::IndiaSahaInstitute}
\item Department of Biology and Geology, Physics and Inorganic Chemistry, Address: C/ Tulip\'an, s/n. 28933 M\'ostoles (Faculty of Experimental Sciences and Technology)\label{AFFIL::SpainReyJuanCarlosUniversity}
\item UCM-ELEC group, EMFTEL Department, University Complutense of Madrid, 28040 Madrid, Spain\label{AFFIL::SpainUCMElectronica}
\item Universidade Tecnol\'ogica Federal do Paran\'a, Av. Sete de Setembro, 3165 - Rebou\c{c}as CEP 80230-901 - Curitiba - PR - Brasil\label{AFFIL::BrazilUTFPR}
\item Macroarea di Scienze MMFFNN, Universit\`a di Roma Tor Vergata, Via della Ricerca Scientifica 1, 00133 Rome, Italy \label{AFFIL::ItalyURomaTorVegata}
\item Universit\"at Hamburg, Institut f\"ur Experimentalphysik, Luruper Chaussee 149, 22761 Hamburg, Germany\label{AFFIL::GermanyUHamburg}
\item Kavli Institute for Particle Astrophysics and Cosmology, Stanford University, Stanford, CA 94305, USA\label{AFFIL::USAStanford}
\item Kavli Institute for Particle Astrophysics and Cosmology, Department of Physics and SLAC National Accelerator Laboratory, Stanford University, 2575 Sand Hill Road, Menlo Park, CA 94025, USA\label{AFFIL::USASLAC}
\item IPARCOS Institute, Faculty of Physics (UCM), 28040 Madrid, Spain\label{AFFIL::SpainIPARCOSInstitute}
\item School of Allied Health Sciences, Kitasato University, Sagamihara, Kanagawa 228-8555, Japan\label{AFFIL::JapanUKitasato}
\item Department of Physics, Yamagata University, Yamagata, Yamagata 990-8560, Japan\label{AFFIL::JapanUYamagata}
\item The Henryk Niewodnicza\'nski Institute of Nuclear Physics, Polish Academy of Sciences, ul. Radzikowskiego 152, 31-342 Cracow, Poland\label{AFFIL::PolandIFJ}
\item University of Bia{\l}ystok, Faculty of Physics, ul. K. Cio{\l}kowskiego 1L, 15-245 Bia{\l}ystok, Poland\label{AFFIL::PolandUBiaystok}
\item Kanagawa University, Faculty of Engineering, 3-27-1 Rokukakubashi, Kanagawa-ku, Yokohama-shi, Kanagawa 221-8686, Japan\label{AFFIL::JapanUKanagawaFacofEngineering}
\item Charles University, Institute of Particle \& Nuclear Physics, V Hole\v{s}ovi\v{c}k\'ach 2, 180 00 Prague 8, Czech Republic\label{AFFIL::CzechRepublicUPrague}
\item Institute for Space{\textemdash}Earth Environmental Research, Nagoya University, Furo-cho, Chikusa-ku, Nagoya 464-8601, Japan\label{AFFIL::JapanUNagoyaISEE}
\item Kobayashi{\textemdash}Maskawa Institute for the Origin of Particles and the Universe, Nagoya University, Furo-cho, Chikusa-ku, Nagoya 464-8602, Japan\label{AFFIL::JapanUNagoyaKMI}
\item Graduate School of Technology, Industrial and Social Sciences, Tokushima University, Tokushima 770-8506, Japan\label{AFFIL::JapanUTokushima}
\item University of Pisa, Largo B. Pontecorvo 3, 56127 Pisa, Italy \label{AFFIL::ItalyUPisa}
\item INAF - Osservatorio Astronomico di Padova, Vicolo dell'Osservatorio 5, 35122 Padova, Italy\label{AFFIL::ItalyOPadova}
\item INAF - Osservatorio Astronomico di Padova and INFN Sezione di Trieste, gr. coll. Udine, Via delle Scienze 208 I-33100 Udine, Italy\label{AFFIL::ItalyOandINFNTrieste}
\item Institute for Theoretical Physics and Astrophysics, Universit\"at W\"urzburg, Campus Hubland Nord, Emil-Fischer-Str. 31, 97074 W\"urzburg, Germany\label{AFFIL::GermanyUWurzburg}
\item Dipartimento di Scienze Fisiche e Chimiche, Universit\`a degli Studi dell'Aquila and GSGC-LNGS-INFN, Via Vetoio 1, L'Aquila, 67100, Italy\label{AFFIL::ItalyUandINFNAquila}
\item Departamento de F{\'\i}sica, Facultad de Ciencias B\'asicas, Universidad Metropolitana de Ciencias de la Educaci\'on, Avenida Jos\'e Pedro Alessandri 774, \~Nu\~noa, Santiago, Chile\label{AFFIL::ChileUMCE}
\item Instituto de Estudios Astrof{\'\i}sicos, Facultad de Ingenier{\'\i}a y Ciencias, Universidad Diego Portales, Av. Ej\'ercito Libertador 441, 8370191 Santiago, Chile\label{AFFIL::ChileUniversidadDiegoPortales}
\item Departamento de Astronom{\'\i}a, Universidad de Concepci\'on, Barrio Universitario S/N, Concepci\'on, Chile\label{AFFIL::ChileUdeConcepcion}
\item Hiroshima Astrophysical Science Center, Hiroshima University, Higashi-Hiroshima, Hiroshima 739-8526, Japan\label{AFFIL::JapanHASC}
\item Gifu University, Faculty of Engineering, 1-1 Yanagido, Gifu 501-1193, Japan \label{AFFIL::JapanUGifu}
\item Laboratory for High Energy Physics, \'Ecole Polytechnique F\'ed\'erale, CH-1015 Lausanne, Switzerland\label{AFFIL::SwitzerlandEPFLausanne}
\item Departamento de F{\'\i}sica, Universidad de Santiago de Chile (USACH), Av. Victor Jara 3493, Estaci\'on Central, Santiago, Chile\label{AFFIL::ChileUniversidaddeSantiagodeChile}
\item Main Astronomical Observatory of the National Academy of Sciences of Ukraine, Zabolotnoho str., 27, 03143, Kyiv, Ukraine\label{AFFIL::UkraineObsNASUkraine}
\item Faculty of Space Technologies, AGH University of Krakow, Aleja Mickiewicza 30, Krak\'ow 30-059, Poland\label{AFFIL::PolandAGHCracowSTC}
\item Academic Computer Centre CYFRONET AGH, ul. Nawojki 11, 30-950, Krak\'ow, Poland\label{AFFIL::PolandCYFRONETAGH}
\item Observat\'orio Nacional, Rio de Janeiro - RJ, 20921-400, Brazil\label{AFFIL::BrazilNationalObservatoryRiodeJaneiro}
\item Warsaw University of Technology, Faculty of Electronics and Information Technology, Institute of Electronic Systems, Nowowiejska 15/19, 00-665 Warsaw, Poland\label{AFFIL::PolandWUTElectronics}
\item Department of Physical Sciences, Aoyama Gakuin University, Fuchinobe, Sagamihara, Kanagawa, 252-5258, Japan\label{AFFIL::JapanUAoyamaGakuin}
\item School of Physics and Astronomy, Sun Yat-sen University, Zhuhai, China\label{AFFIL::ChinaUSunYatsen}
\item Curtin University, Kent St, Bentley WA 6102, Australia\label{AFFIL::AustraliaUCurtin}
\item Institute for Nuclear Research and Nuclear Energy, Bulgarian Academy of Sciences, 72 boul. Tsarigradsko chaussee, 1784 Sofia, Bulgaria\label{AFFIL::BulgariaINRNEBAS}
\item Port d'Informaci\'o Cient{\'\i}fica, Edifici D, Carrer de l'Albareda, 08193 Bellaterrra (Cerdanyola del Vall\`es), Spain\label{AFFIL::SpainPIC}
\item INAF - Osservatorio Astronomico di Cagliari, Via della Scienza 5, I-09047 Selargius (CA), Italy\label{AFFIL::ItalyINAFCagliari}
\item Kapteyn Astronomical Institute, University of Groningen, Landleven 12, 9747 AD, Groningen, The Netherlands\label{AFFIL::NetherlandsUGroningen}
\item Departamento de F{\'\i}sica, Universidad T\'ecnica Federico Santa Mar{\'\i}a, Avenida Espa\~na, 1680 Valpara{\'\i}so, Chile\label{AFFIL::ChileDepFisUTecnicaFedericoSantaMaria}
\end{enumerate}
\end{scriptsize}


\bsp	
\label{lastpage}
\end{document}